\documentclass[defaultstyle,12pt]{thesis}
\usepackage{algpseudocode}
\usepackage{algorithm}
\usepackage{physics}
\usepackage{dcolumn} 
\usepackage{bm}       
\usepackage{amsmath,amssymb,amsfonts,amsthm}
\usepackage{mathtools}
\usepackage[mathscr]{eucal}
\usepackage{xcolor}
\usepackage{tikz}
\usetikzlibrary{matrix,arrows,decorations.pathmorphing}
\usepackage[utf8]{inputenc}
\usepackage{xparse}
\usepackage{hyperref}
\usepackage{cleveref}
\usepackage{listings}
\usepackage{color}
\usepackage{comment}
\usepackage{dsfont}
\usepackage{url}
\usepackage{stmaryrd} % knuth notation
\usepackage{graphicx,xcolor}
\hypersetup{colorlinks=true,citecolor=blue,linkcolor=blue,filecolor=blue,urlcolor=blue}
\providecommand{\ignore}[1]{}

\newif\ifcmnt
\cmnttrue
\ifdefined\cmntsoff\cmntfalse\fi

\ifcmnt

\newtheorem{thm}{Theorem}[section]
\newtheorem{prop}[thm]{Proposition}
\newtheorem{lem}[thm]{Lemma}
\newtheorem{conj}[thm]{Conjecture}
\newtheorem{cor}[thm]{Corollary}

\theoremstyle{definition}
\newtheorem{definition}[thm]{Definition}
\newtheorem{remark}[thm]{Remark}
\newtheorem{example}[thm]{Example}
\numberwithin{equation}{section}

\renewcommand\bra[1]{{\langle{#1}|}}
\renewcommand\ket[1]{{|{#1}\rangle}}
\newcommand{\id}{\mathrm{id}}

\newcommand{\cX}{\mathcal{X}}
\newcommand{\cY}{\mathcal{Y}}

\newcommand{\cP}{\mathcal{P}}

\title{Some Problems Concerning Quantum Channels and Entropies}

\author{Mohammad A. Alhejji}{}

\otherdegrees{B.A., University of Pennsylvania, 2017 \\
	      M.S., University of Pennsylvania, 2017}

\degree{Doctor of Philosophy}		%  #1 {long descr.}
	{Ph.D., Physics}		%  #2 {short descr.}

\dept{Department of}			%  #1 {designation}
	{Physics}		%  #2 {name}

\advisor{Dr.}				%  #1 {title}
	{Emanuel Knill}			%  #2 {name}

\reader{Graeme Smith}	
\readerThree{Josh Combes}
\readerFour{Scott Glancy} 
\readerFive{Ana Maria Rey}

\abstract{  

       This thesis describes contributions to three mathematical problems in quantum information theory. The first concerns the continuity properties of conditional entropy, which gives optimal rates for several information processing tasks including classical source compression with side information and quantum state merging. We prove a tight uniform continuity bound for the conditional Shannon entropy in terms of total variation distance. This bound may be interpreted as a continuity bound for the conditional entropy of classical states. We also include observations about a conjectured continuity bound for the conditional entropy of general quantum states. The second relates to a general class of optimization problems inspired by the following situation. Suppose the entropy of a given mixture of states is to be minimized by choosing each state subject to constraints. If the spectrum of each state is fixed, we expect that to reduce the entropy of the mixture, we should make the states less distinguishable in some sense. This situation arises for example in the recently introduced spin alignment problem, which is the main motivation for our investigation. In its original form, each state in the mixture is constrained to be a freely chosen state on a subset of \(n\) qubits tensored with a fixed state $Q$ on each of the qubits in the complement. According to the spin alignment conjecture, the entropy of the mixture is minimized by choosing the freely chosen state in each term to be a tensor product of projectors onto a fixed maximal eigenvector of $Q$, which maximally ``aligns'' the states in the mixture. We generalize the spin alignment problem in several ways. First, instead of minimizing entropy, we consider maximizing arbitrary unitarily invariant convex functions such as Fan norms and Schatten norms. To formalize and generalize the conjectured required alignment, we define \textit{alignment} as a preorder on tuples of self-adjoint operators induced by majorization. We prove the generalized conjecture for Schatten norms of integer order, for the case where the freely chosen states are constrained to be classical, and for the case where only two states contribute to the mixture and $Q$ is proportional to a projector. The last case fits into a more general situation where we give explicit conditions for maximal alignment. The spin alignment problem has a natural “dual” formulation, versions of which have further generalizations that we introduce. The third is about quantum channel simulation in the specific case where the channel to be simulated is a quantum erasure channel, a task we dub quantum erasure simulation. We show that quantum erasure simulation is closely related to probabilistic quantum error correction. We examine a conjecture put forward by M.B. Hastings that says that erasure channels with erasure probability exceeding a half cannot be used to simulate better erasure channels. We use a combinatorial argument to show that the conjecture is true in the case where the encoding channel is an isometric embedding into a stabilizer code. We also show that it is true if the number of channel uses is not greater than four.
	}

\dedication[Dedication]{	% NEVER use \OnePageChapter here.
	To my family.
	}

\acknowledgements{	\OnePageChapter	% *MUST* BE ONLY ONE PAGE!
	I thank my advisors Manny Knill and Graeme Smith for their guidance and support over the course of my graduate career. This thesis was significantly improved because of their input. I am solely responsible for any infelicities in it. 
    
    I would also like to thank Scott Glancy and Felix Leditzky for their mentorship and generosity. I acknowledge my other collaborators who are affiliates of NIST and JILA including Joshua Levin, Yanbao Zhang, Krister Shalm, Eli Halperin, Shawn Geller, Alex Kwiatkowski, Arik Avagyan, and Vikesh Siddhu. 

    I am grateful to my parents Abdulkareem and Aljohara for their love and patience, and to my siblings Khalid, Faris, Joud, Rayan, and Faisal for entertaining me and engendering much-needed levity. 

    I would like to thank the constellation of friends who helped me get through this time. Among them are: Abdullah alshaffi, Mohammad Alnamlah, Tung Chau, Melissa Diamond, Elizabeth Dresselhaus, Kaila Fluss, Neha Manohar, Shankar Ram, Julia Slater, Jose Valencia, and Maxwell Urmey. 

    Finally, I do not believe it is possible for me to include everybody who ought to be included. To insulate myself, I declare that I am grateful for people. 
	}

\ToCisShort	% use this only for 1-page Table of Contents

\LoFisShort	% use this only for 1-page Table of Figures
\LoTisShort	% use this only for 1-page Table of Tables
\emptyLoT	% use this if there is no List of Tables

\begin{document}

\chapter{Introduction and overview}
\label{chapter: 1}
% introduction part here
At the current moment, the physics research community is, by and large, captivated by the prospect of using quantum mechanics to process information. This prospect fascinates people outside this community, even those who are not researchers. Underlying this fascination is the expectation that quantum-based information processing systems will outperform currently available ones. That is, it is expected that fully considering and utilizing quantum theory in the design and operation of information processing systems is \textit{advantageous} when it comes to some relevant tasks. This so-called quantum advantage was demonstrated rigorously in some cases. An example is Shor's quantum algorithm for integer factoring in the field of computer science \cite{Shor1994}, which offers a super-polynomial speedup over presently known classical algorithms. Another is certified randomness generation and secret key distribution based on quantum protocols in the field of cryptography \cite{Acin2016, Scarani2009}. In other cases, however, the basis for claims of quantum advantage is heuristics. This is especially true for quantum algorithms, where the reasons for the existence or lack of quantum advantage remain an active area of research. In all cases, many of the exciting applications of quantum information science, both within the context of research and without, seem to be decades away. In contrast, I would argue that theoretical quantum information science has reached a mature stage, at least at the level of basic research. I believe that one viable path towards making progress at such a stage is by using open problems, usually left unresolved due to their difficulty, as inspiration for establishing novel analysis tools. Underlying this belief is the hope that the usefulness of these tools extends beyond their origin.

Quantum information science is multidisciplinary. It includes disciplines such as quantum information theory, quantum computer science, quantum cryptography, quantum metrology, and many others. The contents of this thesis concern problems that arose in lines of inquiry in quantum information theory. Quantum information theory is the study of the quantum-physical limits on rates of protocols for faithful storage, transmission, and interconversion of information. It generalizes classical information theory, which grew out of Shannon's landmark work in the $1940$s \cite{Shannon1948}, and like classical information theory, it has proven to be a worthwhile endeavor with practical implications.

It is typical in information theory that the fundamental limits on rates for asymptotically error-free protocols are given by entropic formulas, which are mathematical expressions involving the entropies of states of systems under consideration. An example of this is entanglement distillation from pure states. Suppose that $\ket{\Theta}_{A B}$ is a bipartite quantum state of the spin of two spin $1/2$ particles $A$ and $B$. The state may exhibit entanglement, the property that it cannot be written as a product of two states of $A$ and $B$. The amount of entanglement, however, may not be maximal. Maximally entangled states serve as a powerful resource in quantum information science. For example, a maximally entangled state can be converted to any bipartite state of the same dimension using only local operations and classical communication (LOCC). So, we may ask if it is possible to use LOCC to convert some number $n$ of copies of the state $\ket{\Theta}_{A B}$ to $k$ copies of a maximally entangled state, say the EPR pair singlet state $\ket{\Psi^{-}}_{A B} = \frac{1}{\sqrt{2}} (\ket{0}_{A} \otimes \ket{1}_{B} - \ket{1}_{A} \otimes \ket{0}_{B})$. Furthermore, we may find acceptable a protocol whose output is $\epsilon$-close to $\ket{\Psi^{-}}_{A B}^{\otimes k}$ according to some suitable distance measure for some sufficiently small $\epsilon$. In such a case, the ratio $\frac{k}{n}$ is the rate of the protocol and $\epsilon$ is the error of the protocol. Given $R > 0$, does there exist a sequence of protocols with an asymptotic rate $R$ and asymptotically vanishing error? If yes, then $R$ is said to be an achievable rate for faithful entanglement distillation using the state $\ket{\Theta}_{A B}$. The complete answer to this information-theoretic question was given by Bennett et al. in \cite{Bennett1996}. Specifically, they showed that a rate $R$ is achievable if and only if it does not exceed the von Neumann entropy of the marginal state $\Theta_{A}$. These are two statements. The first is that every $R > 0$ equal or lower than the entropy of $\Theta_{A}$ is achievable. This is an example of what is referred to as an achievability result or a \textit{direct} statement in information theory. The second statement is that no $R$ above the entropy of $\Theta_{A}$ is achievable. That is, if the rate is too high, then the error must be bounded away from zero. This is an example of a \textit{converse} statement in information theory.

Quantum channels serve as a powerful model for describing the evolution of quantum systems. They can be used to model what happens to quantum systems as they travel across space, such as photon loss in optical fibers. They are frequently used in the modeling of error processes in architectures for quantum computers. Associated with each quantum channel $\mathcal{N}$ is a collection of \textit{capacities}. These are numbers giving the fundamental limits on rates for faithful communication of various kinds of information through the channel \cite{Smith2010}. An example is the quantum capacity $\mathcal{Q}(\mathcal{N})$. It is the highest rate at which quantum information can be transmitted through the channel with fidelity arbitrarily close to one. Fidelity here refers to the entanglement fidelity \cite{Barnum2000}. The direct quantum coding theorem, proven for example in \cite{Devetak2005}, says that for a given channel $\mathcal{N}$ the coherent information $I^{\text{(coh)}} (\mathcal{N})$ is an achievable rate for faithful quantum communication. $I^{\text{(coh)}} (\mathcal{N})$ is given by a maximization of an entropic formula over the channel input. Conversely, as was realized earlier \cite{Barnum1998}, any sequence of protocols for quantum communication with rate exceeding $\lim_{n \rightarrow \infty} \frac{1}{n} I^{\text{(coh)}} (\mathcal{N}^{\otimes n})$ must have fidelity bound away from one. Taken together, the two imply a mathematical expression for the quantum capacity
\begin{align}
\label{eq: quantum capacity}
\mathcal{Q} (\mathcal{N}) = \lim_{n \rightarrow \infty} \frac{1}{n} I^{\text{(coh)}} (\mathcal{N}^{\otimes n}).
\end{align}
The expression on the right-hand side is known as a regularized or multi-letter formula. At a first glance, it is not obvious if there is an algorithm that can be used to approximate the limit to some arbitrary precision for general quantum channels. In fact, it is known that the monotonic sequence $(\frac{1}{n} I^{\text{(coh)}} (\mathcal{N}^{\otimes n}))_{n \in \mathbb{N}}$---a sequence of optimal values for optimization problems over states of larger and larger dimension---can have arbitrarily many zeros before becoming positive \cite{Cubitt2015}. The issue of computability of capacities arises as well in the setting of classical communication, both private and otherwise, through quantum channels where only regularized formulas are known \cite{Smith2010}. In contrast, efficient algorithms have long been known and used for computing the capacities of classical channels \cite{Arimoto1972, Blahut1972}. These facts motivate the study of approximating and optimizing entropic formulas over relevant classes of quantum states, which I undertake in this thesis.

I also study quantum erasure channels and whether they can be simulated with a given resource channel, to be used possibly multiple times. The quantum erasure channel is one of the most well-understood and often-used error models for quantum information processing. Just like its classical analog, it models situations where errors are accompanied by flags that allow receivers to locate the erroneous information-carrying components. Also like its classical analog, it is ``the pedagogical noisy channel \textit{par excellence}" \cite{Elias1974}. Quantities that are often difficult to compute such as the quantum capacity, private capacity, or classical capacity have simple formulas in the case of the erasure channel. Moreover, quantum erasure, as classical erasure, can be used to define the notion of code distance in quantum coding theory \cite{rains1998}. However, comparatively little is known about it in the high noise regime where the erasure probability exceeds $\frac{1}{2}$ and the quantum capacity vanishes. As explained above, it is not possible to use such a channel to communicate quantum information at a non-vanishing rate with fidelity arbitrarily close to 1. But, supposing the channel is used, we might ask: how far away from 1 is the fidelity? A channel is said to satisfy the strong converse property if it is known that the fidelity of any sequence of protocols whose rate exceeds the capacity of the channel tends to zero. This property precludes situations where some non-maximal amount of infidelity is tolerable. It is not known if quantum erasure channels have a strong converse for quantum communication. However, the question is related to the notions of probabilistic quantum error correction and quantum erasure simulation. In particular, given a quantum erasure channel with zero quantum capacity, is it possible to use it to simulate a better quantum erasure channel? Clearly, it is not possible to simulate a channel with nonzero quantum capacity. What if the simulated channel has zero quantum capacity but a lower probability of error than the resource channel? If the answer is no, then the fidelity of quantum communication through such channels using this particular class of protocols cannot be made to slow down in its approach to zero. I collect in this thesis a number of observations concerning these questions.

% The contributions of this thesis
The main contributions of this thesis are as follows. In Ch.~\ref{chapter: 3}, we prove Thm.~\ref{thm: tight uni bound for equivocation}, which gives a tight uniform continuity bound for the conditional Shannon entropy in terms of total variation distance. This bound was found useful by machine learning researchers \cite{diaz2019theoretical}. A simple generalization of the proof of Thm.~\ref{thm: tight uni bound for equivocation} yields a tight uniform continuity bound for the conditional entropy of quantum-classical states in terms of trace distance \cite{Wilde2020}. Following an argument in \cite{Winter_2016}, Wilde used this generalized bound to get sharp variation estimates for the entanglement of formation, the regularization of which gives the entanglement cost. In Ch.~\ref{chapter: 4}, we prove Thm.~\ref{thm: spin alignment for Schatten norms of integer order} and Thm.~\ref{thm: two projector fan norm}. The proofs of both theorems crucially rely on the overlap lemma (Lem.~\ref{overlap lemma}). The first theorem gives a resolution for the spin alignment problem where the objective function is a Schatten norm of integer order.  The second gives a characterization of maximal alignment (see Def.~\ref{def: partial alignment}) of two projectors $P_{1}, P_{2}$ of specified ranks that are subject to a constraint of the form $\tr( | P_{1} P_{2} | ) \leq c$. We use this characterization to resolve the spin alignment problem for Fan norms of mixtures of two projectors. In Ch.~\ref{chapter: 5}, we prove Thm.~\ref{thm: deterministic codes do not help}. It says that isometric encodings into a particular class of codes, which we call deterministic and which include stabilizer codes, cannot be used with a quantum erasure channel with zero quantum capacity to simulate a better quantum erasure channel.

\chapter{Preliminaries}
\label{chapter: 2}

\section{Basic notions}
\label{sec: ch2sec1}
For a natural number $n \in \mathbb{N}$, the set \(\{1,\ldots, n\)\} is denoted by $[n]$ and its power set is denoted by $2^{[n]}$. 
We use the symbols $A, B$, etc as well as subsets of $[n]$ to refer to physical systems. Vectors, operators, and other mathematical objects are sometimes written with subscripts to associate them with physical systems. In general, we do not write the curly braces of sets in subscripts.

In quantum theory, physical phenomena are described using complex Hilbert spaces. We restrict our attention to the case where such spaces have finite dimensions. However, the problems we consider can be suitably generalized to the infinite-dimensional case.

A complex inner product space is a vector space over $\mathbb{C}$ equipped with an inner product. A Hilbert space is an inner product space that is complete with respect to the metric induced by the inner product. A finite-dimensional inner product space over $\mathbb{C}$ is always a Hilbert space. Hereinafter, we use the term inner product space to mean a complex Hilbert space.  We denote such spaces by $\mathcal{H},\mathcal{K}$, etc. We denote the statement that $\mathcal{C}$ is a subspace of $\mathcal{H}$ by $\mathcal{C} \leq \mathcal{H}$. The image of a subspace $\mathcal{C}$ under a linear map $L$ is denoted by $L \mathcal{C}$.

We use bra-ket notation throughout and adopt the physicists' convention that inner products are linear in the second argument and anti-linear in the first. The dimension of $\mathcal{H}$ is denoted with $d_{\mathcal{H}}$. As is common in the quantum information science literature, we refer to a physical system modeled based on a $d$-dimensional space as a \textit{qudit}. If $d = 2$, we use the term \textit{qubit}, and if $d =3$, we use the term \textit{qutrit}. We denote the algebra of bounded operators on $\mathcal{H}$ by $\mathcal{B} (\mathcal{H})$. The identity operator on $\mathcal{H}$ is denoted by $\mathds{1}_{\mathcal{H}}$. For $R \in \mathcal{B} (\mathcal{H})$, $\sigma(R)$ is the real $d_{\mathcal{H}}$-dimensional vector whose entries are the singular values of $R$ ordered non-increasingly. We denote the orthogonal complement to the kernel of an operator $R \in \mathcal{B} (\mathcal{H})$ by $\text{supp} (R)$. We denote the real space of self-adjoint or Hermitian operators on $\mathcal{H}$ by $\mathcal{S} (\mathcal{H})$. A self-adjoint operator is commonly referred to as an \textit{observable}. For $S \in \mathcal{S} (\mathcal{H})$, $\lambda(S)$ is the real $d_{\mathcal{H}}$-dimensional vector whose entries are the eigenvalues of $S$ ordered non-increasingly. We denote the set of positive semi-definite operators on $\mathcal{H}$ by $\mathcal{P} (\mathcal{H})$. The statement $S \in \mathcal{P} (\mathcal{H})$ is denoted by $S \geq 0$. For $S, R \in \mathcal{S} (\mathcal{H})$, $S \geq R$ denotes the statement that $S - R \in \mathcal{P} (\mathcal{H})$. In this thesis, projectors are always self-adjoint projection operators. The projector onto some subspace $\mathcal{C} \leq \mathcal{H}$ is denoted by $P_{\mathcal{C}}$. A \textit{state}, sometimes a quantum state, $\rho: \mathcal{B} (\mathcal{H}) \rightarrow \mathbb{C}$ is a positive, continuous linear functional that satisfies $\rho (\mathds{1}_{\mathcal{H}}) = 1$. In the state $\rho$, the expectation value of an observable $A$ is given by the formula $\rho(A)$. The set of states is a convex set, and in the case of finite dimensional $\mathcal{H}$, it can be identified with the set of unit trace positive semi-definite operators, which we denote by $\mathcal{D} (\mathcal{H})$. The state corresponding to $\phi \in \mathcal{D} (\mathcal{H})$ has the action $\tr ( \phi \; (\cdot) )$. The extrema of the set of states are called pure states and they correspond with rank-$1$ projectors. Sometimes, as is common in the literature, we refer to a unit vector in the support of a pure state as a state. From now on, we make no distinction between the set of states on $\mathcal{H}$ and $\mathcal{D} (\mathcal{H})$. A state that is proportional to the identity is called maximally mixed. A linear map $\mathcal{L}: \mathcal{B} (\mathcal{H}) \rightarrow \mathcal{B} (\mathcal{K})$ is called a super-operator. The identity super-operator on $\mathcal{B} (\mathcal{K})$ is denoted by $\id_{\mathcal{K}}$. The qubit Pauli operators have the matrix representation 
\\
\begin{align}
\label{eq: paulis} I = \begin{pmatrix}
1 & 0\\
0 & 1
\end{pmatrix}, \;
X = \begin{pmatrix} 
0 & 1\\
1 & 0\end{pmatrix}, \;
Y = \begin{pmatrix}
0 & -i\\
i & 0\end{pmatrix},\;
Z = \begin{pmatrix}
1 & 0\\
0 & -1\end{pmatrix}.\; 
\end{align}
\\
The qubit Pauli group is the group generated by the Pauli operators and we denote it with $G_{1}$.

Multiple physical systems are jointly described using tensor products. If the state of a system $A$ acts on $\mathcal{H}_{A}$ and the state of a system $B$ acts on $\mathcal{H}_{B}$, then the state of the \textit{composite} system $A B$ acts on $\mathcal{H}_{A} \otimes \mathcal{H}_{B}$. When considering a bipartite composite system $A B$, the partial trace of a system $B$ is the unique super-operator $\tr_{B} : \mathcal{B} (\mathcal{H}_{A} \otimes \mathcal{H}_{B}) \rightarrow \mathcal{B} (\mathcal{H}_{A})$ that satisfies
\begin{align}
\label{eq: partial trace}
\tr ( \rho_{A B} ( M_{A} \otimes \mathds{1}_B ) ) = \tr ( \tr_{B} ( \rho_{A B} ) M_{A} )
\end{align}
for all states $\rho_{A B}$ and observables $M_{A}$. Given a bipartite state $\rho_{A B}$, the state of subsystem $A$ is given by the partial trace $\rho_{A} := \tr_{B} (\rho_{A B})$. A state of a bipartite system is \textit{maximally entangled} if it is pure and one of its two partial traces is a maximally mixed state. The weight of a tensor product operator is the number of factors in it that are not equal to the identity. The $n$-qubit Pauli group, denoted by $G_{n}$, is the group acting on $(\mathbb{C}^{2})^{\otimes n}$ generated by all $n$-fold tensor products of qubit Pauli operators.

 Evolution and change in quantum theory are captured by linear maps. A super-operator $\mathcal{L} $ is positive if $\phi \geq 0 \implies \mathcal{L} (\phi)\geq 0$. $\mathcal{L}$ is completely positive if $\id_{\mathcal{K}} \otimes \mathcal{L}$ is positive for all $\mathcal{K}$. In order to preserve the positivity of composite system states, physical super-operators are required to be completely positive. Furthermore, physical super-operators are generally taken to be trace-preserving. A completely positive trace-preserving super-operator is called a quantum channel or sometimes simply a channel. A completely positive super-operator that is also trace non-increasing, in that it does not increase the trace of positive semi-definite operators, is called a quantum instrument. These generalize measurements in quantum theory. The only kind of measurement considered in this thesis is non-destructive and projective. A set of mutually orthogonal projectors $\{P_{i}\}_{i=1}^{\ell}$ is called a projector valued measure if $\sum_{i=1}^{\ell} P_{i} = \mathds{1}$. Given $\rho \in \mathcal{D} (\mathcal{H})$ and a projector valued measure $\mathcal{P} = (P_{i})_{i=1}^{\ell}$, the operator $\sum_{i = 1}^{\ell} P_{i} \rho P_{i} \otimes \ket{i}\bra{i}_{C}$ captures the process of measuring $\mathcal{P}$. For $i \in [\ell]$, the probability of measuring $i$ is $\tr(\rho P_{i})$ by the Born rule. The state of the system after measuring $i$ is given by $\frac{P_{i} \rho P_{i}}{\tr(\rho P_{i})}$.

 We make use of three characterizations of completely positive super-operators. The first was given by Choi in \cite{CHOI1975}. He showed that a super-operator $\mathcal{L} : \mathcal{B} (\mathcal{H}) \rightarrow \mathcal{B} (\mathcal{K} )$ is completely positive if and only if there exists a set of operators $\{L_{i}\}_{i=1}^{k}$ each with domain $\mathcal{H}$ and codomain $\mathcal{K}$ such that 
\begin{align}
\label{eq: cp maps kraus}
\mathcal{L} (\cdot) = \sum_{i = 1}^{k} L_{i} \; ( \cdot ) \; L_{i}^{*}.
\end{align}
This generally non-unique expression is called a Kraus decomposition of $\mathcal{L}$ and each $L_{i}$ is called a Kraus operator. $\mathcal{L} \sim \{L_{i}\}_{i=1}^{k}$ denotes the statement that $\mathcal{L}$ has a Kraus decomposition via the operators $\{L_{i}\}_{i=1}^{k}$. Different Kraus decompositions of the same completely positive super-operator are related by an isometry. That is, $\mathcal{L} \sim  \{L_{i}\}_{i=1}^{k}$ and $\mathcal{L} \sim \{ \Tilde{L}_{j}\}_{j=1}^{\ell}$ with $k \leq \ell$ if and only if there exists an isometry $m: \mathbb{C}^{k} \rightarrow \mathbb{C}^{\ell}$ with coefficient matrix $(m_{j, i})_{i=1, j=1}^{i=k, j=\ell}$ such that 
\begin{align}
\label{eq: Kraus freedom}
\Tilde{L}_{j} = \sum_{i=1}^{k} m_{j, i} L_{i}.
\end{align}
If $\mathcal{L} \sim  \{L_{i}\}_{i=1}^{k}$ and $\sum_{i=1}^{k} L^{*}_{i} L_{i} \leq \mathds{1}_{\mathcal{H}}$, then $\mathcal{L}$ is trace non-increasing, and if $\sum_{i=1}^{k} L^{*}_{i} L_{i} = \mathds{1}_{\mathcal{H}}$, then it is trace-preserving. Hence, a channel admits a Kraus decomposition with only one Kraus operator if and only if the operator is an isometry. In such a case, the channel is called isometric. If its domain and codomain are equal, it is called unitary. A channel $\mathcal{N}$ is mixed-unitary if $\mathcal{N} \sim \{ \sqrt{p_{i}} V_{i} \}_{i}$ for some probability measure $(p_{i})_{i}$ and tuple of unitary operators $(V_{i})_{i}$. The number $p_{i}$ is interpreted as the probability of the unitary $V_{i}$ occurring.

The second characterization is due to Stinespring \cite{Stinespring1955}. It applies more generally, but we will only need it for the case of quantum channels acting on $\mathcal{B} (\mathcal{H})$ for some finite-dimensional $\mathcal{H}$. A super-operator $\mathcal{N} : \mathcal{B} (\mathcal{H}) \rightarrow \mathcal{B} (\mathcal{K}_{O})$ is a quantum channel if and only if there exists an auxiliary space $\mathcal{K}_{E}$ and an isometry $U_{\mathcal{N}} : \mathcal{H} \rightarrow \mathcal{K}_{O} \otimes \mathcal{K}_{E}$ such that 
\begin{align}
    \mathcal{N} (\cdot) = \tr_{{E}} \circ \; U_{\mathcal{N}} \; (\cdot) \; U_{\mathcal{N}}^{*}.
\end{align} 
The isometric channel $\mathcal{U}_{\mathcal{N}} \sim \{ U_{\mathcal{N}} \}$ is called an isometric extension of $\mathcal{N}$ and the system $E$ is interpreted as an environment of the channel. The channel to the environment $\mathcal{N}^{c} := \tr_{O} \circ \; \mathcal{U}_{\mathcal{N}}$ is called a complementary channel to $\mathcal{N}$.

The third characterization is via Choi isomorphisms \cite{CHOI1975}. These are closely related to the more natural Jamiołkowski isomorphisms \cite{JAMIOLKOWSKI1972} but the latter do not concern us. Let $\mathcal{H}_{R}$ be isomorphic to $\mathcal{H}_{A}$ and let $\ket{\beta}_{R A}$ be proportional to a maximally entangled state and have norm $\sqrt{d_{A}}$. A Choi isomorphism  $\sigma_{\beta}$ maps the super-operator $\mathcal{N} : \mathcal{B} (\mathcal{H}_{A}) \rightarrow \mathcal{B} (\mathcal{K})$ to the ``Choi operator" 
\begin{align}
\label{eq: choi jamoilkowski}
    \sigma_{\beta} (\mathcal{N}) := \id_{R} \otimes \mathcal{N} (\ket{\beta}\bra{\beta}_{R A}) \in \mathcal{B} (\mathcal{H}_{R} \otimes \mathcal{K}).
\end{align}
$\mathcal{N}$ is completely positive if and only if $\sigma_{\beta} (\mathcal{N}) \geq 0$, and trace-preserving if and only if $\tr_{A} (\sigma_{\beta} (\mathcal{N})) = \mathds{1}_{R}$.

A useful tool we use in Ch.~\ref{chapter: 5} is the Schur complement of a block matrix. For operators $A \in \mathcal{S} (\mathcal{H})$, $B : \mathcal{H} \rightarrow \mathcal{K}$ and $C \in \mathcal{S} (\mathcal{K})$, consider the block operator 
\begin{align}
 \label{eq: block operator}
X = 
\begin{pmatrix}
A 
  & & B^* \\
  B& &
C
\end{pmatrix} \in \mathcal{S} ( \mathcal{H} \oplus \mathcal{K}).
\end{align}
If $C$ is invertible, then the Schur complement of $C$ of $X$, denoted with $X / C$, is equal to $A - B^* C^{-1} B$. If $C$ is invertible, then $X$ is positive semi-definite if and only if $C$ and $X/C$ are positive semi-definite \cite{CARLSON1986}.

\section{Majorization} 
\label{sec: ch2sec2}
Here is a brief review of the relevant parts of majorization theory \cite{ANDO1994, Marshall2011, Bhatia1997}. For $x \in \mathbb{R}^{d}$, $x^{\downarrow}$ and $x^{\uparrow}$ denote the $d$-dimensional vectors with the entries of $x$ ordered non-increasingly and non-decreasingly, respectively. Given $\mathcal{H}$, we distinguish an ordered orthonormal basis $\{ \ket{ q_{i} } \}_{i=1}^{ d_{\mathcal{H}}}$ and call it a computational basis. For $S \in \mathcal{S} (\mathcal{H})$, $S^{\downarrow} = \sum_{i = 1}^{d_{\mathcal{H}}} \lambda_{i} (S) \ket{q_{i}}\bra{q_{i}}$. Given $x, y \in \mathbb{R}^{d}$, $x$ is said to majorize $y$ if
\begin{align} 
\label{eq: majorization cond}
    \sum_{i = 1}^{k} x_{i}^{\downarrow} \geq \sum_{i = 1}^{k} y_{i}^{\downarrow}
\end{align}
for each $k \in [d -1]$ and $\sum_{i = 1}^{d} x_{i}^{\downarrow} = \sum_{i = 1}^{d} y_{i}^{\downarrow}$. We denote this statement with $x \succeq y$. A majorization relation can be defined on $\mathcal{S}(\mathcal{H})$ spectrally. For $B, C \in \mathcal{S}(\mathcal{H})$, if $\lambda(B) \succeq \lambda(C)$, then $B$ is said to majorize $C$, and we denote that with $B \succeq C$. Majorization induces a preorder on $\mathbb{R}^{d}$ and $\mathcal{S}(\mathcal{H})$, respectively. Given a set, a majorant, if it exists, is an element of the set that majorizes all elements in the set.

There are several useful characterizations of majorization, two of which are relevant to this thesis. The first is based on doubly-stochastic processing (see Theorem II.1.10 on page 33 in \cite{Bhatia1997}): $x$ majorizes $y$ if and only if there exist a tuple of permutation matrices $(P_{i})_{i}$ and a probability measure $(p_{i})_{i}$ such that 
\begin{align} 
\label{eq: cl stoch majorization}
 y = \sum_{i} p_{i} P_{i} x. 
\end{align}
In words, $x$ majorizes $y$ if and only if $y$ belongs to the convex hull of the orbit of $x$ under the action of the symmetric group. That is, $y = A x$ for some doubly-stochastic matrix $A$. Analogously, $B \succeq C$ if and only if
\begin{align}
\label{eq: q stoch majorization}
C = \sum_{i} p_{i} U_{i} B U_{i}^{*}
\end{align}
for a tuple of unitary operators $(U_{i})_{i}$ and a probability measure $(p_{i})_{i}$. Put differently, $C$ is the value of a mixed-unitary quantum channel at $B$.

The second characterization is based on unjust transfers, which are related to the better-known $T$-transforms \cite{Marshall2011}. For $i,j \in [d]$, and $t \in [0,1]$, a $T$-transform is a linear map with action
 \begin{align}
\label{alg: T-transforms}
(x_{1}, \ldots, x_{i}, \ldots, x_{j}, \ldots, x_{d}) \mapsto (x_{1}, \ldots, (1-t) x_{i} + t x_{j}, \ldots , (1-t) x_{j} + t x_{i}, \ldots, x_{d}).
 \end{align}
 A $T$-transform is a convex combination of the identity and a transposition. $T$-transforms are also known as Robin Hood operations \cite{Arnold2011}. The statement $x \succeq y$ holds if and only if $y$ can be obtained from $x$ by a finite number of $T$-transforms (see Lemma B.1 on page 32 in \cite{Marshall2011}). A transfer is a move on real vectors that takes a non-negative amount from one entry and gives it to another entry. An unjust transfer is a transfer where the receiving entry is not smaller than the giving entry to start. That is, whether or not a transfer is unjust depends on the vector it affects. In particular, unjust transfers are not linear maps like $T$-transforms. An unjust transfer on $x \in \mathbb{R}^{d}$ can be specified by an amount $\varepsilon \geq 0$ and two entries $i, j \in [d]$ such that $j > i$. It implements the changes
\begin{align}
\label{alg: unjust transfer}
x^{\downarrow}_{\; i} \gets x^{\downarrow}_{\; i} + \varepsilon, \quad x^{\downarrow}_{\; j} \gets x^{\downarrow}_{\; j} - \varepsilon. 
\end{align}
In words, the $i$th largest entry of $x$ increases by $\varepsilon$ and while $j$th largest one decreases by the same amount.

 \begin{prop}
 \label{prop: T-transforms and unjust transfers are equivalent}
 Let $x, y \in \mathbb{R}^{d}$. Then, $y = T x$ for some $T$-transform $T$ if and only if $x^\downarrow$ can be obtained from $y^\downarrow$ with an unjust transfer. 
\end{prop}
\begin{proof}
Since both $T$-transforms and unjust transfers act on pairs of coordinates, we may assume without loss of generality that $d = 2$. Denote $x= (x_{1}, x_{2})$ and $y=(y_{1}, y_{2})$. 

Suppose that there is a \(T\)-transform \(T\) such that \(y=Tx\). Then, for some \(t\in[0,1]\)
\begin{align}
\label{eq: T-transform between pairs}
(y_{1}, y_{2}) = ( (1-t)x_{1} + t x_{2},  (1-t) x_{2} + t x_{1} ).
\end{align}
 Because $y^{\downarrow}_{1}$ is a convex combination of $x_{1}$ and $x_{2}$, we have $x^{\downarrow}_{1} \geq y^{\downarrow}_{1}$. Let $\tilde{\varepsilon} := x^{\downarrow}_{1} - y^{\downarrow}_{1}$, and observe that $x^{\downarrow}_{1} = y^{\downarrow}_{1} + \tilde{\varepsilon}$ and $x^{\downarrow}_{2} = x^{\downarrow}_{1} + x^{\downarrow}_{2} - x^{\downarrow}_{1} = y^{\downarrow}_{1} + y^{\downarrow}_{2} -x^{\downarrow}_{1} = y^{\downarrow}_{2} - \tilde{\varepsilon}$, where we used the fact that $x_{1} + x_{2} = y_{1} + y_{2}$. 

Conversely, suppose there exists an unjust transfer of $\varepsilon \geq 0$ that takes $y^{\downarrow}$ to $x^{\downarrow}$. That is, $x^{\downarrow}_{1} = y^{\downarrow}_{1} + \varepsilon$ and $x^{\downarrow}_{2} = y^{\downarrow}_{2} - \varepsilon$. Observe that for $\Tilde{t} := \frac{\varepsilon}{(y^{\downarrow}_{1} -y^{\downarrow}_{2}) + 2 \varepsilon}$, 
\begin{align}
\label{eq: T-transform from unjust transfer}    
y^{\downarrow}_{1} = (1-\tilde{t}) x^{\downarrow}_{1} + \tilde{t} x^{\downarrow}_{2}, \quad y^{\downarrow}_{2} = (1-\tilde{t}) x^{\downarrow}_{2} + \tilde{t} x^{\downarrow}_{1}.
\end{align}
Hence, there exists a $T$-transform $T'$ such that $T' x^\downarrow = y^\downarrow $. Without loss of generality, we take $x = x^\downarrow$. If $y \neq y^\downarrow$, we may compose a transposition with $T'$ to get another $T$-transform $T$ that satisfies $T x = y$. \end{proof}

From this proposition, we conclude that $x \succeq y$ if and only if it is possible to get to $x$ from $y$ with a finite number of permutations and unjust transfers. In the particular case of $x, y \in \mathbb{R}^{2}$, $x \succeq y$ if and only if $x$ can be obtained from $y$ by an unjust transfer, or an unjust transfer and a transposition. This characterization may be extended to self-adjoint operators as follows. For $B, C \in \mathcal{S} (\mathcal{H})$, $B \succeq C$ if and only if there exists a finite number of permutations and unjust transfers that take $\lambda(C)$ to $\lambda(B)$.

Let $f$ be a real-valued function such that $\text{dom} (f)$ is contained in either $\mathbb{R}^{d}$ or $\mathcal{S} (\mathcal{H})$. It is Schur-convex if for all $a, b \in \text{dom} (f)$, $a \succeq b$ implies $f(a) \geq f(b)$. If $-f$ is Schur-convex, then $f$ is said to be Schur-concave. Let $F$ be a real-vector-valued function such that $\text{dom} (F)$ is contained in either $\mathbb{R}^{d}$ or $\mathcal{S} (\mathcal{H})$. $F$ is called strictly isotone if for all $a, b \in \text{dom} (F)$ such that $a \succeq b$, it holds that $F(a) \succeq F(b)$.

\section{Quantum error correction}
\label{sec: ch2sec3}

Quantum error correction originates from the works of Shor \cite{Shor1995} and Steane \cite{Steane1996}. Its foundations were laid by Knill and Laflamme in \cite{KL_1997} and Gottesman in \cite{Gottesman1997}. We point to \cite{Ball2023} for an accessible introduction to the mathematics of quantum error correcting codes. The book \cite{lidar_brun_2013} gives a modern overview of the field of quantum error correction and its relationship to current efforts to build scalable quantum information processing systems. In the following subsections, we give a basic review of the quantum error correction conditions and stabilizer codes.

\subsection{The quantum error correction conditions}
\label{subsec: ch2sec2sub1}

Given a quantum channel $\mathcal{N}: \mathcal{B} (\mathcal{H}) \rightarrow \mathcal{B} (\mathcal{K})$, understood to model the noisy evolution of a quantum-physical system, is it possible to invert its action on the system such that some quantum information may be transmitted through $\mathcal{N}$ without error? This inversion, if possible, must be implementable by a quantum channel. That is, there must exist a channel $\mathcal{R}$ such that $\mathcal{R} \circ \mathcal{N} = \id_{\mathcal{H}}$. The overwhelming majority of the channels used for modeling physics are not invertible in this way. However, for some such $\mathcal{N}$, there might be a subspace $\mathcal{C} \leq \mathcal{H}$, called a \textit{code}, such that $\mathcal{N}|_{\mathcal{B} (\mathcal{C})}$ is invertible by a quantum channel. Such a subspace is called $\mathcal{N}$-correcting. If $d_{\mathcal{C}} \geq 2$, then some quantum information may be transmitted through $\mathcal{N}$ without error. The procedure is as follows. Let $ U_{\mathcal{C}} : \mathbb{C}^{d_{\mathcal{C}}} \rightarrow \mathcal{H}$ be an isometry whose image is $\mathcal{C}$ and let $\mathcal{U}_{\mathcal{C}}$ be the associated isometric channel. Then, it holds that $\mathcal{U}_{\mathcal{C}} = \mathcal{R} \circ \mathcal{N} \circ \mathcal{U}_{\mathcal{C}}$. Since $\mathcal{U}_{\mathcal{C}}$ is invertible on $\mathcal{C}$, a quantum system of dimension $d_{\mathcal{C}}$ may be transmitted through $\mathcal{N}$ without error.

In \cite{KL_1997}, Knill and Laflamme give necessary and sufficient conditions for a code $\mathcal{C}$ to be $\mathcal{N}$-correcting for some channel $\mathcal{N} \sim \{ N_{i} \}_{i=1}^{k}$ in terms of the projector onto the code and the Kraus operators of the channel. Namely, $\mathcal{C}$ is $\mathcal{N}$-correcting if and only if for all $i, j \in [k]$, there exists a matrix of coefficients $( \alpha_{i, j} )_{i, j = 1}^{k}$ such that
\begin{align}
\label{eq: error correction conditions}
P_{\mathcal{C}} N_{j}^{*} N_{i} P_{\mathcal{C}} = \alpha_{i, j} P_{C}. 
\end{align}
The coefficient matrix $\alpha = ( \alpha_{i, j} )_{i, j = 1}^{k}$ corresponds to a self-adjoint operator, a positive semi-definite operator in fact. By the spectral theorem for self-adjoint operators, $\alpha = U D U^{*}$ for some unitary $U = (u_{i, j})_{i, j =1}^{k}$ and diagonal matrix $D = (d_{i, j})_{i=1, j=1}^{k}$ with real coefficients. For $i \in [k]$, define the operator $\Tilde{N}_{i} = \sum_{j=1}^{k} u_{j, i} N_{j}$. Then $\mathcal{N} \sim \{ \Tilde{N}_{i} \}_{i=1}^{k}$ and for each $i, j \in [k]$
\begin{align}
\label{eq: diagonal error correction conditions}
P_{\mathcal{C}} \Tilde{N}_{j}^{*} \Tilde{N}_{i} P_{\mathcal{C}} = D_{i, j} P_{C}.
\end{align}
Since the off-diagonal entries of $D$ are zero, for $i \neq j$, $\Tilde{N}_{i}$ and $\Tilde{N}_{j}$ map the code to orthogonal subspaces. Moreover, when restricted to the code, $\Tilde{N}_{j}$, up to a multiplicative factor, acts isometrically. This makes the inversion procedure plain. First, a measurement is performed to determine which of the orthogonal subspaces $\Tilde{N}_{j} \mathcal{C}$ carries the quantum information. After that, the action of the associated $\Tilde{N}_{j}$ is undone.

\subsection{Independent noise models and stabilizer codes}
\label{sec: ch2sec2sub2}

Error processes in physical architectures are commonly studied with independent noise models, sometimes called local error models. These are relatively simple physical evolution models where each qudit experiences noise \textit{independently} from other qudits in the architecture. It is common to simplify further and assume that each qudit is subject to the same error channel $\mathcal{N}$. The overall channel is then the tensor product $\mathcal{N}^{\otimes n}$. An agnostic choice for the channel $\mathcal{N}$ is to choose it to be a depolarizing channel. A depolarizing channel models the situation where an arbitrary error operator occurs with some probability. To illustrate, we use the qubit depolarizing channel as an example. For $p \in [0, 1]$, the qubit depolarizing channel $\Delta_{2,p}$ has the Kraus decomposition
\begin{align}
\label{eq: qubit depolarizing channel def}  
\Delta_{2,p} \sim \{ \sqrt{1 - \frac{3 p}{4}} I,  \sqrt{p} X, \sqrt{p} Y, \sqrt{p} Z \}. 
\end{align}
For $p > 0$, there is no code $\mathcal{C} \leq (\mathbb{C}^{2})^{\otimes n}$ of dimension greater than 1 that is $\Delta_{2,p}^{\otimes n}$-correcting. However, it may possible to find a code that can correct likely error patterns. If $p << 1$, then the probability of more than one error occurring is rather small. In this regime, it seems intuitive that encoding quantum information into a code that can correct a single arbitrary error would in principle be better for protecting the information than a trivial product encoding. This is the basic premise of quantum coding theory.

The performance of a code against independent noise may be inferred, at least in theory, from its code parameters. The length of the code, usually denoted with $n$, may be interpreted as the number of necessary physical qudits. The dimension of the code, usually denoted with $K$, gives the size of the encoded quantum information. It is common to parameterize code sizes with $k = \log_{d}(K)$ when it is an integer. The minimum distance of the code, usually denoted with $d$ (not to be confused with the local dimension which is now denoted with $q$), is related to how many errors the code can correct. Specifically, a code with distance $d$ can correct up to $\lfloor \frac{d-1}{2} \rfloor$ errors. A code $\mathcal{C} \leq (\mathbb{C}^{q})^{\otimes n}$ with these parameters is a $[[ n, K, d ]]_{q}$ code. The rate of the code is $\frac{\log_{d}(K)}{n}$ and the relative distance of the code is $\frac{d}{n}$. Ideally, we would want both fractions to be large. The relationship between them is complicated and depends on $q$. For more details, see \cite{Ball2023}.

Quantum error correcting codes were first demonstrated to exist by Shor \cite{Shor1995} and Steane \cite{Steane1996}. They showed that quantum information may be encoded into the entanglement of several quantum systems to protect it from local errors. One thing that sets quantum coding theory apart from classical coding theory is that quantum information cannot be cloned \cite{Wootters1982}. The no-cloning theorem is an immediate consequence of the linearity of quantum mechanics. It states that there exists no super-operator with action $\rho \mapsto \rho \otimes \rho$. Hence, simple repetition cannot be used to protect quantum information.

Of particular interest is the class of stabilizer (or additive) quantum error correcting codes. This class contains practically relevant codes that are amenable to analysis and is ubiquitously researched in quantum information science. They can be constructed for any prime power local dimension $q$ \cite{Ball2023}. The first stabilizer codes were qubit codes. An $n$-qubit stabilizer code $\mathcal{C} \leq (\mathbb{C}^{2})^{\otimes n}$ is the joint eigenspace with eigenvalue $1$  of an Abelian subgroup $S_{\mathcal{C}}$ of the $n$-qubit Pauli group that does not contain $-\mathds{1}$. $S_{\mathcal{C}}$ is called the stabilizer of the code. The minimum distance of the code is given by the minimum operator weight of $\text{centralizer} (S_{\mathcal{C}}) \setminus S_{\mathcal{C}}$. In fact, any element of $\text{centralizer} (S_{\mathcal{C}}) \setminus S_{\mathcal{C}}$ acts on $\mathcal{C}$ as a non-identity unitary operator. Similar notions hold in the more general case of prime power local dimension.

\chapter{Continuity bounds for conditional entropies}
\label{chapter: 3}
\noindent This chapter is based on the work \cite{Alhejji2020}.

\section{Introduction}
\label{sec: ch3sec1}

In many models of communication systems, the fundamental limits on rates of asymptotically faithful communication are given by entropic formulas. That certain entropic formulas come with such ``operational interpretations" is seen in scholarly works as early as Shannon's paper that laid the foundations for information theory \cite{Shannon1948}. Since communication systems are built upon physical processes, which can only be modeled imperfectly, these formulas should satisfy some continuity properties with respect to some accuracy measure of the model. Suppose for example that a certain physical process can be modeled well by a discrete memoryless channel. In practice, only a finite number of experiments can be run probing this physical process, and so the channel that best models it can only be inferred up to a certain accuracy level. If two channels are close to each other, in that both model a certain process well, but had wildly different values for some formula with an operational interpretation, such as a communication capacity, then the formula would be practically useless. This is one reason to study the continuity properties of entropies. We discuss several other reasons in the next section.

The following questions arise naturally in light of the preceding discussion. Given two quantum states that are $\epsilon$-close according to some distance measure, how different are their values for some entropic formula? Can this variation be bounded from above by a function that depends only on $\epsilon$? Such an upper bound is called a \textit{uniform} continuity bound. Naturally, the looser a bound is, the less insight it gives into the behavior of the function in question. A uniform continuity bound is \textit{tight} (or sometimes optimal) if for each admissible $\epsilon$, the domain of the function contains a pair of points that are $\epsilon$-close such that the absolute value of the difference of the values of the function at those points equals the bound. The main result of this chapter is a tight uniform continuity bound for the conditional Shannon entropy in terms of total variation distance: Thm~\ref{thm: tight uni bound for equivocation}.

The structure of this chapter is as follows. In Sec~\ref{sec: ch3sec2}, we motivate the study of continuity bounds further and lay out the history and state of the art at the time of publication of this material. In Sec~\ref{sec: ch3sec3}, we introduce relevant mathematical notions and state our continuity bound. Sec~\ref{sec: Ch3sec4} contains a proof, which we break down into three steps, of this continuity bound. In Sec~\ref{sec: ch3sec5}, we collect a few observations about a conjectured tight uniform continuity bound for the quantum conditional entropy. We conclude in Sec~\ref{sec: ch3sec6} with some remarks regarding our approach and its place in the literature.

\section{Motivation and history}
\label{sec: ch3sec2}

 There are at least two more reasons to want sharp estimates for the variation of entropic formulas in terms of relevant distance measures. The first reason is that they are quite useful in proofs of converse coding theorems in information theory. For example, if a certain protocol for quantum communication over a quantum channel has arbitrarily high fidelity, then a continuity bound for the conditional von Neumann entropy can be used to show that its rate cannot exceed the regularized coherent information of the channel (see page 670 in \cite{Wilde_2013}). The second reason is their usefulness in settings where practical algorithms for computing optimal rates are known only for a special class of states or channels. It may be worthwhile to have tight bounds on the error incurred by approximating the optimal rates of arbitrary channels or states by the ones of the closest members of the special class. For example, it is known that in the case of degradable quantum channels \cite{Cubitt2008}, the multi-letter formula Eq.~\ref{eq: quantum capacity} simplifies to a computable ``single-letter" formula. In \cite{Sutter_2017}, Sutter et al., following ideas in \cite{Leung_2009}, use continuity bounds for entropic quantities to give \textit{computable} estimates for the quantum capacity of $\epsilon$-degradable channels. These are channels that are $\epsilon$-close in diamond norm distance to the set of degradable channels.

 There have been a number of relatively recent works in the classical information theory literature studying the continuity of the Shannon entropy with respect to various distance measures on the probability simplex. In \cite{Zhang_2007}, Zhang proves a tight uniform continuity bound for the Shannon entropy in terms of total variation distance. His proof is based on maximal couplings of pairs of random variables. Sason further studied how maximal couplings can be used to derive continuity bounds for the Shannon entropy in \cite{Sason_2013}. In particular, he proved a refined continuity bound for it that is parameterized by both the total variation distance and $\ell_{\infty}$-norm distance. In \cite{Ho_2010}, Ho and Yeung study the breakdown of continuity of the Shannon entropy when the cardinality of the alphabet becomes infinite.

In conjunction, a remarkable number of works appeared in the quantum information theory literature studying the continuity of the von Neumann entropy and related quantities, especially with respect to trace distance. We highlight a few relevant ones here. For a more thorough review, see \cite{Shirkov2022}. The study of the continuity of von Neumann entropy started as early as \cite{Fannes_1973}. In \cite{Audenaert_2007}, Audenaert proves a tight uniform continuity bound for the von Neumann entropy with respect to trace distance. In \cite{Hanson_2018}, among other results, Hanson and Datta prove tight \textit{local} continuity bounds for the von Neumann entropy. Local continuity bounds here refer to upper bounds for variation that are quantum state dependent.

For both the Shannon and von Neumann entropies, several proofs of tight uniform continuity bounds have been presented \cite{Zhang_2007, Ho_2010, Pinelis_2018, Audenaert_2007, Winter_2016,Hanson_2017}. On the other hand, uniform bounds, which are not tight but are independent of the size of the conditioning system, were proven for the conditional Shannon and von Neumann entropies in \cite{Alicki_2004, Winter_2016}.

\section{Mathematical preliminaries}
\label{sec: ch3sec3}

In this section, we establish notation and state the main result of this chapter. Let $\cX := \{1,2,$$..., |\cX|\}$ and $\cY := \{1,2,$$..., |\cY|\}$ for some $|\cX|, |\cY| \in \mathbb{N} \setminus \{1\}$. Let $\cP_{\cX \times \cY}$ denote the probability simplex with $|\cX| |\cY|$ extremal points. The binary entropy function $h: [0,1] \rightarrow [0,1]$ is defined by the action: $x \mapsto h(x) = -x \log x - (1- x) \log (1-x)$. In this section, logarithms are meant in base 2 and entropies are measured in bits. The Shannon entropy of a random variable $X \sim p_{X}$ is defined as \cite{Cover_1991}
\begin{align}
\label{eq: Shannon entropy}
    H(X) := - \sum_{i \in \cX} p_{X}(i) \log{p_{X}(i)}.
\end{align}
Similarly, for two jointly distributed random variables $(X,Y) \sim p_{XY}$, the joint Shannon entropy is given by
\begin{align}
    H(XY) := - \sum_{i \in \cX} \sum_{j \in \cY} p_{XY}(i,j) \log{p_{XY}(i,j)}, 
\end{align}
and the conditional Shannon entropy $H(X|Y)$, also known as the equivocation in $X$ given $Y$, is defined as
\begin{align}
    H(X|Y) :&= H(XY) - H(Y) \label{eq: equiv1} \\
    &=\sum_{j \in \cY} p_{Y}(j) H(X|Y=j) \label{eq: equiv2} \\
    &= - \sum_{i \in \cX} \sum_{j \in \cY}  p_{XY}(i,j) \label{eq: equiv3} \log{\frac{p_{XY}(i,j)}{p_{Y}(j)}}.
\end{align}
When $p_{Y} (j) \neq 0$, the quantity $H(X | Y = j)$ is the Shannon entropy of the random variable $X_{|Y = j} \sim p_{X Y} /p_{Y}(j)$. Otherwise, it is defined to be zero. Given two probability measures $p_{XY},q_{X'Y'} \in \cP_{\cX \times \cY}$, the total variation distance between them is given by
\begin{align}
\label{eq: TV distance}
    \text{TV} (p_{XY}, q_{X'Y'}) =  \frac{1}{2} || p_{XY} - q_{X'Y'}||_{1} = \frac{1}{2}  \sum_{i \in \cX} \sum_{j \in \cY} | p_{XY}(i,j) - q_{X'Y'}(i,j)|.
\end{align}
A bona fide metric on the probability simplex, the total variation distance captures the distinguishability of any two probability measures \cite{Fuchs_1999}. Specifically, it gives the optimal probability of success for a protocol whose task is discriminating between the two measures based on a single sample.

\begin{thm}
\label{thm: tight uni bound for equivocation}
Let $\epsilon \in [0,1 - \frac{1}{|\cX|}]$ and suppose that $(X,Y) \sim p_{XY}$ and $(X',Y') \sim q_{X'Y'}$. If $p_{XY}$ and $q_{X'Y'}$ satisfy $\text{TV} (p_{XY}, q_{X'Y'}) \leq \epsilon$, then 
\begin{align}
\label{eq: bound for cond ent}
&|H(X|Y) - H(X'|Y')| \leq
    \epsilon \log{(|\cX| -1)} + h(\epsilon),
\end{align}
where $h$ is the binary entropy function. 
\end{thm}
Moreover, the inequality is tight. This means that for all $\epsilon \in [0,1 -\frac{1}{|\cX|}]$, there exists a pair of probability measures on $\cX \times \cY$ such that the total variation distance between them is $\epsilon$ and the inequality Eq.~\ref{eq: bound for cond ent} is satisfied with equality. An example of such a pair is given in Eq.~\ref{eq: tightness of bound}. Notice that in general, it holds that
\begin{align}
\label{eq: estimates for cond}
|H(X|Y) - H(X'|Y')| \leq \log{|\cX|}.
\end{align}
So, the case where $\epsilon > 1 - \frac{1}{|\cX|}$ is simply resolved by choosing $p_{XY}$ and $q_{X' Y'}$ such that
\begin{align}
\label{eq: p and q saturating for large epsilon}
p_{X Y} (1, 1) = 1, \;  q_{X' Y'} (i, 1) = \frac{1}{|\cX|} \; \forall i \in \cX.     
\end{align}

\section{Proof of the continuity bound for the conditional Shannon entropy}
\label{sec: Ch3sec4}

The mechanics of the proof are based on $G$-majorization, which refers to an ordering induced by a group $G$ on a real vector space \cite{Francis_2014, Marshall2011}. It is a generalization of majorization, where the group in question is the symmetric group. The majorization ordering can be defined as follows. Given two vectors, we say that one is higher than the other in the majorization ordering whenever the latter is in the convex hull of the orbit of the former under the action of the symmetric group, where the action is in terms of the natural permutation representation. A pair of vectors that can be ordered in this way is said to satisfy a majorization relation. We say that the higher vector majorizes the other and refer to it as a majorant.  It is clear that not every pair satisfies a majorization relation. Since the Shannon entropy is concave and invariant under the action of the symmetric group, if two probability measures satisfy a majorization relation, then the majorant has less entropy. That is, the Shannon entropy is a Schur-concave function.

The conditional Shannon entropy is not Schur-concave. While it is concave, it is not invariant under the action of the symmetric group.
\begin{example}
Let $|\cX| = |\cY| = 2$ and consider a probability measure $\nu_{X Y}$ defined by $\nu_{X Y} (1,1) = \nu_{X Y} (2,2) = \frac{1}{2}$. Conditional on $Y$, there is no entropy in $X$. Yet, there exists a permutation, say $(1, 1) \leftrightarrows (1,2)$, that takes $\nu_{X Y}$ to a probability measure where the entropy of $X$ conditional on $Y$ is equal to $1$ bit. 
\end{example}
However, the conditional Shannon entropy is invariant under the action of a proper subgroup of the symmetric group. We use this invariance to prove the present result. Consider the set $\{\nu_{XY}(i,j) \: |\:  i \in \cX, j \in \cY\}$, where $\nu_{XY}$ is a bivariate probability measure. The conditional entropy associated with $\nu_{XY}$ is invariant under permutations of $j$ indices, as can be seen from Eq.~\ref{eq: equiv2}. Additionally, it is invariant under exchanges of the form $(i_{1},j) \leftrightarrows (i_{2},j)$ for any  $i_{1},i_{2} \in \cX$ and $j \in \cY$. These permutations generate the symmetries of the equivocation in $X$ given $Y$. Denote this subgroup of the symmetric group by $S_{\cX|\cY}$.

We assume without loss of generality that $H(X|Y) \geq H(X'|Y')$. We proceed in three steps.

\subsection{Reordering}

First, we order the components of the bivariate probability measures in a suggestive way. We arrange the components to be in blocks of $|\cX|$ components that share the same $Y$ label. For definiteness, we order the $|\cY|$ blocks such that $q_{Y'} (j)  - p_{Y} (j)$ is non-increasing in $j$. For the block labeled by $j$, we define the set:
\begin{align}
\label{eq: j-set}
    I_{j} = \{ i \: | \: q_{X'Y'} (i,j) \geq p_{XY} (i,j) \},
\end{align}
Within the $j$th block, if both $I_{j}$ and $I_{j}^{c}$ are nonempty, put the elements of $I_{j}$ ahead of those in $I_{j}^{c}$. That is, we permute the components in both vectors simultaneously so that $i < i' $ for all $ i \in I_{j}$ and $ i' \in I_{j}^{c}$. Next, we permute the components so that $q_{X'Y'} (i,j)$ is non-increasing in $i$ for $i \in I_{j}$. We do the same for the components associated with the elements of $I_{j}^{c}$. Note that in addition to preserving total variation distance, all of these operations are in $S_{\cX|\cY}$ and so, they do not affect the equivocations.

\subsection{Walking}
The second step involves an optimized form of a proof technique due to Pinelis (see the third answer here \cite{Pinelis_2018}). We walk the two probability measures across the probability simplex until $H(X'|Y') = 0$. In the process, we take care that the difference of equivocations does not decrease and that the total variation distance does not increase.

We start by zooming into the $j$th block. If $I_{j}$ is not empty, then we make the following replacements on $q_{X ' Y'}$: 
\begin{align}
\label{alg: first replacements on q}
     &q_{X'Y'} (1,j) \gets q_{X'Y'} (1,j) + [q_{X'Y'} (i,j) - 
      p_{XY} (i,j)], \\
     &q_{X'Y'}(i,j) \gets  p_{XY}(i,j),
\end{align}
consecutively for each $i \in I_{j} \setminus{\{1\}}$. By design, these replacements do not affect total variation distance from $p_{X Y}$. To see that $H(X'|Y')$ did not increase, notice that the old probability vector is in the convex hull of the orbit of the new vector under the action of $S_{\cX | \cY}$. Put another way, the probabilities, conditional on $j$, are now no less concentrated than before. This is because for each $i \in I_{j} \setminus \{1\}$, the replacements above correspond with an unjust transfer on the conditional measure $q_{X', Y'}/q_{Y'}(j)$. After these replacements are made, the following inequalities hold:
\begin{align} 
    &q_{X'Y'} (1,j) - p_{XY} (1,j) \geq 0, \label{eq: form}\\
    &p_{XY} (i,j) - q_{X'Y'} (i,j) \: \geq 0, \label{eq: form1}
\end{align}
for each $i \in \cX \setminus{\{1\}}$.

Next, we make $q_{X'Y'} (1, j) = q_{Y'} (j)$ by transferring probability weights in the block from the bottom to the top. Specifically, we make the following replacements:
\begin{align}
     &q_{X'Y'} (1,j) \gets q_{X'Y'} (1,j) + q_{X'Y'}(i,j), \\
     \quad &p_{XY} (1,j) \gets p_{XY} (1,j) + q_{X'Y'}(i,j), \\
     &q_{X'Y'}(i,j) \gets  q_{X'Y'}(i,j) - q_{X'Y'}(i,j), \\
     \quad &p_{XY}(i,j) \gets  p_{XY}(i,j) - q_{X'Y'}(i,j),
\end{align}
consecutively for each $i \in \cX \setminus{\{1\}}$. Note that the transfers are made in both probability measures to ensure that the total variation distance remains the same. Observe what happens when a probability weight $s$, suitably small and non-negative, is taken from outcome ${(i,j)}$ and given to outcome ${(1,j)}$ in both probability measures. The difference of entropies changes by the following amount:
\begin{align*}
    &\{ [ \eta ( q_{X'Y'} (1,j) + s) - \eta ( p_{XY} (1,j) + s)] \\
    & - [ \eta ( q_{X'Y'} (1,j)) - \eta ( p_{XY} (1,j))]\} \\
     &+ \{ [ \eta ( p_{XY} (i,j)) - \eta ( q_{X'Y'} (i,j))] \\
     & - [ \eta ( p_{XY} (i,j) - s) - \eta ( q_{X'Y'} (i,j) - s)]\},
\end{align*}
where $\eta(x)=x\log{x}$. Since it is convex and Eqs.~\ref{eq: form} and \ref{eq: form1} hold, the two differences in curly brackets are not negative. Hence, we can set $q_{X'Y'} (1,j) = q_{Y'} (j)$ and $q_{X'Y'} (i,j) = 0$ for $i \neq 1$. This implies that $q_{Y'} (j) H(X'|Y' = j)$ vanishes.

Now, say $I_{j}$ is empty. Then it holds that $q_{X'Y'} (i,j) - p_{XY} (i,j) < 0$, for all $i \in \cX$. It also holds that $q_{X'Y'} (1,j) \geq q_{X'Y'} (2,j) \geq ... \geq q_{X'Y'} (|\cX|,j)$. Our goal is still to take $q_{Y'} (j) H(X'|Y' = j)$ to zero. To do this, we increase $q_{X'Y'} (1,j)$ by transferring probability weights from the rest of the outcomes in the block. As explained before, such transfers can only decrease $H(X'|Y')$. Furthermore, as long as $q_{X'Y'} (1,j) - p_{XY} (1,j) < 0$, these transfers can be done without affecting the total variation distance. If at any point $q_{X'Y'} (1,j) - p_{XY} (1,j) = 0$, we stop as Eqs.~\ref{eq: form} and \ref{eq: form1} now hold for this block. In such a case, we start the process mentioned in the previous paragraph. Otherwise, we keep going until $q_{X'Y'} (1,j) = q_{Y'} (j)$.

After all blocks have been processed, we can assume without any loss of generality that $q_{X'Y'} (i,j) = 0$ for all $i \neq 1$. With this in mind, we subject both probability measures to a stochastic process that averages over all the blocks. Specifically, a map with action
\begin{align}
    \nu_{XY} (i,j) \mapsto  \mathcal{E} \nu_{XY} (i,j)= \frac{1}{|\cY|} \sum_{j \in \cY} \nu_{XY} (i,j). 
\end{align}
This stochastic map corresponds to a complete loss of the information carried by the variable $Y$. $\mathcal{E}$ is a convex combination of elements in $S_{\cX | \cY}$. This implies that $H(X|Y)$ does not decrease. It does not change the fact that $H(X'|Y') = 0$ since only $q_{X'Y'}(1,j)$ can be nonzero. Recall that the total variation distance between two distributions does not increase under stochastic maps. Note that the outputs of $\mathcal{E}$ are always product distributions with a uniform marginal on $\cY$. As for $\cX$, we have the following marginals
\begin{align}
  q_{X'} (1) &= 1  \: \geq \: p_{X} (1) \: \geq \: 1 - \epsilon, \label{final1}\\
  q_{X'} (i) &=  0  \: \leq \: p_{X} (i), \label{final}
\end{align}
for $i \in \cX \setminus{\{1\}}$.

\subsection{Estimating}
Since the two distributions can be assumed to be product distributions on $\cX \times \cY$, from here one can invoke the bound for the unconditional case \cite{Zhang_2007} to finish the proof. For the sake of completeness, we include the following rather standard estimates. 

Given distributions as in Eq.~\ref{final1} and Eq.~\ref{final}, we can upper-bound $H(X)$ in the following way:
\begin{align}
\label{eq: estimates for tight}
    H(X) &=  -p_{X}(1) \log{p_{X}(1)} 
     - \sum_{i \neq 1} {p_{X}(i)} \log{{{p_{X}(i)}}} \\
    & \leq  -p_{X}(1) \log{p_{X}(1)}  \\
    &\quad - \sum_{i \neq 1} {\frac{(1 - p_{X}(1))}{|\cX| - 1}} \log{{{\frac{(1 - p_{X}(1))}{|\cX| - 1}}}}\\
    & = (1 - p_{X}(1)) \log{(|\cX| - 1)} + h((1 - p_{X}(1)))\\
    & \leq \epsilon \log{(|\cX| - 1)} + h(\epsilon)
\end{align}
where the first inequality follows because the vector $(\frac{1 - p_{X}(1)}{|\cX| - 1})_{i=2}^{|\cX|}$ is majorized by $(p_{X}(i))_{i=2}^{|\cX|}$. The second inequality comes from the fact that the function $x \log (|\cX| - 1) + h(x)$ is monotonically increasing in $x$ for $x \in [0, 1 - \frac{1}{|\cX|}]$. This completes the proof. 

To see that the bound is tight, let $\epsilon \in [0, 1 - \frac{1}{|\cX|}]$ be given and consider the following probability measures.
\begin{align}
\label{eq: tightness of bound}
    q_{X'Y'}(1,1) &= 1 \\ 
    p_{XY}(1,1) &= 1 - \epsilon \quad \text{and} \quad p_{XY}(i,1)= \frac{\epsilon}{|\cX|-1}, 
\end{align}
for $i \in \cX \setminus{\{1\}}$. Evidently, the two are separated by $\epsilon$ in total variation distance and the associated equivocations saturate the bound.

\section{Observations about the continuity of the conditional von Neumann entropy}
\label{sec: ch3sec5}

Before discussing the continuity of quantum entropies, we recall some basic notions. Let $\mathcal{H}_{A}$ be an inner product space with finite dimension $d_{A}$. The von Neumann entropy of a state $\rho \in \mathcal{D} (\mathcal{H}_{A})$ is given by
\begin{align}
\label{eq: von Neumann ent}
H(A)_{\rho} := -\tr (\rho \log \rho).
\end{align}
It is equal to the Shannon entropy of a probability measure whose entries are equal to the eigenvalues of $\rho$. Consider a tensor product $\mathcal{H}_{A} \otimes \mathcal{H}_{B}$ of two inner product spaces $\mathcal{H}_{A}$ and $\mathcal{H}_{B}$ with dimensions $d_{A}$ and $d_{B}$, respectively. The conditional von Neumann entropy of $A$ given $B$ associated with a bipartite state $\rho_{A B} \in \mathcal{D} (\mathcal{H}_{A} \otimes \mathcal{H}_{B})$ is given by
\begin{align}
\label{eq: von Neumann cond ent}
H(A|B)_{\rho} := H(A B)_{\rho} - H (B)_{\rho},
\end{align}
where the second term is the von Neumann entropy of the marginal state $\rho_{B} = \tr_{A} (\rho_{A B})$. Unlike the conditional Shannon entropy, the conditional von Neumann entropy can be negative. Its negativity is a sufficient condition for the state to be entangled. Finally, the quantum generalization of total variation distance: given two state operators $\rho$ and $\sigma$, their trace distance is given by 
\begin{align}
\label{eq: trace distance}
d(\rho, \sigma)_{\tr} := \frac{1}{2} || \rho - \sigma||_{1} = \frac{1}{2} \tr( | \rho - \sigma|).
\end{align}

The current state-of-the-art continuity bound for the conditional von Neumann entropy is due to Winter \cite{Winter_2016} who optimized a proof technique due to Alicki and Fannes \cite{Alicki_2004}. We reproduce it below to contrast it with a conjectured tight bound Eq.~\ref{eq: Wilde conjecture}. Given $\rho_{A B}, \sigma_{A B} \in \mathcal{D} (\mathcal{H}_{A} \otimes \mathcal{H}_{B})$ such that $d_{\tr} (\rho_{A B}, \sigma_{A B}) \leq \epsilon$ for $\epsilon \in [0,1]$, it holds that 
\begin{align}
 \label{eq: Winter bound}
 | H (A | B)_{\rho} - H (A | B)_{\sigma} | \leq 2 \epsilon \log d_{A} + (1 + \epsilon) h( \frac{\epsilon}{1 + \epsilon} ).
\end{align}
Clearly, this continuity bound is uniform and independent of $d_{B}$. This independence becomes important in cases where a bound on the dimension of the conditioning system is not known. An example is squashed entanglement which is a bipartite entanglement measure \cite{Christandl2004}.

We restrict to the case where $d_{A} = d_{B} = d$. In this case, Wilde conjectured the following tighter continuity bound. Given $\rho_{A B}, \sigma_{A B} \in \mathcal{D} (\mathcal{H}_{A} \otimes \mathcal{H}_{B})$ such that $d_{\tr} (\rho_{A B}, \sigma_{A B}) \leq \epsilon$ for $\epsilon \in [0,1 - \frac{1}{d^2}]$, it holds that 
\begin{align}
\label{eq: Wilde conjecture}
 | H (A | B)_{\rho} - H (A | B)_{\sigma} | \leq^{?} \epsilon \log (d^{2} - 1) + h( \epsilon ).
\end{align}
In a similar fashion to the classical case, the restriction on $\epsilon$ comes from the fact that if it were larger, it is easy to find two states whose conditional entropies saturate a rather trivial bound. Specifically, for any two bipartite quantum states $\rho$ and $\sigma$, 
\begin{align}
\label{eq: triv upper bound for difference of quantum entropies}
 | H (A | B)_{\rho} - H (A | B)_{\sigma} | \leq 2 \log d_{A}.
\end{align}
Two such states are the pure state $\ket{\Phi}\bra{\Phi}_{A B}$, where $\ket{\Phi}_{A B}$ is maximally entangled, and $(1 - \epsilon) \ket{\Phi}\bra{\Phi}_{A B} + \frac{\epsilon}{d^{2} - 1} ( \mathds{1}_{A B} - \ket{\Phi}\bra{\Phi}_{A B})$. These two states are mentioned as an example in \cite{Winter_2016}.

We believe that Eq.~\ref{eq: Wilde conjecture} holds. We prove it in special cases in the remainder of this section. We start with the case where both $\rho_{A B}$ and $\sigma_{A B}$ are diagonal in a maximally entangled basis. An orthonormal basis $\{ \ket{\phi_{i}}_{A B} \}_{i=1}^{d^{2}}$ of $\mathcal{H}_{A} \otimes \mathcal{H}_{B}$ is said to be a maximally entangled basis if $\ket{\phi_{i}}_{A B}$ is maximally entangled for each $i \in [d^{2}]$. 

\begin{prop}
 \label{prop: two entangled state bases}
Let $\epsilon \in [0,1 - \frac{1}{d^2_{A}}]$ and suppose that $\rho_{A B}, \sigma_{A B} \in \mathcal{B} (\mathcal{H}_{A} \otimes \mathcal{H}_{B})$ are quantum states that satisfy $d_{\tr} (\rho_{A B}, \sigma_{A B}) \leq \epsilon$. If $\rho_{A B}$ and $\sigma_{A B}$ are diagonal in a maximally entangled basis, then Eq.~\ref{eq: Wilde conjecture} holds for $\rho_{A B}$ and $\sigma_{A B}$.
\end{prop}
\begin{proof}
Since both are diagonal in an entangled basis, $H(B)_{\rho} = H(B)_{\sigma} = \log d$. Hence, 
\begin{align}
\label{eq: estimates for ent rho and sig}
| H (A | B)_{\rho} - H (A | B)_{\sigma} | =| H (A B)_{\rho} - H (A  B)_{\sigma} | \leq \epsilon \log (d^{2} - 1) + h( \epsilon ),
\end{align}
where the inequality follows from the continuity bound for the von Neumann entropy in \cite{Audenaert_2007}. \end{proof}

We can generalize this result to the case where only the state with the smaller value for conditional entropy is diagonal in a maximally entangled basis. This generalization is based on the observation that the conditional von Neumann entropy does not decrease under decoherence in a maximally entangled basis. 
\begin{lem}
\label{lem: decoherence in maximally entangled basis is A-unital}
Let $E = \{ \ket{\phi_{i}}_{A B} \}_{i=1}^{d^{2}}$ be a maximally entangled basis. For all $\rho_{A B} \in \mathcal{D} (\mathcal{H}_A \otimes \mathcal{H}_{B})$, it holds that 
\begin{align}
H(A |B )_{\rho} \leq H(A | B)_{\text{diag}_{E}(\rho)},
\end{align}
where $\text{diag}_{E}(\rho)$ denotes the pinching of $\rho$ in $E$. 
\end{lem}
\begin{proof}
Per \cite{Vempati2022}, it suffices to show that pinching in $E$ preserves the algebra $\mathds{1}_{A} \otimes \mathcal{B} (\mathcal{H}_{B})$. Let $\tau_B \in \mathcal{B} (\mathcal{H}_{B})$ and observe that
\begin{align}
\label{eq: a-unital equalities}
   \text{diag}_{E} ( \mathds{1}_{A} \otimes \tau_B ) &= \sum_{i = 1}^{d^{2}} \ket{\phi_{i}}\bra{\phi_{i}}_{A B} \bra{\phi_{i}}  \mathds{1}_{A} \otimes \tau_B \ket{\phi_{i}} \\
   &= \sum_{i = 1}^{d^{2}} \ket{\phi_{i}}\bra{\phi_{i}}_{A B} \tr ( \ket{\phi_{i}}\bra{\phi_{i}}_{A B} \mathds{1}_{A} \otimes \tau_B ) \\
   &= \sum_{i = 1}^{d^{2}} \ket{\phi_{i}}\bra{\phi_{i}}_{A B} \tr ( \frac{\tau_B}{d} ) = \tr ( \frac{\tau_B}{d} ) \mathds{1}_{A} \otimes \mathds{1}_{B}.
\end{align}
Since $\tau_{B}$ was arbitrary, this completes the proof.\end{proof}

This lemma and Prop.~\ref{prop: two entangled state bases} can be used to prove the following proposition.
\begin{prop}
Let $\epsilon \in [0,1 - \frac{1}{d^2}]$ and suppose that $\rho_{A B}, \sigma_{A B} \in \mathcal{B} (\mathcal{H}_{A} \otimes \mathcal{H}_{B})$ are quantum states that satisfy $d_{\tr} (\rho_{A B}, \sigma_{A B}) \leq \epsilon$. Suppose that $H(A|B)_{\rho} \geq H(A|B)_{\sigma}$ and that $\sigma_{A B}$ is diagonal in a maximally entangled basis $E$. Then, Eq.~\ref{eq: Wilde conjecture} holds for $\rho_{A B}$ and $\sigma_{A B}$. 
\end{prop}
\begin{proof}
Consider the following estimates
\begin{align}
\label{eq: estimates for cond ent one max}
| H (A | B)_{\rho} - H (A | B)_{\sigma} | &= H (A | B)_{\rho} - H (A | B)_{\sigma} \\
&\leq   H (A | B)_{\text{diag}_{E}(\rho)} - H (A | B)_{\sigma} \\
&=  H (A | B)_{\text{diag}_{E}(\rho)} - H (A | B)_{\text{diag}_{E}(\sigma)}.  
\end{align}
The first equality is by assumption, the first inequality is by Lem.~\ref{lem: decoherence in maximally entangled basis is A-unital}, and the last equality follows from the fact that $\sigma_{A B} = \text{diag}_{E} (\sigma_{A B})$. Clearly, $\text{diag}_{E}(\rho_{A B})$ is diagonal in a maximally entangled basis. Moreover, trace distance is monotonically non-increasing under quantum channels and so $d_{tr} (\text{diag}_{E} (\sigma), \text{diag}_{E} (\rho)) \leq \epsilon$. From here, the result can be proven by applying Prop.~\ref{prop: two entangled state bases}. \end{proof}

\section{Concluding remarks}
\label{sec: ch3sec6}
We have presented a proof of a tight uniform continuity bound for the conditional Shannon entropy. The bound is independent of the alphabet size of the conditioning system. However, we have assumed in the proof that the conditioning system has finite support. We proved a conjectured bound for the conditional von Neumann entropy in the special case where the two systems have the same dimension and the state with lower conditional entropy is diagonal in a maximally entangled basis. 

In \cite{Wilde2020}, Wilde generalized our bound Eq.~\ref{eq: bound for cond ent} to quantum-classical states. He used this generalized bound to improve a uniform continuity bound for the entanglement of formation originally given by Winter in \cite{Winter_2016}. In \cite{Jabbour2022}, using an approach based on our proof techniques, Jabbour and Datta prove tight uniform continuity bounds for Arimoto-Rényi conditional entropies for classical and quantum-classical states. 

Our proof of Thm.~\ref{thm: tight uni bound for equivocation} depends crucially on the invariance of $H(X|Y)$ under the action of $S_{\cX | \cY}$. More generally, entropic quantities of interest are invariant under the action of some subgroup of the symmetric group. For example, the mutual information $I(X;Y) = H(X) + H(Y) - H(XY)$ is invariant under
"local" permutations of indices, i.e., tensor products of permutations of $X$ and $Y$ labels. We believe such symmetries provide a path toward a better understanding of the behaviors of the corresponding quantities in terms of continuity and beyond. 

\chapter{The spin alignment problem}
\label{chapter: 4}
\section{Introduction}
\label{sec: ch4 intro}

Minimizing dispersion in a communication
signal is often necessary in both the theory and practice of
information processing. Informally, the term dispersion is used here so that the more dispersion in a signal the more information content it carries.  
How dispersion is quantified depends on context. For example, in a setting where sources are taken to generate signals in an independent
and identically-distributed manner and processes are assumed to be
memoryless, a useful measure of dispersion is the Shannon entropy for classical signals associated with probability distributions
\cite{Shannon1948}, or analogously the von Neumann entropy for quantum
signals associated with quantum states
\cite{Schumacher1995}. By this point, myriads of measures of dispersion have
been introduced and extensively studied
\cite{Renyi1960, Tsallis1988, Chehade2019}. A sensible requirement for a
measure of dispersion is that it does not decrease under processes
where labels designating outcomes are confused. Such a
process is represented classically by a doubly-stochastic transition matrix; the
quantum generalization of which is a mixed-unitary channel. Put in other words, a measure of
dispersion must be a Schur-concave function.  For this chapter, we
switch from concave to convex and primarily focus on the set of convex Schur-convex functions, which is a proper subset of the set of Schur-convex functions. 
To be precise, we study how unitarily invariant convex functions behave under mixing of
signals. The value of these functions at a mixture of signals depends
on the spectra of the individual signals supported in the mixture as
well as the alignment or overlap of the signals. It is our aim to shed
light on the latter kind of dependence.

Maximizing unitarily invariant convex functions is less ubiquitous in quantum information than minimizing them, and is generally a more difficult task. This difficulty merits more analysis and research. Quantum channel minimum output entropy is an example of an information-theoretic quantity that requires such a minimization. A reason to consider it is the equivalence of additivity of minimum output entropy to additivity of Holevo information \cite{Shor2004}, the regularization of which is a formula for the classical capacity \cite{Schumacher1997}. Additivity here is understood to be under tensor products. Despite an existence proof due to Hastings \cite{Hastings2009}, thus far there are no known explicit examples of quantum channels acting on finite-dimensional spaces whose minimum output entropy is not additive. Additionally, such minimization is relevant in broadcast channel scenarios with privacy concerns such as quantum communication. Ideally, the entropy at the environment or adversary is kept at a minimum. Depending on the channels in question, the optimal states may be mixtures, as in the case of platypus channels \cite{Leditzky2022a}.

We are interested in situations where we can unambiguously say that a signal $a$ contains less dispersion than a signal $b$. That is, $f(b) \leq f(a)$ for all Schur-convex function $f$. Luckily, we need not argue such an inequality for every function in this class. It suffices to check that $a$ majorizes $b$. In this sense, of all the ways of measuring dispersion, majorization is the most stringent \cite{Marshall2011, Bhatia1997}. Majorization theory, which is nearly a century old \cite{Hardy1929}, is the basis of Nielsen's celebrated theorem for pure bipartite state conversion using local operations and classical communication \cite{Nielsen1999}. This is one example among many of majorization giving insights into information processing.

Here, we elucidate what we mean by alignment and how it is relevant to the behavior of dispersion under mixing. Let $\lambda(\cdot)$ denote the vector of eigenvalues (with multiplicity) of a self-adjoint operator ordered non-increasingly. Self-adjoint operators $S_{1},\ldots, S_{\ell}$ acting on $\mathbb{C}^{d}$ are said to be \textit{perfectly aligned} if there exists an orthonormal basis $\{ \ket{\phi_{i}} \}_{i=1}^{d}$ such that $S_{1} = \sum_{i =1}^{d} \lambda_{i} (S_{1}) \ket{\phi_{i}}\bra{\phi_{i}}, \: \ldots \:    , S_{\ell} = \sum_{i =1}^{d} \lambda_{i} (S_{\ell}) \ket{\phi_{i}}\bra{\phi_{i}}$. This definition is motivated by Ky Fan's majorization relation \cite{Fan1949}, which in the case of states $\rho_{1},\ldots, \rho_{\ell}$ and a probability measure $p = (p_{i})_{i=1}^{\ell}$ reads
\begin{align} 
\label{Fan's majorization relation}
\lambda(\sum_{i=1}^{\ell} p_{i} \rho_{i}) \preceq \sum_{i=1}^{\ell} p_{i} \lambda (\rho_{i}).
\end{align}
Here \(b\preceq a\) means \(b\) is majorized by \(a\). These inequalities are simultaneously saturated if and only if the states are perfectly aligned. Hence, to minimize the dispersion in a mixture of states subject to spectrum constraints, we should choose the states to be perfectly aligned.

Now, consider a situation where that was not possible. That is, a situation where each state in the mixture is required to satisfy extra constraints that preclude a choice where all the states are perfectly aligned. We might ask if it is possible to conclude a majorization relation as above for all $p$ for some admissible choice of states. The following example is meant to illustrate such a situation.

\begin{example}
\label{ex: toy example}
Let $\tau$ be a state of non-maximal rank, $\ket{\alpha} \in \text{supp}(\tau)$, and $\ket{e} \in \text{ker}(\tau)$. Consider the set of parameterized unit vectors $\ket{v_{\gamma}} := \sqrt{\gamma} \ket{\alpha} + \sqrt{1 - \gamma} \ket{e}$ for $\gamma \in [0,1]$.
For $s \in [0,1]$, let $\tau_{\gamma, s} := s \tau + (1-s) \ket{v_{\gamma}}\bra{v_{\gamma}}$. Suppose we wish to minimize a Schur-concave function, the von Neumann entropy for example, of the mixture $\tau_{\gamma, s}$ for some given $s$ over choices of $\gamma$. If $\ket{\alpha}$ were a maximal eigenvector of $\tau$, then we can see that $\gamma = 1$ is optimal via Eq.~\ref{Fan's majorization relation}. But what if for all choices of $\gamma$, $\ket{v_{\gamma}}\bra{v_{\gamma}}$ is not perfectly aligned with $\tau$? It seems natural that the choice $\gamma = 1$ is still optimal because that is where the two states are as overlapped or aligned as possible. More strongly, we might conjecture that $\tau_{\gamma, s} \preceq \tau_{1,s}$ holds for all $s \in [0,1]$.
\end{example}

We use the spin alignment problem as a case study of these notions of alignment with the aim of refining Fan's majorization relation. Introduced by Leditzky et al. in \cite{Leditzky2022a}, the problem arose in the context of proving additivity of the coherent information of a class of quantum channels with peculiar communication properties. For a state $Q$ and a probability measure $\mu$, they considered $n$-qudit states of the form
\begin{align} 
\label{intro alignment operator}
    \sum_{I \subseteq [n]} \mu_{I} \rho_{I} \otimes Q^{\otimes I^{c}},
\end{align}
where $\rho_{I}$ are states of the qudits in $I$. Their task was to pick the tuple of states $(\rho_{I})_{I \subseteq [n]}$ so that the overall operator has minimum von Neumann entropy. It is not too difficult to see that an optimal choice has to be one where states in the tuple are pure (see Lem.~\ref{lem: pure states sufficiency} below). However, depending on $\mu$, there may be no such choice where the summands in Eq.~\ref{intro alignment operator} are perfectly aligned. Leditzky et al. conjectured that the von Neumann entropy is minimized by choosing the tuple of states $(\ket{q_{1}}\bra{q_{1}}^{\otimes I})_{I \subseteq [n]}$, where $\ket{q_{1}}$ is a maximal eigenvector of $Q$. With this tuple, the summands appear to overlap maximally. They showed that if this conjecture is true, then their channels have additive coherent information. For more information on these channels, see \cite{Leditzky2022a} and the companion paper \cite{Leditzky2022b}.

The structure of this chapter is as follows. In Section~\ref{sec:prelim}, we recall notation and relevant mathematical background. We formally define and motivate the class of spin alignment problems in Section~\ref{sec: spin alignment def}. In Section~\ref{sec:results}, we provide our results, which include resolutions for various instances of spin alignment. In Section~\ref{sec:``dual" problem}, we introduce a general class of optimization problems that is ``dual" to spin alignment. In Section~\ref{sec:conclusion}, we conclude and discuss open problems and follow-up lines of inquiry. Sections ~\ref{sec:app A} and ~\ref{sec:app B} contain proofs of technical lemmas.

\section{Preliminaries}
\label{sec:prelim}

We recall some notation and relevant mathematical notions. $\sigma(\cdot)$ and $\lambda(\cdot)$ denote the ordered vectors of singular values and eigenvalues of a self-adjoint operator, respectively. For $n \in \mathbb{N}$, the set {\(\{1,\ldots, n\)\}} is denoted by $[n]$ and its power set is denoted by $2^{[n]}$. We use the convention that $[0]$ denotes the empty set. Given a measure $\mu$ on a countable set, its support is denoted by $\text{supp}(\mu)$. Vectors $v_{1}, \ldots, v_{s} \in \mathbb{R}^{d}$ are said to be similarly ordered if there exists a permutation matrix $A$ satisfying $v_{1}^{\downarrow} = A v_{1}, \ldots, v_{s}^{\downarrow} = A v_{s}$. This is analogous to perfect alignment for self-adjoint operators. Neither similar ordering nor perfect alignment is transitive. A family of projectors is perfectly aligned if and only if their supports are nested. That is, the family forms a chain in the lattice of projectors. For $B \in \mathcal{S}(\mathbb{C}^{d})$ and $i \in [d]$, $P_{\geq i} (B)$ denotes the projector onto the direct sum of eigenspaces corresponding to the $i$ largest eigenvalues (with multiplicity) of $B$. $P_{\geq 1} (B),\ldots, P_{\geq d} (B) = \mathds{1}_{d}$ are perfectly aligned, and they furnish the ``staircase" decomposition:
\begin{align}
\label{eq: staircase decomp}
B = \sum_{i=1}^{d-1} (\lambda_{i}(B) - \lambda_{i+1}(B)) P_{\geq i} (B) + \lambda_{d}(B)  \mathds{1}_{d}.
\end{align}
The coefficients in this decomposition are non-negative if and only if $B \geq 0$. Self-adjoint $B$ and $C$ are perfectly aligned if and only if for all $i \in [d]$, $P_{\geq i} (B) = P_{\geq i} (C)$.

An important tool in our analysis is Fan's maximum principle \cite{Fan1949} for self-adjoint operators. It states that for $T \in \mathcal{S}(\mathbb{C}^{d})$, $k \in [d]$,
\begin{align} 
\label{eq: Fan maximum prinicple}
    \sum_{i = 1}^{k} \lambda_{i} (T) = \max\{\tr(TP) \; | \;  \text{\(P\) is a rank-\(k\) projector}\}.
\end{align}
It follows directly from this principle that
\begin{align} 
\label{eq: Fan majorization rel self-adjoint}
    \sum_{i=1}^{s} S_{i} \preceq \sum_{i=1}^{s} S_{i}^{\downarrow}
\end{align}
for any self-adjoint $S_{1},\ldots, S_{s}$. Moreover, the inequalities above are simultaneously satisfied with equality if and only if the summands are perfectly aligned.

We are interested in maximizing continuous functions over $\mathcal{D}(\mathcal{H}^{\otimes n})$, specifically those that are convex and unitarily invariant. Of particular interest are unitarily invariant norms. Generically denoted by $||| \cdot |||$, these are operator norms that satisfy $||| \cdot  ||| = ||| U (\cdot)  V ||| $ for all unitaries $U$ and $V$. For an operator $R \in \mathcal{B}(\mathbb{C}^{d})$, its Schatten $p$-norm for $p \in [1, \infty)$ is defined as 
\begin{align}
\label{eq: p-Schatten norm}
    || R ||_{p} := \tr (|R|^{p})^{\frac{1}{p}}, 
\end{align}
where $|R|$ denotes the absolute value $\sqrt{R^{*} R}$. Schatten norms satisfy $\lim_{p \rightarrow \infty} || \cdot ||_{p} = || \cdot ||$, where $|| \cdot ||$ is the operator norm. Both Rényi entropies and Tsallis entropies \cite{Renyi1960, Tsallis1988} are monotonic functions of Schatten $p$-norms for $p \in (1, \infty)$. For $R \in \mathcal{B}(\mathbb{C}^{d})$, $k \in [d]$, the Fan norm of order $k$ is 
\begin{align}
\label{eq: k-Fan norm} 
    || R ||_{(k)} := \sum_{i = 1}^{k} \sigma_{i} (R).
\end{align}
In words, it is the sum of the $k$ largest singular values of $R$. For positive semi-definite $R$, $|| R ||_{(k)}$ is the sum of its $k$ largest eigenvalues. Both Schatten norms and Fan norms belong to the class of unitarily invariant norms. For any $A,B \in \mathcal{B}(\mathbb{C}^{d})$, $|| A ||_{(k)} \leq || B ||_{(k)}$ for all $k \in [d]$ implies $||| A ||| \leq ||| B |||$ for every unitarily invariant norm $||| \cdot |||$. This is Fan's dominance theorem (see Theorem IV.2.2 on page 93 in \cite{Bhatia1997}).

\section{Spin alignment problems and the alignment ordering}
\label{sec: spin alignment def}

We define an instance of spin alignment as follows. 
\begin{definition} (Spin alignment problem).
    Let a local dimension $d$, a number of qudits $n$, a probability measure $\mu$, and a state $Q = \sum_{i=1}^{d} \lambda_{i} (Q) \ket{q_{i}}\bra{q_{i}} \in \mathcal{D}(\mathbb{C}^{d})$ be given. If $f$ is a continuous, unitarily invariant, convex function whose domain includes $\mathcal{D}((\mathbb{C}^{d})^{\otimes n})$, then the spin alignment problem associated with $(d, n, \mu, Q, f)$ is to maximize $f( \sum_{I \subseteq [n]} \mu_{I} \rho_{I} \otimes Q^{\otimes I^{c}})$ over all state tuples $(\rho_{I})_{I \subseteq [n]}$. The operator argument of $f$
    \begin{align} 
\label{alignment operator}
    \sum_{I \subseteq [n]} \mu_{I} \rho_{I} \otimes Q^{\otimes I^{c}}
\end{align}
is called an \textit{alignment operator}. 
\end{definition}

Since $f$ is continuous and the set of alignment operators is compact, an optimal point always exists. For concave functions, such as the von Neumann entropy, the associated spin alignment problem is a minimization problem. The orthonormal basis $(\ket{q_{i}})_{i = 1}^{d}$ and bases arising from taking tensor powers of it are called \textit{computational} bases. Operators that are diagonal in these bases are called \textit{classical}. We refer to assertions that the state tuple $(\ket{q_{1}}\bra{q_{1}}^{\otimes I})_{I \subseteq [n]}$ is optimal for a spin alignment problem as spin alignment conjectures. We mostly focus on the dependence of the problems on the state $Q$ and the objective function $f$ and implicitly take the remaining parameters to be arbitrary. For example, when we speak of the spin alignment problem associated with a state $Q$, we mean the collection of all spin alignment problems with state $Q$.

We conjecture that an alignment operator with state tuple $(\ket{q_{1}}\bra{q_{1}}^{\otimes I})_{I \subseteq [n]}$ majorizes alignment operators with other state tuples for the same $Q$ and $\mu$. That is, it has the largest possible value of the order $k$ Fan norm for each $k \in [d^{n}]$. By the doubly stochastic processing characterization of majorization, if this conjecture is true, then every spin alignment conjecture is true. We call this the \textit{strong} spin alignment conjecture. In the case where $Q$ is pure, the conjecture follows directly from Eq.~\ref{eq: Fan majorization rel self-adjoint}. So, we focus on the cases where $Q$ is mixed.

Spin alignment problems can be extended to cases where the fixed state operators are not the same in each term in the sum. For example, we can consider the more general case where the alignment operators are of the form 
\begin{align}
\label{eq: general alignment operator}
\sum_{I \subseteq [n]} \mu_{I} \rho_{I} \otimes Q_{I^{c}},
\end{align}
and for $I^{c} \subseteq [n], Q_{I^{c}} = \bigotimes_{j \in I^{c}} Q_{I^{c},j}$ such that for each $ j \in [n]$, the states in the tuple $(Q_{I^{c}, j})_{I^{c} \subseteq [n]}$ are perfectly aligned.
In this work, we consider only the case where $ Q_{I^{c}} = Q^{\otimes I^{c}}$ for some fixed qudit state $Q$, though many of our results apply in the case of these more general alignment operators.

\begin{lem}
\label{lem: pure states sufficiency}
For any spin alignment problem, it suffices to restrict to state tuples of pure states. 
\end{lem}
\begin{proof}
Notice that the alignment operator is a linear function of the state tuple, so convex combinations of state tuples correspond to convex combinations of alignment operators. By convexity of the objective function, there exists an extremal point that is optimal. The extremal points of state tuples are pure-state tuples. 
\end{proof}

\begin{remark}
    If the spin alignment conjecture for a function $g$ is true, then it also holds that $f = h \circ g$ attains its maximum over alignment operators at $(\ket{q_{1}}\bra{q_{1}}^{\otimes I})_{I \subseteq [n]}$ whenever $h$ is monotonically non-decreasing. If $h$ is monotonically non-increasing, then $f$ attains a minimum there. Examples of this include the Schatten norms and the corresponding 
     Rényi entropies.
\end{remark}

In addition to showing the additivity of coherent information for platypus channels, proving the strong spin alignment conjecture would formalize intuitions we have about how dispersion behaves under mixing. Namely, to minimize dispersion in a mixture, the optimal procedure ought to be one where each signal has the lowest dispersion possible and the signals collectively overlap as much as possible. This can be considered a refinement of Fan's majorization relation Eq.~\ref{Fan's majorization relation} because in the spin alignment problem, while each summand has minimum dispersion by a pure state choice, we are not free to choose the bases so that they are perfectly aligned.

When operators cannot be chosen to be perfectly aligned to achieve minimum dispersion, we need a way of comparing the alignment of different tuples of operators. The next definition provides such a way while restricting the tuples to have matching spectra.
\begin{definition} (Alignment)
    Let $\mathcal{T} = (T_{1}, T_{2}, \ldots, T_{\ell})$ and $\mathcal{R} = (R_{1}, R_{2}, \ldots, R_{\ell})$ be two tuples of self-adjoint operators such that for all $i \in [\ell], \lambda(T_{i}) = \lambda(R_{i})$. Then, $\mathcal{T}$ is \textit{more aligned} than $\mathcal{R}$ if for all probability measures $(p_{i})_{i \in [\ell]}$, the following majorization relation holds
    \begin{align}
    \label{def: partial alignment}
    \sum_{i =1}^{\ell} p_{i} T_{i}  \succeq  \sum_{i =1}^{\ell} p_{i} R_{i}. 
    \end{align}
\end{definition}
Given any set of $\ell$-tuples of operators such that coordinate-wise the tuples have equal spectra, alignment can be used to construct a preorder on the set. We are interested in cases where there exists a maximal element according to this ordering. The basic example is one where the set in question contains a tuple of perfectly aligned operators. The strong spin alignment conjecture may be phrased as follows. The tuple $(\ket{q_{1}}\bra{q_{1}}^{\otimes I} \otimes Q^{\otimes I^{c}})_{I \subseteq [n]}$ is more aligned than every state tuple of the form $(\ket{\psi_{I}}\bra{\psi_{I}}_{I} \otimes Q^{\otimes I^{c}})_{I \subseteq [n]}$.

A key reason to consider alignment as opposed to perfect alignment is that the latter does not behave well with respect to tensor multiplication. Specifically, the coordinate-wise tensor product of two perfectly aligned tuples is not necessarily perfectly aligned. Even worse, for self-adjoint $A, B$, and $C$, the perfect alignment of $A$ and $B$ does not imply the perfect alignment of $C \otimes A$ and $C \otimes B$. 
\begin{example}
\label{ex: tensor product and perfect alignment}
Consider $D = \ket{1}\bra{1} + \frac{1}{2} \ket{2}\bra{2} + \frac{1}{3} \ket{3}\bra{3}$. It is clear that $D$ and $\ket{1}\bra{1}$ are perfectly aligned. However, $D \otimes D$ and $D \otimes \ket{1}\bra{1}$ are not perfectly aligned. 
\end{example}
In contrast, $(C \otimes A^{\downarrow}, C \otimes B^{\downarrow})$ is more aligned than $(C \otimes A, C \otimes B)$. This is a simple consequence of Eq.~\ref{eq: Fan majorization rel self-adjoint} and the following lemma.

\begin{lem}
\label{lem: tensor product and majorization}
Let $A, B \in \mathcal{S}(\mathcal{H})$ and $C \in \mathcal{S} (\mathcal{K})$. If $A \succeq B$, then $C \otimes A \succeq C \otimes B$.
\end{lem}
\begin{proof}
This follows from the doubly stochastic characterization of majorization. $A \succeq B$ is equivalent to the existence of a mixed-unitary channel $\mathcal{N}$ such that $B = \mathcal{N} (A)$. The channel $\id \otimes \mathcal{N} 
$ is mixed-unitary and $\id \otimes \mathcal{N} (C \otimes A) = C \otimes B$.\end{proof}

If we restrict the first factors to be positive semi-definite, we can prove the following stronger statement.
\begin{prop}
\label{prop: alignment with commuting factor}
    Let $C_{1}, \ldots , C_{\ell} \in  \mathcal{P}(\mathcal{K})$ be mutually commuting. For every $A_{1}, \ldots, A_{\ell} \in \mathcal{S}(\mathcal{H})$, $(C_{i} \otimes A^{\downarrow}_{i})_{i=1}^{\ell}$ is more aligned than $(C_{i} \otimes A_{i})_{i=1}^{\ell}$.
\end{prop}
\begin{proof}
Let $p = (p_{i})_{i=1}^{\ell}$ be an arbitrary probability measure.  Let $\{ \ket{j} \}_{j = 1}^{d_{\mathcal{K}}}$ be an orthonormal basis for $\mathcal{K}$ where $C_{1}, \ldots. C_{\ell}$ are simultaneously diagonal. For each $i \in [\ell]$, let $\sum_{j=1}^{d_{\mathcal{K}}} c_{j | i} \ket{j}\bra{j}$ be the corresponding spectral decomposition of $C_{i}$. Observe that
\begin{align}
\label{eq: conditional mixed-unitary}
\sum_{i = 1}^{\ell} p_{i} C_{i} \otimes A^{\downarrow}_{i} = \sum_{i = 1}^{\ell} p_{i} \sum_{j=1}^{d_{\mathcal{K}}} c_{j|i} \ket{j}\bra{j} \otimes A^{\downarrow}_{i} = \sum_{j=1}^{d_{\mathcal{K}}} \ket{j}\bra{j} \otimes (\sum_{i = 1}^{\ell} p_{i} c_{ j| i} A^{\downarrow}_{i}).
\end{align}
For each $j \in [d_{\mathcal{K}}]$, we know from Eq.~\ref{eq: Fan majorization rel self-adjoint} that $\sum_{i = 1}^{\ell} p_{i} c_{j|i} A^{\downarrow}_{i} \succeq \sum_{i = 1}^{\ell} p_{i} c_{j|i} A_{i}$. Hence, there exists a mixed-unitary channel $\mathcal{N}_{j}$ that takes the former to the latter. The global mixed-unitary channel $\sum_{j = 1}^{d_{\mathcal{K}}}  \ket{j}\bra{j} (\cdot) \ket{j}\bra{j} \otimes \mathcal{N}_{j}$ takes $\sum_{i = 1}^{\ell} p_{i} C_{i} \otimes A^{\downarrow}_{i}$ to $\sum_{i = 1}^{\ell} p_{i} C_{i} \otimes A_{i}$. \end{proof}

The strong spin alignment conjecture subsumes a conjecture about maximizing unitarily invariant norms at the output of a tensor product of depolarizing channels. For $q \in [\frac{-1}{d^{2} -1}, 1]$, the qudit depolarizing channel $\Delta_{d, q}$ has the action: 
\begin{align}
\label{eq: depolarizing channel def}  
\Delta_{d, q} ( \cdot ) := q \id (\cdot) + (1- q) \frac{\mathds{1}}{d} \tr( \cdot ).
\end{align}
In \cite{King2003}, to show additivity of the Holevo information of depolarizing channels, King proves that for any $p \in [1,\infty)$, the Schatten $p$-norm of the output of a tensor product of depolarizing channels is maximized at a pure product input. It is reasonable to extrapolate and suspect that this holds for all unitarily invariant norms. The basic idea is that any entanglement in the input manifests as excess dispersion at the output. For $i \in [n]$, let $q_{i} \in [0,1]$ and notice that 
\begin{align}
\label{eq: depolarizing to alignment} 
\bigotimes_{i=1}^{n} \Delta_{d, q_{i}} (\rho_{[n]}) = \sum_{I \subseteq [n]} \nu_{I} \rho_{I} \otimes (\frac{\mathds{1}}{d})^{\otimes I^{c}},
\end{align}
where \( \nu_{I}= \prod_{i\in I} q_{i} \prod_{j\in I^{c}} (1-q_{j})\) and $\rho_{I}$ denotes the marginal state on $I$ of $\rho_{[n]}$. This is an alignment operator with $Q = \frac{\mathds{1}}{d}$ arising from a state tuple satisfying an instance of the quantum marginal problem \cite{Schilling2014}. Since the conjectured optimal point $(\ket{q_{1}}\bra{q_{1}}^{\otimes I})_{I \subseteq [n]}$ satisfies an instance of the quantum marginal problem, it follows that a proof of the strong spin alignment conjecture for $Q = \frac{\mathds{1}}{d}$ implies a resolution for this problem. This problem is analogous to the now-solved problem of maximizing unitarily invariant output norms for single-mode Gaussian channels. Specifically, it was shown in \cite{Mari2014} that coherent states are optimal.

\section{Results}
\label{sec:results}

\subsection{First observations}
\label{subsec: 4.4.1}

First, we show that the spin alignment conjecture is true for the case where the objective function is $\lambda_{1}$. 
\begin{prop} \label{maximum eigenvalue case}
     $\lambda_{1} ( \sum_{I \subseteq [n]} \mu_{I} \rho_{I} \otimes Q^{\otimes I^{c}})$ is maximized at $(\rho_{I})_{I \subseteq [n]} = (\ket{q_{1}}\bra{q_{1}}^{\otimes I})_{I \subseteq [n]}$.
\end{prop}
\begin{proof}
    This follows from Eq.~\ref{eq: Fan maximum prinicple}. Specifically, 
    \begin{align}
    \label{eq: inequalities for spectral radius}
        \lambda_{1} ( \sum_{I \subseteq [n]} \mu_{I} \rho_{I} \otimes Q^{\otimes I^{c}}) \leq  \sum_{I \subseteq [n]} \mu_{I} \lambda_{1}(\rho_{I} \otimes Q^{\otimes I^{c}})
        \leq  \sum_{I \subseteq [n]} \mu_{I} \lambda_{1}(Q^{\otimes I^{c}}).
    \end{align}
    At $(\ket{q_{1}}\bra{q_{1}}^{\otimes I})_{I \subseteq [n]}$, both inequalities are satisfied with equality. \end{proof}
This proves one of $d^{{n}} - 1$ conditions in Eq.~\ref{eq: majorization cond} necessary to prove the strong spin alignment conjecture. The difficulty in proving the remaining ones lies in the fact that summands of alignment operators with pure-state tuples cannot be made perfectly aligned in general.

Next, for \(Q\) of rank greater than \(1\), we prove a reduction to cases where $\lambda_{1} (Q) = \lambda_{2} (Q)$. Again, spin alignment problems with pure $Q$ can be simply resolved by using Eq.~\ref{eq: Fan majorization rel self-adjoint}. 
\begin{prop} \label{reduction to flatter spectrum}
Suppose $Q = \sum_{i=1}^{d} \lambda_{i} (Q) \ket{q_{i}}\bra{q_{i}} \in \mathcal{D} (\mathcal{H})$ has rank $\geq 2$. Define
\begin{align}
\label{def: tilde Q for mixed Q}
    \tilde{Q} = \frac{1}{1 - (\lambda_{1} (Q) - \lambda_{2} (Q))} \Big(\lambda_{2} (Q) (\ket{q_{1}}\bra{q_{1}} + \ket{q_{2}}\bra{q_{2}}) + \sum_{i=3}^{d} \lambda_{i} (Q) \ket{q_{i}}\bra{q_{i}}\Big). 
\end{align}
If for all $\tilde{\mu}$, $f( \sum_{I \subseteq [n]} \tilde{\mu}_{I} \rho_{I} \otimes \tilde{Q}^{\otimes I^{c}})$ is maximized at $(\rho_{I})_{I \subseteq [n]} = (\ket{q_{1}}\bra{q_{1}}^{\otimes I})_{I \subseteq [n]}$, then for all $\mu$, $f( \sum_{I \subseteq [n]} \mu_{I} \rho_{I} \otimes Q^{\otimes I^{c}})$ is maximized there as well. 
\end{prop} 
\begin{proof} 
Denote $\varepsilon := \lambda_{1} (Q) - \lambda_{2} (Q)$. Observe that 
$Q = (1- \varepsilon) \tilde{Q} + \varepsilon \ket{q_{1}}\bra{q_{1}}$. That is, $Q$ lies on the line segment between $\tilde{Q}$ and $\ket{q_{1}}\bra{q_{1}}$. For any probability measure $\mu$, we can express
\begin{align}
\label{eq: expanded Q into tilde Q and q_1}
    \sum_{I \subseteq [n]} \mu_{I} \rho_{I} \otimes Q^{\otimes I^{c}} = \sum_{K \subseteq [n]} \tilde{\mu}_{K} \tilde{\rho}_{K} \otimes \tilde{Q}^{\otimes K^{c}},
\end{align}
where each factor of \(Q\) on the left-hand side was replaced
  by \((1- \varepsilon) \tilde{Q} + \varepsilon \ket{q_{1}}\bra{q_{1}}\) and \( \mu_{I} \rho_{I} \otimes \varepsilon^{|J|} (1-\varepsilon)^{|(I \cup J)^{c}|} \ketbra{q_{1}}^{\otimes J}\) for \(I\cup J=K\) were summed to obtain \(\tilde{\mu}_{K}\tilde{\rho}_{K}\). By assumption,
\begin{align}
\label{estimates for Q and Q tilde alignment problems}
f(\sum_{I \subseteq [n]} \mu_{I} \rho_{I} \otimes Q^{\otimes I^{c}}) &= f(\sum_{K \subseteq [n]} \tilde{\mu}_{K} \tilde{\rho}_{K} \otimes \tilde{Q}^{\otimes K^{c}})\\
&\leq f(\sum_{K \subseteq [n]} \tilde{\mu}_{K} \ket{q_{1}}\bra{q_{1}}^{\otimes K} \otimes \tilde{Q}^{\otimes K^{c}})  = f(\sum_{I \subseteq [n]} \mu_{I} \ket{q_{1}}\bra{q_{1}}^{\otimes I} \otimes Q^{\otimes I^{c}}), 
\end{align} 
where we reversed the process used to obtain the right-hand side
of Eq.~\ref{eq: expanded Q into tilde Q and q_1} for \((\rho_{I})_{I \subseteq [n]} =(\ketbra{q_{1}}^{\otimes I})_{I \subseteq [n]}\) to arrive at the last equality. \end{proof}
\begin{remark}
For $d = 2$, this reduction implies that $Q = \frac{\mathds{1}}{2}$ may be assumed without loss of generality. One can therefore apply the argument in Sec.~6.2 of~\cite{Leditzky2022a} to show that the strong spin alignment conjecture holds for \(n=d=2\). 
\end{remark}

The last result of this subsection relates to problem instances where the union of the support of the probability measure $\mu$ is not $[n]$.  That is, those instances where $\mu$ is such that is there exists nonempty $K^{c} \subseteq [n]$ such that $\sum_{I \subseteq [n]} \mu_{I} \rho_{I} \otimes Q^{\otimes I^{c}} =  (\sum_{J \subseteq [n]\setminus K^{c}} \mu_{J} \rho_{J} \otimes Q^{\otimes J^{c}}) \otimes Q^{\otimes K^{c}}$. If the objective function is multiplicative under tensor product, such as Schatten norms \cite{AUBRUN_2011}, then we may without loss of generality trace over the qudits in $K^{c}$ and consider the optimization problem on the rest. Not all Fan norms are multiplicative under tensor product. However, to prove the strong spin alignment conjecture, it may be assumed without loss of generality that the union of the support of \(\mu\) is \([n]\). 
\begin{prop}
\label{prop: union of mu support is maximal}   
   For every $n_{1}, n_{2} \in \mathbb{N}$, if the strong spin alignment conjecture holds for $\sum_{I \subseteq [n_{1}]} \mu_{I} \rho_{I} \otimes Q^{\otimes I^{c}}$, then it holds for $\sum_{I \subseteq [n]} \mu_{I} \rho_{I} \otimes Q^{\otimes I^{c}} \otimes Q^{\otimes n_2}$.
\end{prop}
\begin{proof} This follows from Lem.~\ref{lem: tensor product and majorization}. \end{proof}

\subsection{Classical spin alignment}
\label{subsec: class spin alignment}

We prove a classical version of the strong spin alignment conjecture. That is, we prove it with the extra assumption that $(\rho_{I})_{I \subseteq [n]}$ is such that $\rho_{I}$ is diagonal in the computational basis for each $I \subseteq [n]$.

\begin{prop} \label{diagonal spin alignment}
    For every probability measure $\mu$ and  tuple of classical states $(\rho_{I})_{I \subseteq [n]}$,
    \begin{align}
    \label{rel: classical spin alignment}
             \sum_{I \subseteq [n]} \mu_{I} \rho_{I} \otimes Q^{\otimes I^{c}} \preceq   \sum_{I \subseteq [n]} \mu_{I} \ket{q_{1}}\bra{q_{1}}^{\otimes I} \otimes Q^{\otimes I^{c}}, 
    \end{align}  
\end{prop}
\begin{proof}
    By Lem.~\ref{lem: pure states sufficiency}, without loss of generality, the state tuple $(\rho_{I})_{I \subseteq [n]}$ may be taken to consist of pure classical states. For $I \subseteq [n]$, we denote $\rho_{I} = \ket{t_{I}}\bra{t_{I}}$, where $t_{I}$ is a string of length $|I|$ over the alphabet $[d]$. For example, $ \ket{1 2 3}\bra{1 2 3}  := \ket{q_{1}}\bra{q_{1}} \otimes  \ket{q_{2}}\bra{q_{2}} \otimes  \ket{q_{3}}\bra{q_{3}}$. 
    
    The idea of the proof is to start with an arbitrary classical assignment $(t_{I})_{I \subseteq [n]}$ and flip to 1 sequentially to improve the assignment making use of the transitivity of the majorization ordering. We start by considering the first qudit. We write the alignment operator in the form 
    \begin{align}
    \label{eq: decompostion of alignment operator}
        \sum_{I \subseteq [n]} \mu_{I} \ket{t_{I}}\bra{t_{I}} \otimes Q^{\otimes I^{c}} = Q \otimes B + \ket{1}\bra{1} \otimes C + \sum_{j = 2}^{d} \ket{j}\bra{j} \otimes D_{j}, 
    \end{align}
    where \(B\), \(C\), and \(D_{j}\) can be expressed as
    \begin{align}
        B &= \sum_{I \subseteq [n] \setminus \{1\} } \nu_{I} \ket{t'_{I}} \bra{t'_{I}} \otimes Q^{I^{c}},\\
        C &=  \sum_{I \subseteq [n] \setminus \{1\} } w_{I} \ket{t''_{I}} \bra{t''_{I}} \otimes Q^{I^{c}}, \\
        D_{j} &=  \sum_{I \subseteq [n] \setminus \{1\} } \kappa_{j,I} \ket{t_{j,I}} \bra{t_{j,I}} \otimes Q^{I^{c}}.
    \end{align}   
    The measures $\nu, w,$ and $\kappa_{j}$ are positive, and the operators $B, C,$ and $D_{j}$ are positive semi-definite and classical. We consider each \(j \in [d] \setminus \{1\}\) in sequence and flip the first entries of those \(t_{I}\) starting with \(j\) to \(1\), causing the change $\ket{j} \bra{j} \otimes D_{j} \gets \ket{1} \bra{1} \otimes D_{j}$. That the operator arising from this change is an alignment operator compatible with $Q$ and $\mu$ can be seen by noting that the change can be implemented by the following substitutions in $(t_{I})_{I \in \subseteq [n]}$:
     \begin{align}
         \ket{j}\bra{j} \otimes \ket{t_{j,I}} \bra{t_{j,I}} \gets \ket{1}\bra{1} \otimes \ket{t_{j,I}} \bra{t_{j,I}}
     \end{align} 
      for each $I \subseteq [n]\setminus \{1\}$.  For a string $r$ of length $n-1$ over $[d]$, let $b_{r}, c_{r},$ and $d_{j, r}$ be the eigenvalues for the eigenstate $\ket{r}$ of $B, C,$ and $D_{j, r}$ before the change. The change in the spectrum can be implemented by composing transfers of the form: $d_{j, r}$ is taken from the eigenvalue of $\ket{j}\otimes\ket{r}$ and given to the eigenvalue of $\ket{1} \otimes \ket{r}$ for each string $r$. These transfers may be implemented in parallel. For a string $r$, those two entries in the spectrum before the change are
      \begin{align}
      \label{eq: old entires}
      (q_{1} b_{r} + c_{r}, b_{r} q_{j} + d_{j,r}),  
      \end{align}
      while after, they are
      \begin{align}
      \label{eq: new entires}
      (q_{1} b_{r} + c_{r} + d_{j, r}, b_{r} q_{j}). 
      \end{align}
      To see that the spectrum after the transfer majorizes the one before it, observe that
    \begin{align}
        q_{1} b_{r} + c_{r} + d_{j,r} \geq \max{(q_{1} b_{r} + c_{r},q_{j} b_{r} + d_{j,r})}. 
    \end{align}
    That is, the larger value of the pair did not decrease after the transfer. Hence, the transfer is either unjust or its composition with a transposition is unjust. By transitivity of majorization, the same process may now be done to the remaining qudits. \end{proof}

\subsection{Schatten norms of integer order}
\label{subsec: Schatten norms of integer order}

We prove the spin alignment conjecture for Schatten norms of integer order. By extension, this implies that any monotonic function of these norms, such as the corresponding Rényi entropies, is optimized at the conjectured optimal state tuple $(\ket{q_{1}}\bra{q_{1}}^{\otimes I})_{I \subseteq [n]}$. The proof relies on the fact that overlaps between summands in an alignment operator are simultaneously maximized at the conjectured optimal state tuple. Here overlap means the absolute value of the trace of any product of such summands. In fact, the overlap lemma Lem.~\ref{overlap lemma} in Sec.~\ref{sec:app A} establishes the stronger statement that \textit{every} unitarily invariant norm of such a product is maximized at $(\ket{q_{1}}\bra{q_{1}}^{\otimes I})_{I \subseteq [n]}$.

\begin{cor} 
\label{cor: ovelap of absolute value of trace}
     Let $I_{1},\ldots,I_{\ell}$ be subsets of $[n]$. For any tuple $(R_{I_{j}})_{j=1}^{\ell}$ of operators with unit trace norm, the following inequality holds
    \begin{align} 
    \label{ineq: absolute value of trace}
        \left| \tr( \;\prod_{j=1}^{\ell}  R_{I_{j}} \otimes Q^{\otimes I_{j}^{c}}) \right| \leq   \tr( \;\prod_{j=1}^{\ell} \ket{q_{1}}\bra{q_{1}}^{\otimes I_{j}} \otimes Q^{\otimes I_{j}^{c}}). 
    \end{align}
\end{cor}
\begin{proof}
The result follows from an application of Lem.~\ref{overlap lemma} in the following chain of inequalities:
  \begin{align}
\label{ineq: chain of estimates for trace of absolute value.}
    \left| \tr( \;\prod_{j=1}^{\ell} R_{I_{j}} \otimes Q^{\otimes I_{j}^{c}})\right| &\leq \tr( \left| \;\prod_{j=1}^{\ell} R_{I_{j}} \otimes Q^{\otimes I_{j}^{c}}\right|) \nonumber\\
    &\leq \tr( \left|\;\prod_{j=1}^{\ell} \ket{q_{1}}\bra{q_{1}}^{\otimes I_{j}} \otimes Q^{\otimes I_{j}^{c}}\right|) \hspace{.25in} \textrm{by Lem.~\ref{overlap lemma}}\nonumber\\
    &=  \tr( \;\prod_{j=1}^{\ell} \ket{q_{1}}\bra{q_{1}}^{\otimes I_{j}} \otimes Q^{\otimes I_{j}^{c}}). 
\end{align} \end{proof}

\begin{thm}
\label{thm: spin alignment for Schatten norms of integer order}
For arbitrary $m \in \mathbb{N}$, the following inequality holds
\begin{align}
    || \sum_{I \subseteq [n]} \mu_{I} \rho_{I} \otimes Q^{\otimes I^{c}} ||_{m} \leq || \sum_{I \subseteq [n]} \mu_{I} \ket{q_{1}}\bra{q_{1}}^{\otimes I} \otimes Q^{\otimes I^{c}} ||_{m}
\end{align}
\end{thm}
 \begin{proof}
   Notice that $ || \sum_{I \subseteq [n]} \mu_{I} \rho_{I} \otimes Q^{\otimes I^{c}} ||_{m}^{m}$ is a linear combination of traces of
   products of operators of the form \(\rho_{I} \otimes Q^{\otimes I^{c}}\) for some $I \subseteq [n]$. By Corollary (\ref{cor: ovelap of absolute value of trace}), the absolute value of each term is maximized at
   $(\ket{q_{1}}\bra{q_{1}}^{\otimes I})_{I \subseteq [n]}$. By the triangle inequality for the absolute value and the fact that the terms in the sum have equal phase at $(\ket{q_{1}}\bra{q_{1}}^{\otimes I})_{I \subseteq [n]}$, 
   \begin{align}
        || \sum_{I \subseteq [n]} \mu_{I} \rho_{I} \otimes Q^{\otimes I^{c}} ||_{m}^{m} \leq || \sum_{I \subseteq [n]} \mu_{I} \ket{q_{1}}\bra{q_{1}}^{\otimes I} \otimes Q^{\otimes I^{c}} ||_{m}^{m}.
   \end{align}
   The claim follows from the fact that raising to the $m$th power is monotonically non-decreasing on $[0,\infty)$. \end{proof}

\subsection{The case where $\text{supp} (\mu) \leq 2$ and $Q$ is proportional to a projector}
\label{subsec: two projectors}

We now prove the strong spin alignment conjecture for alignment operators of the form
\begin{align} 
\label{spin alignment with mu 2}
(1 - t) \ket{\psi_{{1}}}\bra{\psi_{{1}}}_{I_{1}} \otimes Q^{\otimes I_{1}^{c}} + t \ket{\psi_{{2}}}\bra{\psi_{{2}}}_{I_{2}}  \otimes Q^{\otimes I_{2}^{c}}, 
\end{align}
where $t \in [0,1]$, $I_{1}, I_{2} \subseteq [n]$, and $Q$ is proportional to a projector. Since we have established that we may take $I_{1} \cup I_{2} = [n]$ without loss of generality (see Lem.~\ref{prop: union of mu support is maximal}), the cases where $I_{1} \subseteq I_{2}$ or $I_{1} \supseteq I_{2}$ are simply resolved by Eq.~\ref{eq: Fan majorization rel self-adjoint}. To address the other cases, we consider the more general problem of the alignment of two projectors subject to a certain overlap constraint.

Let $\mathcal{K}$ be given. For each tuple $(r_{1}, r_{2}, c)$, where $r_{1}, r_{2} \in [d_{\mathcal{K}}] \cup \{0\}$ and $c \geq 0$, let $G_{r_{1}, r_{2}, c}$ denote the set of pairs of projectors $(P_1, P_2)$ satisfying the conditions:  $\text{rank}(P_{1}) = r_{1}$, $\text{rank}(P_{2}) = r_{2}$, and $\tr(|P_{1} P_{2}|) \leq c$.

\begin{lem}
\label{lem: non-empty criterion for feasible sets.}
$G_{r_{1}, r_{2}, c}$ is nonempty if and only if 
$r_{1} + r_{2} - d_{\mathcal{K}} \leq \lfloor c \rfloor$.
\end{lem}
\begin{proof}
Suppose $r_{1} + r_{2} - d_{\mathcal{K}} > \lfloor c \rfloor$. If $P_{1}$ and $P_{2}$ are projectors satisfying $\text{rank} (P_{1}) = r_{1}$ and $\text{rank} (P_{2}) = r_{2}$, then from basic linear algebra 
\begin{align}
\label{ineq: basic linear algebra dim rel}
    \dim( \text{supp}(P_{1}) \cap \text{supp}(P_{2})) \geq r_{1} + r_{2} - d_{\mathcal{K}} > \lfloor c \rfloor.
\end{align}
Hence, $P_{1} P_{2}$ acts as the identity on a subspace of dimension at least $\lceil c \rceil$ and so $\tr(|P_{1} P_{2}|) > c$. This implies $G_{r_{1}, r_{2}, c}$ is empty.

If on the other hand $r_{1} + r_{2} - d_{\mathcal{K}} \leq \lfloor c \rfloor$, let $P$ be a projector of rank equal to $\max(r_{1} + r_{2} - d_{\mathcal{K}}, 0)$. Consider the two projectors $P_{1} = P + R_{1}$ and $P_{2} = P + R_{2}$, where $R_{1}, R_{2}$ are projectors satisfying $R_{1} \perp P, R_{2} \perp P, R_{1} \perp R_{2}$, and $\text{rank}(R_{1}) = r_{1} - \text{rank}(P), \text{rank}(R_{2}) = r_{2} - \text{rank}(P)$. Then $P_{1} P_{2} = P$, $\tr(|P_{1} P_{2}|) = \tr(P) \leq \lfloor c\rfloor$, and so $(P_{1}, P_{2}) \in G_{r_{1}, r_{2}, c}$. \end{proof}

From now on, we take for granted that $r_{1} + r_{2} - d_{\mathcal{K}} \leq \lfloor c \rfloor$. Moreover, since $\tr(|P_{1} P_{2}|) \leq \min(\tr(P_{1}), \tr(P_{2}))$ for any two projectors $P_{1}$ and $ P_{2}$, we take $c \leq \min(r_{1}, r_{2})$.  Otherwise, the constraint on the product may as well not be considered.

The crucial ingredient of our analysis is a characterization of the angle between two subspaces due to Camille Jordan \cite{Jordan1875}. This characterization, commonly referred to as Jordan's lemma, allows us to reason about the relationship between the eigenvalues of a sum of two projectors and the singular values of their product. 
\begin{thm} 
\label{thm: two projector fan norm}
There exists $(P^{\text{opt}}_{1}, P^{\text{opt}}_{2}) \in G_{r_{1}, r_{2} ,c}$ that is maximal in the alignment ordering on $G_{r_{1}, r_{2} ,c}$, and it satisfies $\tr(|P^{\text{opt}}_{1} P^{\text{opt}}_{2}|) = c$. 
Moreover, if $c$ is an integer, then $[P^{\text{opt}}_{1}, P^{\text{opt}}_{2}] = 0$.
\end{thm}

\begin{proof}
  The statement is equivalent to 
  \begin{align}
      s_{1} P_{1}^{\text{opt}} + s_{2} P_{2}^{\text{opt}}  \succeq s_{1} P_{1} + s_{2} P_{2} 
  \end{align}
  for all $s_{1}, s_{2} \geq 0$ and $(P_{1}, P_{2}) \in G_{r_{1}, r_{2} ,c}$. We show existence by explicit construction. Let arbitrary $s_{1}, s_{2} \geq 0$ be given and consider an arbitrary pair $(P_{1},P_{2}) \in G_{r_{1}, r_{2},c}$. We modify the pair such that it remains in $G_{r_{1}, r_{2} ,c}$ and the corresponding Fan norms do not decrease until we reach an optimal pair.

We can decompose \(\mathcal{K}\) into minimal subspaces invariant under both \(P_{1}\) and \(P_{2}\). According to Jordan's lemma, these subspaces are either one- or two-dimensional. Thus, we may write
$\mathcal{K} = \mathcal{K}^{(1)} \oplus
\mathcal{K}^{(2)}$, where $\mathcal{K}^{(1)} = \bigoplus_{i_{1} = 1}^{m_{1}}
\mathcal{K}_{i_{1}}^{(1)}$ is the direct sum of invariant subspaces of dimension 1,
and $\mathcal{K}^{(2)} = \bigoplus_{i_{2} = 1}^{m_{2}}
\mathcal{K}_{i_{2}}^{(2)}$ is the direct sum of invariant subspaces of
dimension 2. Moreover, for each $i_{2} \in
[m_{2}]$,
$P_{1}|_{\mathcal{K}_{i_{2}}}$ and
$P_{2}|_{\mathcal{K}_{i_{2}}}$ are projectors of rank
$1$. Denote the overlap $\tr (| P_{1} P_{2} |) = \tilde{c}_{1} + \tilde{c}_{2}$ where $\tilde{c}_{1} = \tr(|(P_{1} P_{2})|_{\mathcal{K}^{(1)}} |) \;
\text{and} \; \tilde{c}_{2} = \tr(|(P_{1} P_{2})|_{\mathcal{K}^{(2)}} |)$. Observe that
$\tilde{c}_{1}$ is an integer as it equals the dimension of
$\text{supp}(P_{1}|_{\mathcal{K}^{(1)}}) \cap
\text{supp}(P_{2}|_{\mathcal{K}^{(1)}})$.

First, we reduce to the case where $m_{2} \in \{0, 1\}$. For each $i_{2} \in [m_{2}]$, we notate
\begin{align}
    P_{1}|_{\mathcal{K}_{i_{2}}^{(2)}} =: \ket{\alpha_{i_{2}}}\bra{\alpha_{i_{2}}} ,    P_{2}|_{\mathcal{K}_{i_{2}}^{(2)}} =: \ket{\beta_{i_{2}}}\bra{\beta_{i_{2}}}.
\end{align}
Notice that $\tilde{c}_{2} = \sum_{i_{2} = 1}^{m_{2}} | \braket{\alpha_{i_{2}}}{\beta_{i_{2}}}|$. By Lem.~\ref{lemma: majorants}, the vector $(| \braket{\alpha_{i_{2}}}{\beta_{i_{2}}}|)_{i_{2} = 1}^{m_{2}}$ is majorized by the vector $(\underbrace{1, \ldots, 1}_{\lfloor \tilde{c}_{2} \rfloor\; \text{times}}, \tilde{c}_{2} - \lfloor \tilde{c}_{2} \rfloor, 0, \ldots,0)$. There are vectors $\ket{\alpha_{i_{2}}'}, \ket{\beta_{i_{2}}'} \in \mathcal{K}_{i_{2}}^{(2)}$, for $i_{2} \in [m_{2}]$, such that 
\begin{align}
\label{eq: majorizing tuple}
(| \braket{\alpha_{i_{2}}'}{\beta_{i_{2}}'}|)_{i_{2}=1}^{m_{2}} = (\underbrace{1, \ldots, 1}_{\lfloor \tilde{c}_{2} \rfloor\; \text{times}}, \tilde{c}_{2} - \lfloor \tilde{c}_{2} \rfloor, 0, \ldots,0).
\end{align}
 Specifically, for each $1$ we choose the two vectors to be equal and for each $0$ we choose them to be orthogonal. For the remaining entry, we pick the two such that they have overlap $\tilde{c}_{2} - \lfloor \tilde{c}_{2} \rfloor$. By Lemma (\ref{lemma: isotone}), the eigenvalues of $(s_{1} P_{1} + s_{2} P_{2}) |_{\mathcal{K}^{(2)}}$ are a strictly isotone function of the nonzero singular values of $P_{1} P_{2}|_{\mathcal{K}^{(2)}}$, and so we may make the replacements 
\begin{align}
    \ket{\alpha_{i_{2}}} \gets \ket{\alpha_{i_{2}}'} \; , \;
    \ket{\beta_{i_{2}}} \gets \ket{\beta_{i_{2}}'}
\end{align}
for each $i_{2} \in [m_{2}]$. Save for at most one, this replaces each $\mathcal{K}_{i_{2}}^{(2)}$ with two $1$-dimensional invariant subspaces. That is, we may assume without loss of generality that $m_{2} \in \{0,1\}$ and that $\tilde{c}_{2} < 1$. In particular, there is at most one $2$-dimensional subspace where $P_{1}$ and $P_{2}$ do not commute.

Now, suppose $c$ is an integer. $\lambda((s_{1} P_{1} + s_{2} P_{2})|_{\mathcal{K}^{(1)}})$ has $s_{1} + s_{2}$ with multiplicity $\tilde{c}_{1}$ along with the smaller values $s_{1}, s_{2}$ and $0$. If $\tilde{c}_{2} > 0$, then we may adjust the action of the projectors on the sole invariant $2$-dimensional subspace so that they are equal when restricted to it. This does not cause the two projectors to exit $G_{r_{1}, r_{2},c}$ as $c - \tilde{c}_{1}$ is a positive integer. Hence, we may assume without loss of generality that $\tilde{c}_{2} = 0$. Then, the number of one-dimensional invariant subspaces
  with \(s_{1}\) is $r_{1} - \tilde{c}_{1} \geq c - \tilde{c}_{1}$ and the number of one-dimensional invariant subspaces with $s_{2}$ is $r_{2} - \tilde{c}_{1} \geq c - \tilde{c}_{1}$. We pair up $c - \tilde{c}_{1}$ many of the two kinds of invariant one-dimensional subspaces to get $c - \tilde{c}_{1}$ two-dimensional subspaces. We adjust the projectors so that they are equal when restricted to these $c - \tilde{c}_{1}$ subspaces. For each such subspace, this causes a transfer on the spectrum of $s_{1} P_{1} + s_{2} P_{2}$ of the form: 
\begin{align}
 s_{1} \gets 0 \; , \; s_{2} \gets s_{1} + s_{2}.
\end{align}
 Since this is either an unjust transfer or an unjust transfer composed with a transposition, the Fan norms of $s_{1} P_{1} + s_{2} P_{2}$ do not decrease after these transfers. Denote the pair arising after all the transfers are done with $(P_{1}^{\text{opt}}, P_{2}^{\text{opt}})$. Notice that $\tr (P_{1}^{\text{opt}} P_{2}^{\text{opt}}) = c$ and $[P_{1}^{\text{opt}}, P_{2}^{\text{opt}}] = 0$.

On the other hand, consider the case where $c$ is not an integer. If $m_{2} = 0$, then we may take $\tilde{c}_{1} = \lfloor c \rfloor$ by the arguments in the previous paragraph. Consider two invariant $1$-dimensional subspaces where the two projectors are orthogonal. Such a pair exists because $c \leq \lceil c \rceil \leq \min(r_{1}, r_{2})$. Then, we adjust the action of the two projectors on the direct sum of these two subspaces so that $m_{2} = 1$ and $\tilde{c}_{2} = c - \lfloor c \rfloor$. That is, we make the projectors align there as much as possible. From Eq.~\ref{eq: eigenvalue equation for two rank1 projector sum}, this is either an unjust transfer or an unjust transfer composed with a transposition. If $m_{2} = 1$, then we adjust the overlap in the $2$-dimensional invariant subspace until $\tilde{c}_{2} = c - \lfloor c \rfloor$. Denote the pair arising after these adjustments with $(P^{\text{opt}}_{1}, P^{\text{opt}}_{2})$. \end{proof}

By Thm.~\ref{thm: two projector fan norm}, if $c \in \mathbb{N}$, then any pair of commuting projectors in $G_{r_{1}, r_{2} ,c}$ whose product has rank $c$ is optimal. By Lem.~\ref{overlap lemma}, the projectors in the alignment operator Eq.~\ref{spin alignment with mu 2} have overlap at most $(\text{rank}(Q))^{|(I_{1} \cup I_{2})^{c}|}$.
\begin{cor} 
\label{corollary: two term projector fan norms optimality}
   For $I_{1}, I_{2} \subseteq [n]$ and $Q \in \mathcal{D}(\mathcal{H})$ proportional to a projector, the pair $
   (\ket{q_{1}}\bra{q_{1}}^{\otimes {I_{1}}} \otimes Q^{\otimes I_{1}^{c}}, \ket{q_{1}}\bra{q_{1}}^{\otimes {I_{2}}} \otimes Q^{\otimes I_{2}^{c}})$
   is more aligned than $( \ket{\psi_{{1}}}\bra{\psi_{{1}}}_{I_{1}} \otimes Q^{\otimes I_{1}^{c}}, \ket{\psi_{{2}}}\bra{\psi_{{2}}}_{I_{2}}  \otimes Q^{\otimes I_{2}^{c}})$.
\end{cor}

\section{``Dual" problem to spin alignment and its generalizations}
\label{sec:``dual" problem}

We introduce a general class of conjectures that formalize the intuition that under a fixed global spectrum constraint, the least noisy multipartite states are classical. The inspiration came from considering spin alignment problems with $Q \propto \mathds{1}$ and using Fan's maximum principle Eq.~\ref{eq: Fan maximum prinicple}. Specifically, for a positive measure $\nu$ and a state tuple $(\ket{\psi_{I}} \bra{\psi_{I}})_{I \subseteq [n]}$, we consider 

\begin{align}
    \max_{P: \text{rank}(P) = k}  \tr( P_{[n]} \sum_{I \subseteq [n]} \nu_{I} \ket{\psi_{I}} \bra{\psi_{I}} \otimes \mathds{1}^{\otimes I^{c}}) &=    \max_{P: \text{rank}(P) = k} \sum_{I \subseteq [n]}  \nu_{I} \tr( P_{I} \ket{\psi_{I}} \bra{\psi_{I}}) \\ &\leq    \max_{P: \text{rank}(P) = k} \sum_{I \subseteq [n]} \nu_{I} \lambda_{1} (P_{I}).   
\end{align}

 At $(\ket{q_{1}}\bra{q_{1}}^{\otimes I})_{I \subseteq [n]}$, the inequality is satisfied with equality. This maximization problem may be
generalized by letting the optimization variable be quantum states
with a prescribed global spectrum, and by replacing
$\lambda_{1}$ with any continuous, convex,
unitarily invariant function. Moreover, since the objective function is convex,
we may without loss of generality take the convex closure of the set of multipartite states with a prescribed spectrum. We formalize this below.

Let inner product spaces $\mathcal{H}_{1}, ..., \mathcal{H}_{n}$ be given and denote $\mathcal{H}_{I} = \bigotimes_{i \in I} \mathcal{H}_{i}$ for $I \subseteq [n]$. For a probability measure $p \in \mathbb{R}^{\Pi_{i \in [n]} d_{\mathcal{H}_{i}}}$, consider the maximization problem:
\begin{align} 
\label{``dual" program conjecture}
    &\max_{\tau_{[n]} \in \mathcal{D} (\mathcal{H}_{[n]})} \quad \quad \sum_{I \subseteq [n]} f_{I} ( \tau_{I}), \\
    &\text{subject to:} \quad \quad \; \lambda( \tau_{[n]} ) \preceq p, 
\end{align}
where for all $I \subseteq [n]$, $f_{I}$ is continuous, convex, and unitarily invariant on a domain containing $\mathcal{D} (\mathcal{H}_{I})$. We conjecture that there exists a classical optimal point. This is self-evident in the case where $p$ has only one nonzero value because in that case, it is possible to maximize all the functions in the sum simultaneously. More interesting are the cases where the optimization is over strictly mixed states. An example of such a problem in the quantum information literature appears in Ref.~\cite{Wilde2014}. Namely, Eq.~39 there gives an upper bound on the fidelity of entanglement generating pure codes in terms of a sum as above where for each $I \subseteq [n]$, $f_{I} (\cdot) \propto || \cdot ||_{\alpha}^{\alpha}$ for a fixed $\alpha \in [1,2]$.

Below, we show that this ``dual" conjecture holds in the special case where non-constant functions in $(f_{I})_{I \subseteq [n]}$ correspond to members of a partition of $[n]$.

\begin{prop} 
\label{``dual" program with part}
    Let $\mathcal{W} \subseteq 2^{[n]}$ be a partition of $[n]$. For each $I \in \mathcal{W}$, let $f_{I} : \mathcal{D} (\mathcal{H}_{I}) \rightarrow \mathbb{R}$ be a continuous, convex, and unitarily invariant function. Then, the problem
    \begin{align} 
    \label{``dual" program statement for partitioning}
        &\max_{\tau_{[n]} \in \mathcal{D} (\mathcal{H}_{[n]})} \quad \quad \sum_{I \in \mathcal{W}} f_{I} ( \tau_{I}), \\
        &\text{subject to:} \quad \; \;  \lambda( \tau_{[n]} ) \preceq p. 
    \end{align}
    has a classical solution. 
\end{prop}
\begin{proof}
   Let $\tau_{[n]}\in \mathcal{D} (\mathcal{H}_{[n]})$ be an arbitrary quantum state such that $\lambda( \tau_{[n]} ) \preceq p$. For each $I \in \mathcal{W}$, if necessary, a unitary channel $\mathcal{U}_{I} \otimes \id_{I^{c}}$ may be applied so that $\tau_{I}$ is classical. This does not affect the objective function. For each $I \in \mathcal{W}$, let $\mathcal{N}_{\tau, I}$ denote the quantum channel that fully decoheres in a basis of $\tau_{I}$. That is, it has the pinching action $ \mathcal{N}_{\tau, I} (\cdot ) = \text{diag}(\cdot)$. It is clear that $\tau'_{[n]} = \bigotimes_{I \in \mathcal{W}} \mathcal{N}_{\tau, I} (\tau_{[n]})$ is classical, and for each $I \in \mathcal{W}$, $\tau'_{I} = \tau_{I}$. Since pinching is mixed-unitary (see chapter 4 of \cite{Watrous2018}
    ), $\lambda(\tau'_{[n]} ) \preceq  \lambda(\tau_{[n]} ) \preceq p$. Since $\tau_{[n]}\in \mathcal{D} (\mathcal{H}_{[n]})$ was arbitrary, there exists an optimal state that is classical. 
\end{proof}

\section{Concluding remarks}
\label{sec:conclusion}
We generalized and systematically studied spin alignment problems. We gave non-trivial examples of cases where tuples of spectrally constrained self-adjoint operators are more aligned than others. These can be considered refinements of Ky Fan's relation Eq.~\ref{Fan's majorization relation}. We conclude with a non-exhaustive list of problems in this area that we hope will help guide future research.
\begin{enumerate}
    \item (\textbf{Schatten norms of non-integer order}) We used the corollary Cor.~\ref{cor: ovelap of absolute value of trace} to resolve the spin alignment problem for Schatten norms of integer order. It is of interest to know if a similar approach can be used in cases where the order of the norm is not an integer. Suppose that $A_{0}, A_{1}, B_{0}, B_{1} \geq 0$ satisfy $\lambda(A_{0}) = \lambda(B_{0})$, $\lambda(A_{1}) = \lambda(B_{1})$. For a string $s \in \{0, 1\}^{*}$, define the ordered products $s (A_{0}, A_{1}) := \Pi_{i \in [|s|]} A_{s(i)}$ and $s(B_{0}, B_{1}) := \Pi_{i \in [|s|]} B_{s(i)}$. If for all strings $s$, $\tr(s (A_{0}, A_{1})) \geq |\tr(s (B_{0}, B_{1}))|$, does it hold that $|| A_{0} + A_{1} ||_{p} \geq || B_{0} + B_{1} ||_{p}$ for all $p \in [1,\infty)$? If so, then the overlap lemma may be used to prove the spin alignment conjecture for all Schatten norms, and by extension for the von Neumann entropy as well. 

    \item (\textbf{Algebraic approximations of Fan norms}) The Schatten norms of integer order may be said to be algebraic norms, in that they can be defined using polynomials of the absolute value of the operator in question. Since $\lim_{m \rightarrow \infty} || \cdot ||_{m} = || \cdot ||$, they approximate the Fan norm of order $1$ arbitrarily well. This allows for a resolution of the question posed in Ex.~\ref{ex: toy example} in the case where the state $\tau$ is a qutrit state. Clearly, the trace norm is algebraic as well. Do there exist algebraic operator norms that approximate other Fan norms? If so, is there any interesting structure to the corresponding polynomials, especially with regard to spin alignment?

    \item (\textbf{Separable refinement of Ky Fan's relation}) Given $A_{0}, A_{1}, B_{0}, B_{1} \geq 0$, does the following majorization relation
    \begin{align}
    \label{rel: sep Ky Fan}
    \lambda ( A_{0} \otimes B_{0} + A_{1} \otimes B_{1}) \preceq  \lambda ( A^{\downarrow}_{0} \otimes B^{\downarrow}_{0} + A^{\downarrow}_{1} \otimes B^{\downarrow}_{1})    
    \end{align}
    hold? In the special case where $\text{rank} (B_{0}) = \text{rank} (B_{1}) = 1$, we know the answer to be yes by way of a characterization of separable states due to Nielsen and Kempe \cite{Nielsen2001}. Specifically, they showed that separable states contain more dispersion globally than locally. If the relation Eq.~\ref{rel: sep Ky Fan} holds more generally, then it may be used to prove a separable version of the strong spin alignment conjecture.   
\end{enumerate}

\section{The overlap lemma}
\label{sec:app A}

In this appendix, we prove the statement
\begin{lem} 
\label{overlap lemma}
Let $I_{1},\ldots,I_{\ell}$ be a family of subsets of $[n]$. For any unitarily invariant norm $||| \cdot |||$, the maximization problem 
\begin{align}  \label{overlap equation}  
\max_{(T_{I_j})_{j=1}^{\ell}}     ||| \; \;\prod_{j=1}^{\ell} T_{I_{j}} \otimes Q^{\otimes I_{j}^{c}} \; |||,
\end{align}
where the variable tuple $(T_{I_{j}})_{j=1}^{\ell}$ is of operators with unit trace norm, has $(\ket{q_{1}}\bra{q_{1}}^{\otimes I_{j}})_{j=1}^{\ell}$ as a solution. \end{lem}
Notice that the objective function above is convex in each coordinate if the rest are fixed. Hence,  without loss of generality, by the singular value decomposition, we take $(T_{I_{j}})_{j=1}^{\ell} = (\ket{\psi_{j}}\bra{\tilde{\psi}_{j}}_{I_{j}})_{j=1}^{\ell}$. We argue by reducing to the case where $Q \propto \mathds{1}$.

\begin{lem}
\label{lem: overlap of O and identity}
Consider a family of sets $I_{1}, \ldots, I_{\ell} \subseteq [n]$. For each $j \in [\ell]$, let $ O_{I_{j}^{c}} \in \mathcal{B} (\mathcal{H}^{\otimes I_{j}^c})$ be a completely factorizable operator satisfying $|| O_{I_{j}^{c}} || \leq 1$. Then, for any tuple of normalized (in trace norm) operators $(\ket{\psi_{j}}\bra{\tilde{\psi}_{j}}_{I_{j}})_{j=1}^{\ell}$, there exists a tuple of sub-normalized operators $(\ket{\phi_{j}}\bra{\tilde{\phi}_{j}}_{I_{j}})_{j=1}^{\ell}$ such that
\begin{align}
    \prod_{j=1}^{\ell} \ket{\psi_{j}}\bra{\tilde{\psi}_{j}}_{I_{j}} \otimes O_{ I_{j}^{c}}
    = (\prod_{j=1}^{\ell} \ket{\phi_{j}}\bra{\tilde{\phi}_{j}}_{I_{j}} \otimes \mathds{1}^{\otimes I_{j}^{c}}) \; (\mathds{1}_{\cup_{j=1}^{\ell} I_{j}} \otimes \tilde{O}_{(\cup_{j=1}^{\ell} I_{j})^{c}}), 
\end{align}
where $\tilde{O}_{(\cup_{j=1}^{\ell} I_{j})^{c}}$ is the product of all factors acting on the qudits in ${(\cup_{j=1}^{\ell} I_{j})^{c}}$. \end{lem}

\begin{proof}
We proceed via induction on $\ell$. The case with $\ell = 1$ is trivial. By the inductive hypothesis, 
\begin{align*}
(&\ket{\psi_{1}}\bra{\tilde{\psi}_{1}}_{I_{1}} \otimes O_{I_{1}^{c}} \ldots 
\ket{\psi_{\ell-1}}\bra{\tilde{\psi_{\ell-1}}}_{I_{\ell-1}} \otimes O_{I_{\ell-1}^{c}}) 
(\ket{\psi_{\ell}}\bra{\tilde{\psi_{\ell}}}_{I_{\ell}} \otimes O_{I_{\ell}^{c}})\\
=(&\ket{\eta_{1}}\bra{\tilde{\eta}_{1}}_{I_{1}} \otimes \mathds{1}_{I_{1}^{c}} \ldots 
\ket{\eta_{\ell-1}}\bra{\tilde{\eta_{\ell-1}}}_{I_{\ell-1}} \otimes \mathds{1}_{I_{\ell-1}^{c}} \; \mathds{1}_{\cup_{j=1}^{\ell-1} I_{j}} \otimes \tilde{O}_{(\cup_{j=1}^{\ell-1} I_{j})^{c}})
(\ket{\psi_{\ell}}\bra{\tilde{\psi_{\ell}}}_{I_{\ell}} \otimes O_{I_{\ell}^{c}}).
\end{align*}
Denote $J =  I_{\ell} \; \cap \; (\cup_{j = 1}^{\ell-1} I_{j})^{c}$. If $J = \emptyset$, then we set $\ket{\phi_{\ell}}\bra{\tilde{\phi}_{\ell}}_{I_{\ell}} = \ket{\psi_{\ell}}\bra{\tilde{\psi}_{\ell}}_{I_{\ell}}$. Otherwise, let $\tilde{O}_{(\cup_{j=1}^{\ell-1} I_{j})^{c}, J}$ denote the factor in $\tilde{O}_{(\cup_{j=1}^{\ell-1} I_{j})^{c}}$ acting on the qudits in $J$ and set  
\begin{align*}
    \ket{\phi_{\ell}}\bra{\tilde{\phi}_{\ell}}_{I_{\ell}} = \tilde{O}_{(\cup_{j=1}^{\ell-1} I_{j})^{c}, J} \otimes \mathds{1}^{\otimes I_{\ell} \setminus J} \ket{\psi_{\ell}}\bra{\tilde{\psi}_{\ell}}_{I_{\ell}}.
\end{align*}  
Similarly, for each $S \subseteq I_{\ell}^{c}$, let $O_{I_{\ell}^{c}, S}$ denote the factor in $O_{I_{\ell}^{c}}$ acting on the qudits in $S$. If $S$ is empty, $O_{I_{\ell}^{c}, S}$ is taken to equal identity. For each $j \in [\ell-1]$, define $ K_{j} = \{ k \in I_{j} | k \in I_{\ell}^{c}, \forall j' > j, k \notin I_{j'} \}$. For each $j \in [\ell-1]$, we set
\begin{align*}
\ket{\phi_{j}}\bra{\tilde{\phi}_{j}}_{I_{j}} =  
\ket{\eta_{j}}\bra{\tilde{\eta}_{j}}_{I_{j}} O_{I_{\ell}^{c}, K_{j}} \otimes \mathds{1}^{\otimes I_{j} \setminus K_{j}}. 
\end{align*}
The remainder of the factors in $O_{I_{\ell}^{c}}$, if any, can be composed with the remainder of the factors in $\tilde{O}_{(\cup_{j=1}^{\ell-1} I_{j})^{c}}$, if any, and written as $\tilde{O}_{(\cup_{j=1}^{\ell} I_{j})^{c}}$. \end{proof}
The next corollary is an immediate consequence of Lem~\ref{lem: overlap of O and identity}. 
\begin{cor} \label{corollary ineq}
For any  $O_{I_{1}^{c}},\ldots, O_{I_{k}^{c}},$ and $\tilde{O}_{(\cup_{j=1}^{\ell} I_{j})^{c}}$ satisfying the conditions of the previous lemma, it holds that
\begin{align*}
\max_{(\ket{\psi_{j}}\bra{\tilde{\psi}_{j}}_{I_{j}})_{j=1}^{k}}      &||| \prod_{j=1}^{\ell} \ket{\psi_{j}}\bra{\tilde{\psi}_{j}}_{I_{j}} \otimes O_{I_{j}^{c}} ||| \\
     &\leq  \max_{(\ket{\psi_{j}}\bra{\tilde{\psi}_{j}}_{I_{j}})_{j=1}^{k}}    ||| (\prod_{j=1}^{\ell}\ket{\psi_{j}}\bra{\tilde{\psi}_{j}}_{I_{j}} \otimes \mathds{1}^{\otimes I_{j}^{c}}) (\mathds{1}_{\cup_{j=1}^{\ell} I_{j}} \otimes \tilde{O}_{(\cup_{j=1}^{\ell} I_{j})^{c}}) |||, 
\end{align*}
where the optimization variable $(\ket{\psi_{j}}\bra{\tilde{\psi}_{j}}_{I_{j}})_{j=1}^{k}$ is a tuple of normalized operators.
\end{cor}

Next, we prove a lemma that subsumes the case where $Q \propto \mathds{1}$. 
\begin{lem} \label{overlap identity}
Let $I_{1}, I_{2}, \ldots, I_{\ell} \subseteq [n]$ be a family of sets and $\tilde{R}_{(\cup_{j=1}^{\ell} I_{j})^{c}}$ be an arbitrary operator acting on the qudits in ${(\cup_{j=1}^{\ell} I_{j})^{c}}$. The state tuple $(\ket{q_{1}}\bra{q_{1}}^{\otimes I_{j}})_{j=1}^{\ell}$ is a solution of the maximization problem
\begin{align} 
\max_{(\ket{\psi_{j}}\bra{\tilde{\psi}_{j}}_{I_{j}})_{j=1}^{\ell}}    ||| \;  (\prod_{j=1}^{\ell} \ket{\psi_{j}}\bra{\tilde{\psi}_{j}}_{I_{j}} \otimes \mathds{1}^{\otimes I_{j}^{c}}) \mathds{1}_{\cup_{j=1}^{\ell} I_{j}} \otimes \tilde{R}_{(\cup_{j=1}^{\ell} I_{j})^{c}}  \; |||, 
\end{align}
where the variable tuple $(\ket{\psi_{j}}\bra{\tilde{\psi}_{j}}_{I_{j}})_{j=1}^{\ell}$
is of normalized operators.   
\end{lem}

\begin{proof}
We proceed via induction on the number of sets $\ell$. If $\ell=1$, then all choices of $\ket{\psi_{1}}\bra{\tilde{\psi_{1}}}_{I_{1}}$ yield the same objective value, proving the base case. For $\ell >1$, define
$A_{j} := \ket{\psi_{j}}\bra{\tilde{\psi}_{j}}_{I_{j}} \otimes \mathds{1}^{\otimes I_{j}^{c}}$ for $j \in [\ell]$, and $J := I_{1} \cap I_{2}$. If $J = \emptyset$, then $A_{1} A_{2} = \ket{\psi_{1}}\bra{\tilde{\psi}_{1}}_{I_{1}} \otimes \ket{\psi_{2}}\bra{\tilde{\psi}_{2}}_{I_{2}} \otimes \mathds{1}^{\otimes (I_{1} \cup I_{2})^{c}}$ and the number of sets is reduced to $\ell-1$. Otherwise, we consider the Schmidt decompositions:
     \begin{align}
     \ket{\tilde{\psi}_{1}}_{I_{1}} &= \sum_{i} \sqrt{\lambda_{i}} \ket{\gamma_{i}}_{J} \ket{\alpha_{i}}_{I_{1} \setminus J} \;, \;
      \ket{\psi_{2}}_{I_{2}} = \sum_{j} \sqrt{\mu_{j}} \ket{\delta_{j}}_{J} \ket{\beta_{j}}_{I_{2} \setminus J}. 
     \end{align}
     We expand
     \begin{align}
     A_{1} A_{2} = \sum_{i, j} \sqrt{\lambda_{i}} \sqrt{\mu_{j}} \braket{\gamma_{i}}{\delta_{j}} \underbrace{\ket{\psi_{1}}_{I_{1}} \ket{\beta_{j}}_{I_{2} \setminus J} \bra{\alpha_{i}}_{I_{1} \setminus J} \bra{\tilde{\psi_{2}}}_{I_{2}}}_{{B_{i,j}}} \otimes   \mathds{1}^{\otimes (I_{1} \cup I_{2})^{c}}.
     \end{align}  
     Observe that $B_{i', j'}^{*} B_{i, j} = 0$ if $j \neq j'$. Define $C^{2} = (A_{1} A_{2})^{*} A_{1} A_{2}$ and notice that in the expression
     \begin{align}
     C^{2} = (\sum_{i, i'} \sqrt{\lambda_{i}} \sqrt{\lambda_{i'}} \; \underbrace{\bra{\gamma_{i}} (\sum_{j} \mu_{j} \ket{\delta_{j}}\bra{\delta_{j}}) \ket{\gamma_{i'}}}_{\rho_{i', i}} \; \ket{\alpha_{i'}}\bra{\alpha_{i}}_{I_{1} \setminus J}) \otimes \ket{\tilde{\psi_{2}}}\bra{\tilde{\psi_{2}}}_{I_{2}}
     \otimes \mathds{1}^{\otimes (I_{1} \cup I_{2})^{c}}
     \end{align}
     the factor acting on the qudits in $I_{1} \setminus J$ is a Hadamard product of two states 
     \begin{align}
     \rho = \sum_{i', i} \rho_{i', i} \ket{\alpha_{i'}} \bra{\alpha_{i}} \text{ and } \sigma = \sum_{i', i} \sqrt{\lambda_{i'} \lambda_{i}} \ket{\alpha_{i'}} \bra{\alpha_{i}}.
     \end{align}
     By the Schur product theorem, $\rho \odot \sigma \geq 0$. Moreover, by Schur-concavity of the square root and the fact that $\lambda (\cdot) \succeq \text{diag}(\cdot)$, 
     \begin{align}
         \tr (\sqrt{\rho \odot \sigma}) \leq \tr(\sqrt{\text{diag}(\rho \odot \sigma)})) \leq \sum_{i} \sqrt{\lambda_{i}} \sqrt{\rho_{i, i}}
         \leq 1. 
     \end{align}
     Hence, the positive square root  
     \begin{align}
     C = \Tilde{C}_{I_{1} \setminus J} \otimes \ket{\tilde{\psi_{2}}}\bra{\tilde{\psi_{2}}}_{I_{2}}
     \otimes \mathds{1}^{\otimes (I_{1} \cup I_{2})^{c}}
     \end{align} 
     satisfies $\Tilde{C} \geq 0$ and $\tr (\Tilde{C}) \leq 1$. Since unitarily invariant norms satisfy $||| \; \cdot \; ||| = ||| \; | \cdot | \; |||$, $A_{1} A_{2}$ can be replaced by $C$. Since $C$ is a sub-convex combination of terms of the form $
     \ket{c}\bra{c}_{I_{1} \setminus J} \otimes \ket{\tilde{\psi_{2}}}\bra{\tilde{\psi_{2}}}_{I_{2}}
     \otimes \mathds{1}^{\otimes (I_{1} \cup I_{2})^{c}}$
     the statement follows by the triangle inequality and the inductive hypothesis. \end{proof}

Finally, we are ready to put it all together and prove Lem.~\ref{overlap lemma}.
\begin{proof}
     Without loss of generality $(T_{I_{j}})_{j=1}^{\ell} = (\ket{\psi_{j}}\bra{\tilde{\psi}_{j}}_{I_{j}})_{j=1}^{\ell}$. Let $U = \cup_{j=1}^{\ell} I_{j}$ and let $\tilde{Q}_{U^{c}}$ denote the products of factors of $Q$ supported on $U^{c}$. By using the staircase decomposition of $Q$ and expanding sums over products in $U$, the operator product in Eq.~\ref{overlap equation} can be written as a positive sum of terms that have the form
     \begin{align} \label{projector overlap}
     (\prod_{j=1}^{\ell} \ket{\psi_{j}}\bra{\tilde{\psi}_{j}}_{I_{j}} \otimes P_{U \setminus I_{j}}) \otimes \tilde{Q}_{U^{c}} .  
     \end{align}
     where $P_{U \setminus I_{1}}, P_{U \setminus I_{2}}, \ldots, P_{U \setminus I_{\ell}}$ are tensor products of the various perfectly aligned single-qudit projectors decomposing $Q$. Since 
     \begin{align}
     (\prod_{j=1}^{\ell} \ket{q_{1}}\bra{q_{1}}^{\otimes I_{j}} \otimes P_{U \setminus I_{j}}) \otimes \tilde{Q}_{U^{c}}   = 
      (\prod_{j=1}^{\ell} \ket{q_{1}}\bra{q_{1}}^{\otimes I_{j}} \otimes \mathds{1}_{U \setminus I_{j}}) \otimes \tilde{Q}_{U^{c}},
     \end{align}
       by Cor.~\ref{corollary ineq} and Lem.~\ref{overlap identity}, $|||  (\prod_{j=1}^{\ell} \ket{\psi_{j}}\bra{\tilde{\psi}_{j}}_{I_{j}} \otimes P_{U \setminus I_{j}})  \otimes \tilde{Q}_{U^{c}}  |||$ is maximized at $(\ket{q_{1}}\bra{q_{1}}^{\otimes I_{j}})_{j=1}^{\ell}$. Finally, notice that $(\prod_{j=1}^{\ell} \ket{q_{1}}\bra{q_{1}}^{\otimes I_{j}} \otimes P_{U \setminus I_{j}}) \otimes \tilde{Q}_{U^{c}} = \ket{q_{1}}\bra{q_{1}}^{\otimes U} \otimes \tilde{Q}_{U^{c}}$. That is, not only are all the summands maximized at $(\ket{q_{1}}\bra{q_{1}}^{\otimes I_{j}})_{j=1}^{\ell}$, they are also proportional to each other with positive proportionality constants there. Therefore, the statement follows by the triangle inequality for the norm $||| \cdot |||$. 
     \end{proof}

\section{On the relationship between the sum of two projectors and their product}
\label{sec:app B}

In this appendix, we prove lemmas necessary to elucidate the relationship between a non-negative linear combination $s_{1} P_{1} + s_{2} P_{2}$ of two projectors $P_{1}, P_{2} 
\in \mathcal{S}(\mathcal{K})$ and their product $P_{1} P_{2}$. By \cite{Jordan1875}, $\mathcal{K}$ may be decomposed into a direct sum of subspaces, each of dimension at most $2$, that are invariant under the action of both $P_{1}$ and $P_{2}$. Moreover, when restricted to each invariant subspace, the two projectors have rank at most 1. So, we may write $\mathcal{K}$ as an orthogonal direct sum of one- and two-dimensional minimal invariant subspaces
\begin{align}
\mathcal{K} = \bigoplus_{i_{1} = 1}^{m_{1}} 
\mathcal{K}_{i_{1}}^{(1)} \oplus \bigoplus_{i_{2} = 1}^{m_{2}} \mathcal{K}_{i_{2}}^{(2)}, 
\end{align}
where the $K_{i}^{(1)}$ are one-dimensional and the $K_{i}^{(2)}$ are two dimensional. By minimality, the latter contain no proper nonzero invariant subspace.
For each $i_{2} \in [m_{2}]$, we notate
\begin{align}
    P_{1}|_{\mathcal{K}_{i_{2}}^{(2)}} =: \ket{\alpha_{i_{2}}}\bra{\alpha_{i_{2}}} \; , \;    P_{2}|_{\mathcal{K}_{i_{2}}^{(2)}} =: \ket{\beta_{i_{2}}}\bra{\beta_{i_{2}}}.
\end{align}
$P_{1}$ and $P_{2}$ commute if and only if $m_{2} = 0$. The difficulty in reasoning about the eigenvalues of a linear combination of two projectors lies in these subspaces where they do not commute.

When restricted to $\bigoplus_{i_{2} = 1}^{m_{2}} \mathcal{K}_{i_{2}}^{(2)}$, the nonzero singular of $P_{1} P_{2}$ are $(| \braket{\alpha_{i_{2}}}{\beta_{i_{2}}}|)_{i_{2} = 1}^{m_{2}}$. The eigenvalues of the restriction $(s_{1} P_{1} + s_{2} P_{2}) |_{\mathcal{K}_{i_{2}}^{(2)}}$ may be computed as
\begin{align}
\label{eq: eigenvalue equation for two rank1 projector sum}
\frac{1}{2} ( (s_{1} + s_{2}) \pm \sqrt{(s_{1} - s_{2})^{2} + 4 s_{1} s_{2} | \braket{\alpha_{i_{2}}}{\beta_{i_{2}}}|^{2}}).
\end{align}
Hence, the eigenvalues of $s_{1} P_{1} + s_{2} P_{2}$ are a function of the singular values of the product $P_{1} P_2$. We show next that it is in fact a \textit{strictly isotone} function (see page 41 of Ref.~\cite{Bhatia1997}). A strictly isotone function $G$ is one that preserves the majorization ordering in the sense that if $v\succeq w$, then $G(v)\succeq G(w)$.  That is, the less dispersed the singular values of $P_{1} P_{2}$, the less dispersed the eigenvalues of $s_{1} P_{1} + s_{2} P_{2}$. This is a consequence of the fact that for $a , b \geq 0$, the map $x \mapsto \sqrt{a + b x^{2}}$ is convex on $\mathbb{R}_{\geq 0}$. 
\begin{lem} \label{lemma: isotone}
Let $A \subseteq \mathbb{R}$ be convex and $g: A \rightarrow \mathbb{R}_{\geq 0}$ be a convex function. For $t \in \mathbb{R}$, define the mapping $G: A^{m} \rightarrow \mathbb{R}^{2 m}$ with action
\begin{align}
(v_{1},\ldots, v_{m}) \mapsto (t + g(v_{1}), \ldots, t + g(v_{m}) ) \oplus (t - g(v_{1}), \ldots, t - g(v_{m}) ).% = (t \pm g(v_{1}), t \pm g(v_{2}), \ldots, t \pm g(v_{m}) )
\end{align}
Then, $G$ is strictly isotone. \end{lem}
\begin{proof}
Let $v, w \in A^{m}$ be such that $v \succeq w$. Observe that for $k \in [m]$, 
\begin{align}
\sum_{j=1}^{k} G(v)^{\downarrow}_{j} = k t + \sum_{j=1}^{k} g(v)_{j}^{\downarrow} \geq k t + \sum_{j=1}^{k} g(w)_{j}^{\downarrow} = \sum_{j=1}^{k} G(w)^{\downarrow}_{j},
\end{align}
where the inequality follows from the convexity of $g$ and the doubly-stochastic characterization of majorization (see, for example, Theorem II.3.3 on page 41 of \cite{Bhatia1997}).  If $k > m$, then 
\begin{align}
\sum_{j=1}^{k} G(v)^{\downarrow}_{j} &= k t + \sum_{j=1}^{m} g(v)_{j}^{\downarrow} - \sum_{j=1}^{k - m} g(v)_{j}^{\uparrow} = k t + \sum_{j=1}^{2m - k} g(v)_{j}^{\downarrow}\\
&\geq k t + \sum_{j=1}^{2m - k} g(w)_{j}^{\downarrow} = \sum_{j=1}^{k} G(w)^{\downarrow}_{j}.
\end{align}
Since $\sum_{j=1}^{2 m} G(\cdot)_{j} = 2 m t$, $G(v) \succeq G(w)$. \end{proof}

The following lemma exhibits the maximal elements in the majorization order in a superset of the possible tuples of nonzero singular values of $P_{1} P_{2}|_{\bigoplus_{i_{2} = 1}^{m_{2}} \mathcal{K}_{i_{2}}^{(2)}}$. 
\begin{lem} \label{lemma: majorants}
Given $m \in \mathbb{N}, e \geq 0$, elements of the set
\begin{align}
    S_{e} := \{ x \in \mathbb{R}^{m} \; | \; \forall i \in [m], x_{i} \in [0,1], \sum_{i=1}^{m} x_{i} = e\}
\end{align}
The element $(\underbrace{1, \ldots, 1}_{\lfloor e \rfloor\; \text{times}}, e - \lfloor e \rfloor, 0, \ldots,0) \in S_{e}$ majorizes every element of $S_{e}$. 
\end{lem}
\begin{proof}
Let $x \in S_{e}$ be arbitrary. For $k \in [m]$, $k \leq \lfloor e \rfloor$, observe that 
$\sum_{i = 1}^{k} x^{\downarrow}_{i} \leq \sum_{i=1}^{k} 1 = k$. And for $k > \lfloor e \rfloor$, $\sum_{i = 1}^{k} x^{\downarrow}_{i} \leq \sum_{i = 1}^{m} x^{\downarrow}_{i} \leq e$. 
\end{proof}

\chapter{Quantum erasure simulation and probabilistic error correction}
\label{chapter: 5}

\section{Introduction}
\label{ch5: sec 1}

Quantum erasure is a good error model for architectures suffering from leakage. For example, in a trapped-ion architecture, each information-carrying qubit is ideally encoded in a subspace, called the computational subspace, that is spanned by a particular set of electronic states of each ion \cite{Brown2021}. These states are chosen for their long coherence times. However, due to interactions with its environment, the state of an ion may leak out of the computational subspace. This process may be modeled as an erasure because it is possible, at least in principle, to make a measurement to find out if and where an error occurred. See \cite{Grassl1997} for other practical reasons to consider quantum erasure.

\begin{definition} \label{def: erasure} (Quantum erasure channel)
Let $\mathcal{H}$ be an inner product space and denote $\Tilde{\mathcal{H}} := \mathcal{H} \oplus \mathbb{C}$. Let $\iota: \mathcal{B}(\mathcal{H}) \rightarrow \mathcal{B}(\Tilde{\mathcal{H}})$ be an isometric channel and let $\ket{e}$ be a unit vector spanning the kernel of $\iota (\mathds{1}_{\mathcal{H}})$. For $q \in [0,1]$, a quantum erasure channel $\mathcal{E}_{q}: \mathcal{B}(\mathcal{H}) \rightarrow \mathcal{B}(\Tilde{\mathcal{H}})$ acts in the following way:
\begin{align}
    \mathcal{E}_{q} (\cdot) = (1 - q) \iota(\cdot)+ q \tr (\cdot) \ket{e}\bra{e}.
\end{align}
The state $\ket{e}$ is called the error flag and $q$ is called the erasure probability. 
\end{definition}

The structure of this chapter is as follows. In Section~\ref{ch5: sec 2}, we motivate erasure as an error model and review the relevant literature. In Section~\ref{ch5: sec 3}, we recall some notation and mathematical background. In Section~\ref{ch5: sec 4}, we define quantum simulation and show how it is related to probabilistic error correction. We also introduce Hastings' problem. In Section~\ref{ch5: sec 5}, we give a complete characterization of quantum channels that can contribute to a convex decomposition of erasure. In Section~\ref{ch5: sec 6}, we report our latest progress on Hastings' problem. Finally, we conclude in Section~\ref{ch5: sec 7} with some ideas for further research.

\section{Motivation and history}
\label{ch5: sec 2}

The quantum and classical capacities of quantum erasure channels were derived as explicit functions of the erasure probability by Bennett et al. in Ref.~\cite{Bennett1997}. In particular, they proved that 
\begin{align}
\label{eq: quantum capacity for erasure}
\mathcal{Q}(\mathcal{E}_{q}) = \max(0, (1 - 2 q) \log d_{\mathcal{H}}).
\end{align}
As is standard in quantum information, the logarithm base is $2$ and the capacity is measured in qubits. Crucial to their proof is the fact that the quantum capacity of a symmetric channel, which is one that is isometrically equivalent to its complementary channel, is zero. They used this to show that the quantum capacity of a quantum erasure channel vanishes when its erasure probability exceeds $\frac{1}{2}$. We include a proof of the latter statement for completeness. 

\begin{prop}
\label{prop: erasure channel with less than 1/2 transmissivity has no quantum capacity}
    If $q \geq \frac{1}{2}$, then the quantum capacity of $\mathcal{E}_{q}$ is zero.
\end{prop}
\begin{proof}
    First, we reduce to the case where the erasure probability is exactly equal to $\frac{1}{2}$. For $q > \frac{1}{2}$, there exists a channel $\Tilde{\mathcal{E}}_{\frac{1}{2} \rightarrow q}: \mathcal{B} (\Tilde{\mathcal{H}}) \rightarrow \mathcal{B} (\Tilde{\mathcal{H}})$ such that $\Tilde{\mathcal{E}}_{\frac{1}{2} \rightarrow q} \circ \mathcal{E}_{\frac{1}{2}} = \mathcal{E}_{q}$. It is
    \begin{align}
    \label{eq: erasure degrador}
     \Tilde{\mathcal{E}}_{\frac{1}{2} \rightarrow q} (\cdot) = (1-\Tilde{q}) \id_{\Tilde{\mathcal{H}}} (\cdot) + \Tilde{q} \tr (\cdot) \ket{e}\bra{e},  
    \end{align}
    where $\Tilde{q} = 2 ( q - \frac{1}{2})$. This shows that if the quantum capacity of $\mathcal{E}_{q}$ for some $q > \frac{1}{2}$ is nonzero, then the quantum capacity of $\mathcal{E}_{\frac{1}{2}}$ is nonzero.  

    We argue by contradiction. The following isometry gives an isometric extension of $\mathcal{E}_{\frac{1}{2}}$. For all $\ket{\phi} \in \mathcal{H}$, 
    \begin{align}
    \label{eq:  isometric extension of 1/2 erasure}
    U_{\frac{1}{2}} (\ket{\phi}) = \frac{1}{\sqrt{2}} ( \ket{\phi}_{1} \otimes \ket{e}_{2} + \ket{e}_{1} \otimes \ket{\phi}_{2} ).
    \end{align}
    This says in particular that $\mathcal{E}_{\frac{1}{2}}$ is isometrically equivalent to its complementary. That is, it is a symmetric channel. If its  quantum capacity is nonzero, then there exists, in particular, a sequence $(\mathcal{V}_{n}, \mathcal{D}_{n})_{n \in \mathbb{N}}$ of encoding and decoding channels such that $\text{dom}(\mathcal{V}_{n}) = \text{codom}(\mathcal{D}_{n}) =\mathcal{H}$ and
    \begin{align}
    \label{eq: linear cloner}
    (\mathcal{D}_{n})_{1} \otimes (\mathcal{D}_{n})_{2} \circ \mathcal{U}_{\frac{1}{2}}^{\otimes n} \circ \mathcal{V}_{n} (\rho) = \rho \otimes \rho + \xi_{n},
    \end{align}
    for all $\rho \in \mathcal{B} (\mathcal{H})$, and $|| \xi_{n} ||_{1} \rightarrow 0$ as $n \rightarrow \infty$.  This implies that $\lim_{n \rightarrow \infty}  (\mathcal{D}_{n})_{1} \otimes (\mathcal{D}_{n})_{2} \circ \mathcal{U}_{\frac{1}{2}}^{\otimes n} \circ \mathcal{V}_{n}$ is a linear cloning machine. This cannot be. \end{proof}

In \cite{Grassl1997}, Grassl et al. show that the smallest single erasure correcting qubit code has length $4$ and give the example of the $[[4,1,2]]_{2}$ stabilizer code. The authors of \cite{Wilde2014} used a Rényi divergence approach to show that sequences of Haar random codes for entanglement generation at a rate above the capacity of the quantum erasure channel have exponentially decaying entanglement fidelity. 

In \cite{hastings2014notes}, Hastings initiates the study of the problem of simulating quantum erasure channels (see Def.~\ref{def: exact simulation}) using other quantum erasure channels. In particular, he showed that for $m \in \mathbb{N}$, the channel $\mathcal{E}_{1 - \frac{1}{m}}$ cannot be used to simulate a channel $\mathcal{E}_{q}$ for $q < 1 - \frac{1}{m}$. He conjectured that quantum erasure channels with zero quantum capacity cannot be used to simulate better quantum erasure channels. More details and progress on this are provided below. 

In \cite{Kukulski2022}, Kukulski et al. systematically study probabilistic quantum error correction. It is not the first study of probabilistic recovery of quantum information (see \cite{Nayak2006} for example). However, this does seem to be the first place where the optimal probability of recovery is formulated as the value of a semi-definite program parameterized by Kraus operators of the channel in question. We reached similar conclusions independently. Notably, Kukulski et al. give a class of channels for which isometric encodings are sub-optimal for maximizing the probability of recovery.

\section{Preliminaries}
\label{ch5: sec 3}

We recall some notation and mathematical notions. Let $\mathcal{H}_{A}$ be a given inner product space and let $\mathcal{H}_{R} \cong \mathcal{H}_{A}$. Let $\ket{\beta}_{R A} = \sum_{i = 1}^{d_{\mathcal{H}}} \ket{i}_{A} \otimes \ket{i}_{R}$ be an unnormalized maximally entangled state. The value of a super-operator $\mathcal{T} : \mathcal{B}(\mathcal{H}_{A}) \rightarrow \mathcal{B}(\mathcal{H}_{B})$ under the Choi isomorphism $\sigma_{\beta}$ is given by $\sigma_{\beta}(\mathcal{T})_{R B} = (\id_{R} \otimes \mathcal{T}_{A}) (\ket{\beta}\bra{\beta}_{R A})$. From now on, we use $\sigma$ to denote $\sigma_{\beta}$ for some fixed $\ket{\beta}$. For a code $\mathcal{C}$, $\mathcal{U}_{\mathcal{C}}$ denotes an isometry whose image is $\mathcal{C}$, and $P_{\mathcal{C}}$ denotes the projector onto $\mathcal{C}$. For natural $n$, $[ n ] $ denotes the set $\{1, 2, ..., n\}$ and $\binom{[n]}{\ell}$ denotes the family of $\ell$-sets in $[n]$. We argue based on the notion of typicality (see, for example, Ch.14 of \cite{Wilde_2013}) to prove Thm.~\ref{thm: deterministic codes do not help} below. Specifically, we consider strongly typical binary strings relative to a Bernoulli distribution. These strings represent subsets of qudits that are not erased.

\begin{definition}
\label{def: typical weights} (Typical weights)
Given $u \in (0,1), \delta > 0$, and $n \in \mathbb{N}$, the set of $\delta$-typical weights relative to $u$ is 
\begin{align}
\label{eq: typical set}
    T^{(n)}_{u,\delta} = \{ \ell \in [n] \cup \{0\} \; \big| \; |\frac{ \ell }{n} - u| \leq \delta \}. 
\end{align}
 A subset $S \subseteq [n]$ is called $\delta$-typical relative to $u$ if $|S| \in T^{(n)}_{u, \delta}$. If $u \in \{0,1\}$, then by definition only one weight is $\delta$-typical relative to $u$ and it is $0$ if $u = 0$ and $n$ if $u = 1$.  
\end{definition}

Let $u \in [0,1]$. For each $n \in \mathbb{N}$, let $X^{n} = X_{1}, \ldots, X_{n}$ be independent realizations of a binary random variable $X$ with $\text{Pr} (X = 1) = u$. It is a consequence of the law of large numbers that for all $\delta > 0$, $\text{Pr}( X^{n}: |X^{n}| \in  T^{(n)}_{u,\delta}) \rightarrow 1$ as $n \rightarrow \infty$.

\begin{definition} (Exact channel simulation)
\label{def: exact simulation}
    Given a pair of channels $\mathcal{M}_{1}$ and $\mathcal{M}_{2}$, if there exists two channels $\mathcal{T}$ and $\mathcal{R}$ such that 
    \begin{align}
    \label{eq: exact simulation}
        \mathcal{M}_{1} = \mathcal{R} \circ \mathcal{M}_{2} \circ \mathcal{T},
    \end{align}
    then $\mathcal{M}_{1}$ is \textit{exactly} simulable with $\mathcal{M}_{2}$. The channel $\mathcal{M}_{1}$ is the target of the simulation and the channel $\mathcal{M}_{2}$ is the resource. When clear from context, we may say that $\mathcal{M}_1$ can be simulated with $\mathcal{M}_{2}$ if $\mathcal{M}_{1}$ is exactly simulable with $\mathcal{M}_{2}^{\otimes n}$ for some $n \in \mathbb{N}$. 
\end{definition}

In quantum error correction, the goal is to simulate the identity channel by the error channel. Following the terminology of quantum error correction, we use the term encoder to refer to someone with access and control of the input to the resource channel and the term decoder to refer to someone with access to the output of the resource channel. If a channel $\mathcal{M}_{1}$ is exactly simulable with $\mathcal{M}_{2}^{\otimes n}$, then it is exactly simulable with $\mathcal{M}_{2}^{\otimes m}$ for $m \geq n$. The encoding channels $\mathcal{T}$ may be assumed to be isometric at the cost of increasing the number of resource channel uses.

\begin{prop}
\label{prop: isometric encoding is sufficient with more systems}
    If $\mathcal{M}_{1}$ is exactly simulable by $\mathcal{M}_{2}^{\otimes n}$ for some $n \in \mathbb{N}$, then 
    \begin{align}
        \mathcal{M}_{1} = \Tilde{\mathcal{R}} \circ \mathcal{M}_{2}^{\otimes n + t} \circ \mathcal{U},
    \end{align}
    where $\Tilde{\mathcal{R}}$ is a channel, $\mathcal{U}$ is an isometric channel, and $t \leq n + 1$. 
\end{prop}

\begin{proof}
By assumption, there exist channels $\mathcal{T}$ and $\mathcal{R}$ such that $\mathcal{M}_{1} = \mathcal{R} \circ \mathcal{M}_{2}^{\otimes n} \circ \mathcal{T}$. For $\mathcal{T}$, there exists a Kraus decomposition $\{ T_{i} \}_{i = 1}^{t}$ with $t \leq d_{\mathcal{H}} d_{\mathcal{H}}^{ n } $ \cite{CHOI1975}. Let $U_{\mathcal{T}}: \mathcal{H} \rightarrow \mathcal{H}^{\otimes (n + \lceil \log_{d_\mathcal{H}}(t) \rceil)}$ be defined by the action
\begin{align}
\label{eq: isometric extension of T}
    \mathcal{H} \ni \ket{\phi} \mapsto U_{\mathcal{T}} \ket{\phi} := \sum_{i = 1}^{t}  T_{i} \ket{\phi} \otimes \ket{i}_{\text{aux}},
\end{align}
where $\{ \ket{i}_{\text{aux}} \}_{i =1}^{t}$ are orthonormal. Then, $\mathcal{U} \sim \{U_{T}\}$ is an isometric extension of $\mathcal{T}$ \cite{Stinespring1955}. After $\mathcal{M}^{\otimes n + t}$, the systems in $[n + \lceil \log_{d_\mathcal{H}}(t) \rceil] \setminus [n]$ are discarded and then $\mathcal{R}$ is applied to the systems in $[n]$. That is, $\Tilde{\mathcal{R}} := \mathcal{R} \circ \tr_{[n + \lceil \log_{d_\mathcal{H}}(t) \rceil] \setminus [n]}$.
\end{proof}

Other kinds of quantum channel simulation may be considered. For example, we may consider approximate simulation where $\mathcal{R} \circ \mathcal{M}_{2} \circ \mathcal{T}$ is required to be $\epsilon$-close, say in diamond norm distance, to $\mathcal{M}_{1}$. We say that a target channel $\mathcal{M}_{1}$ is \textit{asymptotically faithfully} simulable with a resource channel $\mathcal{M}_{2}$ if there exists a sequence of simulations $(\mathcal{R}_{n} \circ \mathcal{M}_{2}^{\otimes n}  \circ \mathcal{T}_{n})_{n \in \mathbb{N}}$ and a sequence of non-negative numbers $(\epsilon_{n})_{n \in \mathbb{N}}$ such that $\mathcal{R}_{n} \circ \mathcal{M}_{2}^{\otimes n}  \circ \mathcal{T}_{n}$ is $\epsilon_{n}$-close to $\mathcal{M}_{1}$ in diamond norm distance and $\epsilon_{n} \rightarrow 0$ as $n \rightarrow \infty$. Another kind of simulation to consider is resource-assisted simulation, where the encoder and decoder have access to auxiliary joint resources such as classical communication or entanglement. Quantum channel simulation is a special type of resource conversion as studied in the context of quantum resource theory \cite{Devetak2008}.

Following the same reasoning as in the proof of Prop.~\ref{prop: erasure channel with less than 1/2 transmissivity has no quantum capacity}, one can verify that if $ q \leq p$, then  $\mathcal{E}_{p}$ is exactly simulable with $\mathcal{E}_{q}$. If $q \leq \frac{1}{2}$, then the forward coding theorem for quantum erasure channels \cite{Bennett1997, Devetak2005} says that the identity is asymptotically faithfully simulable by $\mathcal{E}_{q}$ at a nonzero rate. Specifically, it says that there is a sequence of codes $(\mathcal{C}_{n})_{n \in \mathbb{N}}$, where $\mathcal{C}_{n} \leq \mathcal{H}^{\otimes n}$, and recovery channels $(\mathcal{R}_{n})_{n \in \mathbb{N}}$ such that $\mathcal{R}_{n} \circ \mathcal{E}_{q}^{\otimes n} \circ \mathcal{U}_{\mathcal{C}_{n}}$ differs from $\id^{\otimes k_n}$ by $\epsilon_{n}$ and $\lim_{n \rightarrow \infty} \frac{k_n}{n} > 0$ and $\lim_{n \rightarrow \infty} \varepsilon_{n} = 0$. Even in the setting of exact simulation, where it is clear that $\id$ is not simulable by $\mathcal{E}_{q}^{\otimes n}$ for any $q > 0$ and $n \in \mathbb{N}$, quantum error correcting codes may be used to improve quantum erasure channels. Here, improve means that the channel may be used to simulate another with a strictly lower erasure probability. Here is a simple example of that.

\begin{example}
\label{ex: 4,1,2 erasure simulation}
Consider the following code in $(\mathbb{C}^{2})^{\otimes 4}$:
\begin{align}
\label{eq: 4,1,2 code}
\mathcal{C}_{[[4,1,2]]_{2}} = \text{span}( \ket{0}_{L} \propto \ket{0000} + \ket{1111}, \ket{1}_{L} \propto \ket{1 0 0 1} + \ket{0 1 1 0}).
\end{align}
This is a $[[4,1,2]]_{2}$ stabilizer code, the smallest code capable of correcting the erasure of a single qubit \cite{Grassl1997}. Notably, it does not correct the erasure of any two qubits. To see this, observe that $Z \otimes Z \otimes I \otimes I$ acts as a logical $Z$ operator when restricted to the code. The same is true for $Z \otimes I \otimes Z \otimes I$, $ I \otimes Z \otimes I \otimes Z$ and $I \otimes I \otimes Z \otimes Z$. So, erasure of qubit pairs $(1,2), (1,3), (2,4)$ and $(3,4)$ cannot be corrected. Similarly, observe that $X \otimes I \otimes I \otimes X$ and $I \otimes X \otimes X \otimes I$ act as a logical $X$. Hence, there is no recovery from the erasure of qubit pairs $(1,4)$ or $(2,3)$.

\begin{figure}[t!]
\label{fig: plot 4,1,2 simulation}
    \centering
    \includegraphics[scale=0.50]{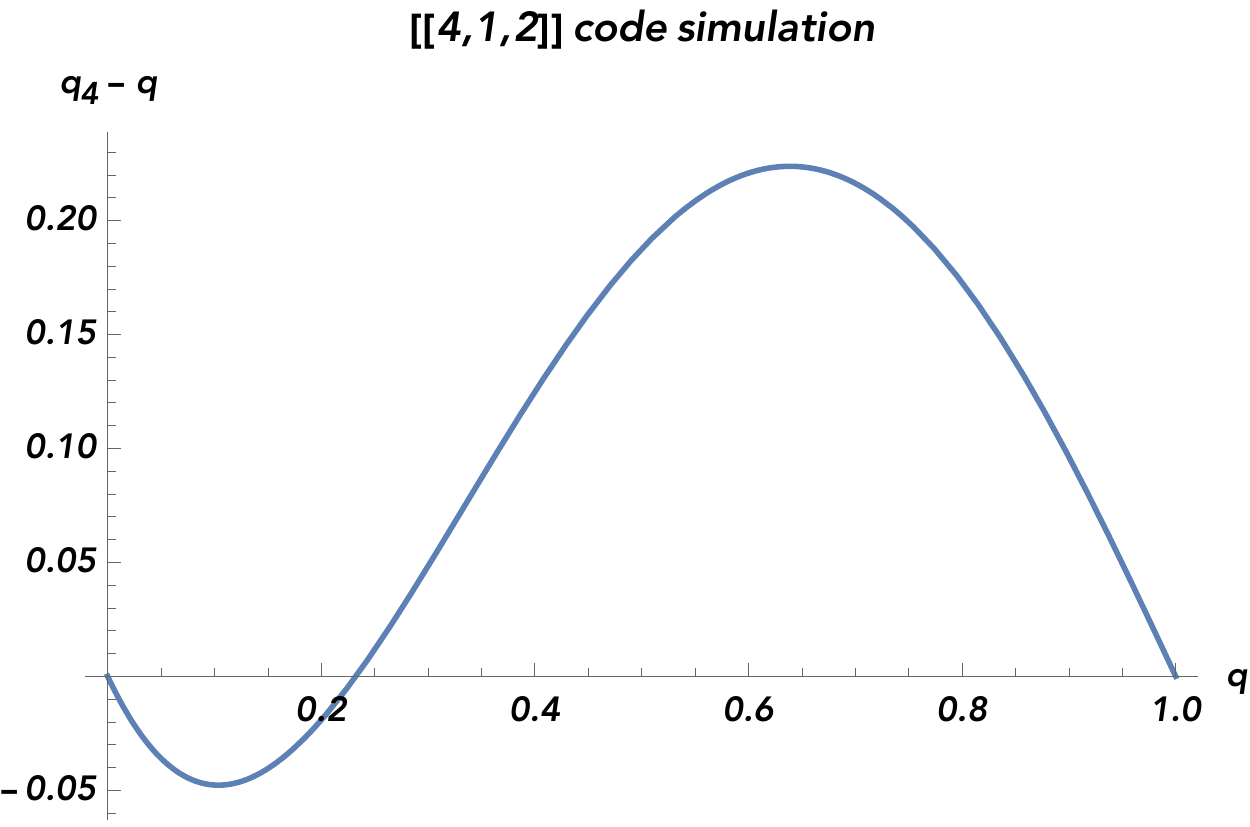}
    \caption{This is a plot of the difference of erasure probabilities $q_4 - q$ vs the erasure probability of the resource channel $q$. As $q$ increases, high-weight erasures become more likely. These are erasures that the code $\mathcal{C}_{[[4,1,2]]_{2}}$ cannot correct.}
    \label{fig:412 simulation}
\end{figure}

This code may be used along with $\mathcal{E}_{q}^{\otimes 4}$ and the following recovery procedure. If there is one error, correct and output the state. Otherwise, output the error flag $\ket{e}$. Let $\mathcal{R}_{[[4,1,2]]_{2}}$ denote the channel implementing this procedure. Then, 
\begin{align}
\label{eq: 4,1,2 simulation}
\mathcal{R}_{[[4,1,2]]_{2}} \circ \mathcal{E}_{q}^{\otimes 4} \circ \mathcal{U}_{\mathcal{C}_{[[4,1,2]]_{2}}} = \mathcal{E}_{q_4},
\end{align}
where $q_{4} = \sum_{k=2}^{4} \binom{4}{k} q^{k} (1 - q)^{4 - k}$. Fig.~\ref{fig: plot 4,1,2 simulation} shows a plot of the difference of erasure probabilities $q_4 - q$ versus $q$. As long as $q$ is small enough, precisely below $\frac{1}{6} ( 5 - \sqrt{13})$, the simulated channel $\mathcal{E}_{q_{4}}$ is better than $\mathcal{E}_{q}$. We show later on that $\mathcal{E}_{q_4}$ is in fact the best erasure channel that $\mathcal{E}_{q}^{\otimes 4}$ can be used to simulate with this code.
\end{example}

Our main concern will be the more interesting case where $q \geq \frac{1}{2}$. Clearly, for such $q$, $\mathcal{E}_{q}^{\otimes n}$ cannot be used to simulate $\mathcal{E}_{p}$ for $p < \frac{1}{2}$ no matter how large $n$ is. Otherwise, $\mathcal{E}_{q}$ would have nonzero quantum capacity. The next question to ask is this: can these channels be improved at all? This question was studied by Hastings \cite{hastings2014notes}. Using an argument based on the no-cloning theorem, he showed that for each $m \in \mathbb{N}$, the channel $\mathcal{E}_{1 - \frac{1}{m}}$ cannot be improved.

\begin{conj}
\label{conj: Hastings} (Hastings' conjecture). For $q \in [\frac{1}{2}, 1]$ and $p < q$, it is not possible to simulate $\mathcal{E}_{p}$ with $\mathcal{E}_{q}$. 
\end{conj}

Hastings' conjecture is the inspiration for this chapter. In the original statement of the conjecture, ``simulation" referred to asymptotically faithful simulation. We are primarily concerned with exact simulation of erasure. That is, we consider all simulations of erasure 
\begin{align} 
\label{eq: sim of erasure}
   \mathcal{E}_{p} =  \mathcal{R} \circ \mathcal{E}_{q}^{\otimes n} \circ \mathcal{T}, 
\end{align}
with the aim of finding non-trivial lower bounds for $p$. Given that we suspect that Conj.~\ref{conj: Hastings} holds, we may take $n$ as large as needed and so by Prop.~\ref{prop: isometric encoding is sufficient with more systems} from now on we take the encoding channels to be isometric. For an isometric encoding channel $\mathcal{U}_{\mathcal{C}_{n}}$ and a decoding channel $\mathcal{R}$, we may expand sums over tensor products and use the linearity of channels to get
\begin{align}
\label{eq: expansion of simu}
     \mathcal{R} \circ \mathcal{E}_{q}^{\otimes n} \circ \mathcal{U}_{\mathcal{C}_{n}} = \sum_{S \subseteq [n]} q^{n - |S|} (1 - q)^{|S|} \mathcal{R} \circ (\iota^{\otimes S} \otimes \ket{e}\bra{e}^{\otimes S^{c}} \tr_{S^{c}}) \circ \mathcal{U}_{\mathcal{C}_{n}}. 
\end{align}
Here, $S^{c} \subseteq [n]$ is the set of erased systems, which following \cite{hastings2014notes}, we call an erasure pattern. We show in Section~\ref{ch5: sec 5} that we may as well take $\mathcal{R} \circ (\iota^{\otimes S} \otimes \ket{e}\bra{e}^{\otimes S^{c}} \tr_{S^{c}}) \circ \mathcal{U}_{\mathcal{C}_{n}}$ to be an erasure channel $\mathcal{E}_{1 - \gamma_{S} (\mathcal{C}_{n})}$. Here, $\gamma_{S} (\mathcal{C}_{n})$ is the optimal probability of recovery of quantum information conditional on the occurrence of erasure pattern $S^{c}$ for the code $\mathcal{C}_{n}$. We show that it can be computed via a semi-definite program in Section~\ref{ch5: sec 4}.
Therefore, the problem becomes finding non-trivial upper bounds for the optimal overall recovery probability
\begin{align}
r_{q, n} = \mathrm{sup}_{\mathcal{C}_{n}} \sum_{S \subseteq [n]}  q^{n - |S|} (1 - q)^{|S|}  \gamma_{S}(\mathcal{C}_{n})
\end{align}
for all $n \in \mathbb{N}$. Since $(r_{q,n})_{n \in \mathbb{N}}$ is monotonically non-decreasing, it suffices to bound from above  $r_{q} = \lim_{n \rightarrow \infty} r_{q, n}$. According to Conj.~\ref{conj: Hastings}, $r_{q} \leq 1-q$ for $q \in [\frac{1}{2}, 1]$.

\section{Erasure simulation and probabilistic error correction}
\label{ch5: sec 4}

In this section, we collect a few observations about probabilistic recovery of quantum information and how it relates to exact simulation of erasure. By probabilistic recovery, we mean recovery procedures where decoding super-operators are trace non-increasing. This scenario was considered by Nayak and Sen in the context of quantum state encryption \cite{Nayak2006}. They state a characterization of what they call invertible quantum operations---those are completely-positive maps with a completely-positive inverse---pointed out to them by Jon Tyson (see Theorem 2.2 in \cite{Nayak2006}). More recently, Kukulski et al. systematically studied this scenario in \cite{Kukulski2022} and noted that the optimal recovery probability for a given encoding map is a solution to a semi-definite program parameterized by the error channel. We arrived at this observation independently.

Since we established in Prop.~\ref{prop: isometric encoding is sufficient with more systems} that we may consider isometric encoding channels without loss of generality, we formulate the conditions for probabilistic recovery using codes. For a more general formulation, see \cite{Kukulski2022}.

\begin{definition}
\label{def: probabilistic rec}
Let a channel $\mathcal{M}: \mathcal{B}(\mathcal{H}) \rightarrow \mathcal{B}(\mathcal{K})$ be given. We say that $\mathcal{C} \leq \mathcal{H}$ is a  $(\mathcal{M}, t)$-correcting code  if there is a completely-positive, trace non-increasing $\mathcal{R}: \mathcal{B} (\mathcal{K}) \rightarrow \mathcal{B} (\mathcal{H}) $ such that 
\begin{align}
 \mathcal{R} \circ \mathcal{M} \circ (P_{\mathcal{C}} (\cdot) P_{\mathcal{C}}) = t P_{\mathcal{C}} (\cdot) P_{\mathcal{C}},
\end{align} 
where the proportionality constant $t$ is the probability of recovery.
\end{definition}

If the action of the channel on the code is not one-to-one, then recovery with positive probability is not possible. The converse statement is not true. For example, for $0 < p < \frac{4}{3}$, the partially-depolarizing qubit channel $\Delta_{2,p}$ (Eq.~\ref{eq: qubit depolarizing channel def}) is one-to-one on $\mathcal{B} (\mathbb{C}^{2})$. However, if $p >0$, then the maximum probability of recovery is zero by Lem.~\ref{lem: cp reovery for authomorphisms}. The following proposition gives necessary and sufficient conditions for probabilistic recovery in terms of the code and Kraus operators of the error channel.

\begin{prop}
Let a channel $\mathcal{M}: \mathcal{B}(\mathcal{H}) \rightarrow \mathcal{B}(\mathcal{K})$ with Kraus decomposition $\{ M_{\alpha} \}_{\alpha=1}^{M}$ be given. A code $\mathcal{C} \leq \mathcal{H}$ is $(\mathcal{M}, t)$-correcting if and only if there exist a positive semi-definite operator $\Pi \in \mathcal{B} (\mathcal{K})$ satisfying $\Pi \leq \mathds{1}_{\mathcal{K}}\; $ and an $M \times M$ matrix $T$ satisfying $\tr (T) = t$, such that
\begin{align}
\label{eq: cond error correction cond}
    P_{\mathcal{C}} M_{\beta}^{*} \Pi M_{\alpha} P_{\mathcal{C}} = T_{(\alpha, \beta)} P_{\mathcal{C}}
\end{align}
holds for all $\alpha, \beta \in [M]$. 
\end{prop}

\begin{proof}
For necessity, we argue as in Theorem 10.1 of Ref.~\cite{Nielsen2000}. Suppose that $\mathcal{R} \circ \mathcal{M} ( P_{\mathcal{C}} ( \cdot ) P_{\mathcal{C}}) = t P_{\mathcal{C}} ( \cdot ) P_{\mathcal{C}}$ for some $\mathcal{R} \sim \{ R_{\ell} \}_{\ell =1}^{L}$. By theorem 8.2 of Ref.~\cite{Nielsen2000}, this implies the existence of a set of complex coefficients $\{u_{(\ell, \alpha)} \}_{\alpha=1, \ell=1}^{\alpha = M, \ell = L}$ such that $R_{\ell} M_{\alpha} P_{\mathcal{C}} = u_{(\ell, \alpha)} P_{\mathcal{C}}$ for all $\alpha \in [M]$ and $\ell \in [L]$. Multiplying on the left by $P_{\mathcal{C}} M^{*}_{\beta} R^{*}_{\ell}$ and summing over $\ell$ yields
\begin{align}
    P_{\mathcal{C}} M_{\beta}^{*} (\sum_{\ell} R_{\ell}^{*} R_{\ell} ) M_{\alpha} P_{\mathcal{C}} = (\sum_{\ell} u_{(\ell, \beta)}^{*} u_{(\ell, \alpha)} ) P_{\mathcal{C}} = T_{(\alpha, \beta)} P_{\mathcal{C}}.
\end{align}
Let $\Pi = \sum_{\ell=1}^{L} R_{\ell}^{*} R_{\ell}$ and observe that $\tr(T) = t$. 
For sufficiency, let $\Pi \geq 0$ be given and let $\xi$ be any operator satisfying $\xi^{*} \xi = \Pi$ such as $\sqrt{\Pi}$ for example. Observe that $\mathcal{C}$ satisfies the quantum error correction conditions Eq.~\ref{eq: error correction conditions} for the error super-operator $\xi \mathcal{M} (\cdot) {\xi^{*}}$ and so there exists a recovery channel $\mathcal{R}$ for it. Hence, the completely-positive, trace non-increasing map $\Tilde{\mathcal{R}} = \mathcal{R} \circ \xi (\cdot) \xi^*$ satisfies
$\Tilde{\mathcal{R}} \circ \mathcal{M}(P_{\mathcal{C}} (\cdot) P_{\mathcal{C}})  = \tr(T) P_{\mathcal{C}} ( \cdot ) P_{\mathcal{C}}.$ \end{proof}

We call the positive semi-definite operator $\Pi$ a recovery operator for the channel $\mathcal{M}$ and the code $\mathcal{C}$. On the decoder side, $\sqrt{\Pi} ( \cdot )\sqrt{\Pi}$ can be realized as part of a channel with the aid of an auxiliary qubit. Namely, the decoder may implement the following isometry: 
\begin{align}
\label{eq: isometry for erasure sim recovery}
    \mathcal{K} \ni \ket{\phi} \mapsto \sqrt{\Pi} \ket{\phi} \otimes \ket{0}_{\text{aux}} + \sqrt{\mathds{1}- \Pi} \ket{\phi} \otimes \ket{1}_{\text{aux}} 
\end{align}
 By measuring this auxiliary qubit in the computational basis, the decoder can ascertain whether $\sqrt{\Pi}$ occurred or not. If the result of the measurement is $0$, then a channel may be applied to recover the quantum information. Otherwise, the decoder may declare the occurrence of an error by outputting an error flag. This occurs with probability $1 - \tr(T)$. This proves the following corollary.

 \begin{cor}
     A code $\mathcal{C}$ is $(\mathcal{M}, t)$-correcting for a channel $\mathcal{M}$ if and only if $\mathcal{M}$ can be used to simulate an erasure channel with erasure probability $1 - t$ with $\mathcal{U}_{\mathcal{C}}$ as an encoding channel. 
 \end{cor}

Next, we observe that for a given code and error channel, the set of recovery operators is convex and contains the zero operator. A more general version of the following corollary is on page 4 of Ref.~\cite{Kukulski2022}.
\begin{cor}
\label{cor: semi definite formulation of recovery probability}
    Given a channel $\mathcal{M} \sim \{ M_{k} \}_{k = 1}^{M}$ and a code $\mathcal{C}$, the maximum recovery probability for $\mathcal{M}$ with $\mathcal{C}$ is the value of the following semi-definite program
    \begin{align}
    \label{prog: semidefinite}
    \text{\textbf{max}     } \quad &\tr(T), \\ 
    \text{\textbf{subject to: }} &0 \leq \Pi \leq \mathds{1},\\
    & P_{\mathcal{C}} M_{k'}^{*} \Pi M_{k} P_{\mathcal{C}} = T_{k, k'} P_{\mathcal{C}}, \quad \forall k, k' \in [M].
    \end{align}
    If the value of the program above is $t_{\text{opt}}$, then the best erasure channel that $\mathcal{M}$ can be used to simulate with the code $\mathcal{C}$ is $\mathcal{E}_{1 - t_{opt}}$. 
\end{cor}

Given a channel $\mathcal{M}$, a code $\mathcal{C}$, and a recovery operator $\Pi$, a geometric interpretation of the recovery of quantum information with nonzero probability is obtained by considering a Kraus decomposition of $\mathcal{M}$ where the coefficient matrix $T$---a positive semi-definite matrix---is diagonal. That this decomposition exists follows from theorem 8.2 of Ref.~\cite{Nielsen2000}. Since $\tr(T) > 0$, then for some Kraus operator in this decomposition, say $M_{1}$, it holds that $P_\mathcal{C} M_{1}^{*} \Pi M_{1} P_{\mathcal{C}} = T_{(1, 1)} P_{\mathcal{C}}$ and $T_{(1, 1)} \neq 0$. Hence,  $\sqrt{\Pi} M_{1}|_{\mathcal{C}}$ is invertible. Since $T$ is diagonal, then the image of $\mathcal{C}$ under $\sqrt{\Pi} M_{1}$ is orthogonal to its image under $\sqrt{\Pi} M_{k}$ for $k \neq 1$. This is possible only if the subspace $M_{1} \mathcal{C}$, which necessarily has dimension $\text{dim} (\mathcal{C})$, and the subspace $\sum_{k \neq 1} M_{k} \mathcal{C}$ are  linearly independent. That is, $M_{1} \mathcal{C} \cap \sum_{k \neq 1} M_{k} \mathcal{C} = \{0\}$. To see this, observe that if there is a nonzero vector in the intersection, then the orthogonality relations in Eq.~\ref{eq: cond error correction cond} do not hold. Conversely, if $M_{1}|_{\mathcal{C}}$ is invertible and the intersection is trivial, then there exists an operator $\xi$ such that $\xi M_{1}|_{\mathcal{C}} = t \mathds{1}$, with $t > 0$, and $\text{ker}(\xi) \geq \sum_{k \neq 1} M_{k} \mathcal{C}$. This implies the following corollary.

\begin{cor}
\label{cor: geometric int of prob cor}
A code $\mathcal{C}$ is $(\mathcal{M}, t)$ correcting for some $t > 0$ if and only if $\mathcal{M}$ admits a Kraus decomposition $\{ M_{k} \}_{k = 1}^{M}$ such that for some $k_{0} \in [M]$, the following holds:
\begin{enumerate}
    \item $M_{k_{0}}$ is invertible on $\mathcal{C}$. 
    \item $M_{k_{0}} \mathcal{C} \cap \sum_{k \neq k_{0}} M_{k} \mathcal{C} = \{ 0 \}$. 
\end{enumerate}
\end{cor}

\section{Structure of channels contributing to a convex decomposition of erasure}
\label{ch5: sec 5}

In this section, we characterize the possible ways an erasure channel can be written as a convex combination of channels. Specifically, we give necessary and sufficient conditions for a channel to contribute to a convex decomposition of an erasure channel in terms of its Choi operator. Furthermore, we give an explicit characterization of the extremal points of this set, which include the isometry $\iota$ and the probability one erasure channel $ \ket{e}\bra{e} \tr$. In particular, we show that every non-extremal erasure channel can be written as a convex combination of two channels neither of which is an erasure channel itself. This is a surprising fact, given that at a first glance, it might seem like this set is entirely comprised of erasure channels.

In this section, we suppress the inclusion map $\iota$ and write $\Tilde{\mathcal{H}} = \mathcal{H} \oplus \mathbb{C} \ket{e}$. Given a super-operator $ \mathcal{T}: \mathcal{B}(\mathcal{H}) \rightarrow \mathcal{B}(\Tilde{\mathcal{H}})$, its Choi operator can be written in block form:
\begin{align}
\label{eq: block structure of super-operators}
\sigma(\mathcal{T}) = 
\begin{pmatrix}
A(\mathcal{T})  
  & & B(\mathcal{T})^{*} \\
  B(\mathcal{T}) & &
C(\mathcal{T})
\end{pmatrix}
\end{align}
where $A(\mathcal{T}) \in \mathcal{B} (\mathcal{H}_{R} \otimes \mathcal{H})$, $B(\mathcal{T}) :\mathcal{H}_{R} \otimes \mathcal{H} \rightarrow \mathcal{H}_{R} \otimes \mathbb{C} \ket{e}$ and $C(\mathcal{T}) \in \mathcal{B} (\mathcal{H}_{R} \otimes \mathbb{C} \ket{e})$. If $\sigma(\mathcal{T}) \geq 0$, then it necessarily holds that  $A(\mathcal{T})\geq 0$ and $ C(\mathcal{T}) \geq 0$. For an erasure channel $\mathcal{E}_{\alpha}$, it is clear that $A(\mathcal{E}_{\alpha}) = (1 - \alpha) \ket{\beta} \bra{\beta}, B(\mathcal{E}_{\alpha}) = 0$ and $C(\mathcal{E}_{\alpha}) = \alpha \mathds{1} \otimes \ket{e}\bra{e}_{\mathcal{}} $.

\begin{prop}
Consider a convex decomposition of an erasure channel, 
\begin{align}
\label{eq: convex decomposition of erasure}
    \mathcal{E}_{\alpha} = \sum_{k} r_{k} \mathcal{M}_{k},
\end{align}
where $\alpha \in [0,1]$ and for each $k$, $r_{k} > 0$ and $\mathcal{M}_{k}$ is a channel. With respect to the block structure of Eq.~\ref{eq: block structure of super-operators}, the Choi operators of $\mathcal{M}_{k}$ satisfy
\begin{align}
     A(\mathcal{M}_{k}) &= (1 - \alpha_{k}) \ket{\beta} \bra{\beta}, \label{eq:A}\\
     C(\mathcal{M}_{k}) &= \alpha_{k} \mathds{1} \otimes \ket{e}\bra{e}, \label{eq:C}\\
     B(\mathcal{M}_{k}) &= \eta_{k} \ket{\delta_{k}} \bra{\beta}, \label{eq:B}
\end{align}
where $\alpha_{k} \in [0,1]$, $\ket{\delta_{k}} \in \mathcal{H}_{R} \otimes \mathbb{C} \ket{e}$ has unit norm, and $\eta_{k} \geq 0$ satisfies $\eta_{k} \leq \sqrt{\alpha_{k} (1 - \alpha_{k})}$.
\end{prop}

\begin{proof}
By the linearity of the Choi isomorphism, we have $\sum_{k} r_{k} A(\mathcal{M}_{k}) =  A(\mathcal{E}_{\alpha}) = (1 - \alpha) \ket{\beta}\bra{\beta}$, which has rank at most 1. The result of a positive combination of positive semi-definite operators has rank at most 1 if and only if each operator in the sum is proportional to the result. That is, $A(\mathcal{M}_{k}) = (1 -\alpha_{k}) \ket{\beta} \bra{\beta}$ for some $\alpha_{k} \in [0,1]$. Since each $\mathcal{M}_{k}$ is trace-preserving,
\begin{align}
    \tr_{\Tilde{\mathcal{H}}} (\sigma(\mathcal{M}_{k})) = \tr_{\mathcal{H}} (A(\mathcal{M}_{k})) + \tr_{\mathbb{C} \ket{e} } (C(\mathcal{M}_{k})) = \mathds{1}_{R}.
\end{align}
Therefore, $C(\mathcal{M}_{k}) = \alpha_{k} \mathds{1} \otimes \ket{e} \bra{e}$. If $\alpha_{k} \in \{0,1\}$, then $B(\mathcal{M}_{k}) = 0$. To see this, notice that if $\alpha_{k} = 0$, then $C(\mathcal{M}_k) = 0$. In this case, the off-diagonal blocks have to be zero, or else $\sigma(\mathcal{M}_{\alpha})$ has a negative eigenvalue. The same is true for the case where $\alpha_{k} = 1$ as that implies $A(\mathcal{M}_{k}) = 0$. If $\alpha_{k} \in (0,1)$, then the Schur complement of $\sigma(\mathcal{M}_{k}) / C(\mathcal{M}_{k})$ must be positive semi-definite for $\sigma(\mathcal{M}_{k})$ to be positive semi-definite. That is,
\begin{align}
    (1 - \alpha_{k}) \ket{\beta} \bra{\beta} - \alpha_{k}^{-1} |B(\mathcal{M}_{k})|^{2} \geq 0. 
\end{align}
Hence, $B(\mathcal{M}_{k})$ is an operator whose support is spanned by $\ket{\beta}$ and whose largest singular value is no larger than $\sqrt{\alpha_{k} (1 - \alpha_{k})}$. \end{proof}

It follows from this proposition that $\mathcal{E}_{\alpha}$, with $\alpha \in (0,1)$, can be written as a convex combination of two channels whose Choi states have nonzero off-diagonal block matrices. That is, neither of the two is an erasure channel. For example, for a unit vector $\ket{0} \in \mathcal{H}_{R}$, consider the two channels with the following Choi operators
\begin{align}
\sigma(\mathcal{M}_{\alpha, \pm}) = 
\begin{pmatrix}
 (1 - \alpha) \ket{\beta}\bra{\beta}
  & & \pm \sqrt{\alpha ( 1 - \alpha)}  \ket{\beta} \bra{e} \otimes \bra{0}\\
  \pm \sqrt{\alpha ( 1 - \alpha)} \ket{0}\otimes\ket{e} \bra{\beta} & &
\alpha \mathds{1} \otimes \ket{e} \bra{e}
\end{pmatrix}.
\end{align}
We have $\frac{1}{2} (\mathcal{M}_{\alpha, +} + \mathcal{M}_{\alpha, -}) = \mathcal{E}_{\alpha}$. More notably, the set of channels satisfying Eqs.~\ref{eq:A},~\ref{eq:C} and \ref{eq:B} is the preimage of the set of erasure channels under left composition with decoherence of $\mathcal{H}$ and $\mathbb{C} \ket{e}$. That, composition the channel with Kraus operators $ \mathds{1}_{\Tilde{\mathcal{H}}} - \ket{e} \bra{e}$ and $\ket{e} \bra{e}$. Erasure channels are invariant under left composition with this channel. Hence, if Eq~\ref{eq: sim of erasure} holds, then without loss of generality, it may be assumed that
\begin{align}
\label{eq: erasure simulation for each pattern}
\mathcal{R} \circ (\iota^{\otimes S} \otimes \ket{e}\bra{e}^{\otimes S^{c}} \tr_{S^{c}}) \circ \mathcal{T} = \mathcal{E}_{1 - \gamma_{S}}
\end{align}
for each erasure pattern $S^c \subseteq [n]$.

In the next proposition, we show that the set of extremal channels among channels satisfying Eqs.~\ref{eq:A},~\ref{eq:C} and \ref{eq:B} consists of $\iota$ and channels $\mathcal{M}$ where the Schur complement $\sigma(\mathcal{M}) / C(\mathcal{M})$ vanishes.

\begin{prop}
\label{prop: extremal channel structure}
Suppose that $\mathcal{M} \neq \iota$ is a channel that satisfies \eqref{eq:A}, \eqref{eq:C} and \eqref{eq:B} for some $\alpha, \eta, \ket{\delta}$. Then, $\mathcal{M}$ is an extremal channel if and only if the Schur complement of $C(\mathcal{M})$ vanishes. 
\end{prop}

\begin{proof}
Suppose that the Schur complement of $C(\mathcal{M})$ vanishes. That is, $\eta = \sqrt{\alpha (1 - \alpha)}$. If $\sigma(\mathcal{M}) = \sum_{k} r_{k} \sigma (\mathcal{M}_{k})$, such that for all $k$, $r_{k} > 0$, then each $\mathcal{M}_{k}$ satisfies \eqref{eq:A}, \eqref{eq:C} and \eqref{eq:B} as argued before. Let the relevant parameters be $\alpha_{k}, \eta_{k}$ and $\ket{\delta_{k}}$ for each channel $\mathcal{M}_{k}$. Then it follows from the linearity of $\sigma$ that 
\begin{align}
    \sum_{k} r_{k} \alpha_{k} &= \alpha, \\
    \sum_{k} r_{k} \eta_{k} \ket{\delta_{k}} &= \sqrt{\alpha (1 - \alpha)} \ket{\delta}.
\end{align}
Observe the chain of inequalities
\begin{align}
    \sqrt{\alpha (1 - \alpha) } = ||\sqrt{\alpha (1 - \alpha)} \ket{\delta} ||_{2}
    &= ||\sum_{k} r_{k} \eta_{k} \ket{\delta_{k}} ||_{2} \\  
    &\leq \sum_{k} r_{k} \eta_{k} \\
    &\leq \sum_{k} r_{k}  \sqrt{\alpha_{k} (1 - \alpha_{k})} \\
    &\leq     \sqrt{\alpha (1 - \alpha) }.
\end{align}
Each inequality has to be satisfied with equality. The first inequality is the triangle inequality for $|| \cdot ||_{2}$ and so is saturated if and only if $\ket{\delta_{k}} = \ket{\delta}$ for all $k$. The second inequality is saturated if and only if $\eta_{k} = \sqrt{\alpha_{k} (1 - \alpha_{k})}$ for all $k$. The third inequality is saturated if and only if $\alpha_{k} = \alpha$ all $k$ by the strict concavity of the function $f(t) = \sqrt{ t (1- t)}$ on $(0,1)$. Therefore, $\mathcal{M}_{k} = \mathcal{M}$ for each $k$ and so $\mathcal{M}$ is extremal. 

Conversely, suppose that $\mathcal{M}$ is extremal. Consider the following pair of operators
\begin{align}
\begin{pmatrix}
A(\mathcal{M})
  & & (1 \pm \epsilon) B(\mathcal{M})^*\\
  (1 \pm \epsilon) B(\mathcal{M}) & &
C(M)
\end{pmatrix},
\end{align}
where $\epsilon > 0$. The extremal Choi operator $\sigma(\mathcal{M})$ is an average of these two operators. Under the Choi isomorphism, the two correspond with trace-preserving super-operators as their value under a partial trace of $\Tilde{\mathcal{H}}$ is $\mathds{1}_{R}$. Since $\mathcal{M}$ is an extremal channel, the corresponding super-operators cannot be both completely positive. Therefore, one of the two operators above must have a negative eigenvalue. This implies in particular that $(1 + \epsilon)^{2} \eta^{2} > \alpha (1 - \alpha)$ for all $\epsilon \in (0,1)$. This implies $\eta = \sqrt{\alpha (1 - \alpha)}$ and the Schur complement of $C(\mathcal{M})$ vanishes. \end{proof}

\section{Progress on Hastings' conjecture}
\label{ch5: sec 6}
In this section, we restrict our attention to isometric encoding channels into a sequence of codes $(\mathcal{C}_{n})_{n \in \mathbb{N}}$. As stated before, this comes at no loss of generality for the purposes of Hastings' conjecture as we argue in the large $n$ limit. The output of $\mathcal{E}_{q}^{\otimes n}$ is a convex combination of orthogonal states. The decoder may start with a projective measurement whose outcome corresponds with the occurring erasure pattern. This does not disturb the state and so causes no loss of information. The decoder may then tailor a recovery procedure for each erasure pattern.

\begin{lem}
\label{lem: Support of erasure of S^c is S}
Let a code $\mathcal{C} \leq \mathcal{H}^{\otimes n}$ be given. For $S \subseteq [n]$, let $\sigma_{S^{c}} \in \mathcal{D} (\mathcal{K}^{S^c})$ be given. There exists an optimal recovery operator for $\tr_{S^{c}} \otimes \sigma_{S^c} $ and $\mathcal{C}$ of the form $A_{S} \otimes \mathds{1}_{S^{c}}$.
\end{lem}
\begin{proof}
Let $\Pi \in \mathcal{P} (\mathcal{H}^{\otimes S})$ be an optimal recovery operator for $\tr_{S^{c}}$ and $\mathcal{C}$. Then, $ \Pi_{S} \otimes \mathds{1}_{S^{c}}$ is a recovery operator for $\tr_{S^{c}} \otimes \sigma_{S^c}$ and $\mathcal{C}$. If  $ \Pi_{S} \otimes \mathds{1}_{S^{c}}$ were not optimal, then the decoder may tensor multiply $\sigma_{S^{c}}$ to the output of $\tr_{S^{c}}$ and find a strictly better recovery procedure for $\tr_{S^{c}}$.     
\end{proof}

From now on, $\Pi_{S} (\mathcal{C}) \in \mathcal{P} (\Tilde{\mathcal{H}}^{\otimes S})$ denotes an operator such that $\Pi_{S} (\mathcal{C}) \otimes \mathds{1}_{S^c}$ is an optimal recovery operator---that is, an optimal point for the program~\ref{prog: semidefinite} for a given code $\mathcal{C}$---for the erasure pattern $S^{c} \subseteq [n]$. The channel $\tr_{S^c} \otimes \ket{e}\bra{e}^{S^c}$ admits a Kraus decomposition $\{ \mathds{1}_{S} \otimes (M_i)_{S^c}\}_{i = 1}^{m}$, where $(M_i)_{S^c}: \mathcal{H}^{\otimes S^c} \rightarrow \Tilde{\mathcal{H}}^{\otimes S^c}$ for each $i \in [m]$. It is clear that the optimal recovery operator $\Pi_{S} (\mathcal{C}) \otimes \mathds{1}_{S^c}$ commutes with all of these Kraus operators. From Eq.~\ref{eq: cond error correction cond}, we get
\begin{align}
\label{eq: S cond corr}
  \sum_{i=1}^m  P_{\mathcal{C}} ((\iota^*)^{\otimes S} \otimes (M_i)_{S^c}^*) (\Pi_{S} (\mathcal{C}) \otimes \mathds{1}_{S^c})  (\iota^{\otimes S} \otimes (M_i)_{S^c}) P_{\mathcal{C}} &= \sum_{i=1}^{m} T_{(i, i)} P_{\mathcal{C}} \\
   P_{\mathcal{C}} ((\iota^*)^{\otimes S} \Pi_{S} (\mathcal{C})   (\iota)^{\otimes S} \otimes \mathds{1}_{S^c}) \sum_{i=1}^{m} (\mathds{1}_{S} \otimes (M_i)_{S^c}^{*} (M_i)_{S^c}) P_{\mathcal{C}} &= \gamma_{S} (\mathcal{C}) P_{\mathcal{C}} \\
     P_{\mathcal{C}} ((\iota^*)^{\otimes S} \Pi_{S} (\mathcal{C})   (\iota)^{\otimes S} \otimes \mathds{1}_{S^c})  P_{\mathcal{C}} &= \gamma_{S} (\mathcal{C})  P_{\mathcal{C}},
\end{align}
where $\gamma_{S} (\mathcal{C})$ denotes the optimal recovery probability for the code $\mathcal{C}$. From now on, we suppress the symbol $\iota$. When clear from context, we write $\Pi_{S}$ to mean $\Pi_{S} (\mathcal{C}) \otimes \mathds{1}_{S^c}$ and $\gamma_S$ to mean $\gamma_{S} (\mathcal{C})$ for some code $\mathcal{C} \leq \mathcal{H}^{\otimes n}$. From the third equation in Eqs.~\ref{eq: S cond corr}, we see that when restricted to the code, the action of $\sqrt{\Pi_{S}}$ is, up to a multiplicative factor, isometric. 

Let $T_{A}, T_{B} \subseteq [n]$ be disjoint sets and suppose that both $\Pi_{T_{A}}$ and $\Pi_{T_{B}}$ are optimal recovery operators for some code $\mathcal{C}$. Let decoder $A$ be in control of the systems in $T_{B}$ and decoder $B$ be in control of the systems in $T_{B}$. Both may implement isometries as in Eq~.\ref{eq: isometry for erasure sim recovery} to recover the quantum information. If upon measuring their auxiliary qubit, decoder $A$ finds that they can recover the quantum information, then it must be the case that decoder $B$ cannot recover with any positive probability. Otherwise, the two can construct a linear cloning machine. The same holds with $A$ and $B$ reversed. Hence, it must be the case that $\mathcal{C} \leq \text{ker} ( \Pi_{T_{A}}  \Pi_{T_{B}})$. This has consequences for the optimal recovery probabilities associated with these patterns. 
\begin{prop}
\label{prop: partitioning no-cloning.}
Let $\{S_{i}\}_{i \in I}$ be a partitioning of $[n]$. Then, it holds that 
\begin{align} \label{partition eq}
   1 \geq \sum_{i \in I} \gamma_{S_{i}} .  
\end{align}
\end{prop}
\begin{proof}
First observe that $\{ \Pi_{S_{i}} \}_{i \in I}$ is a set of mutually commuting positive semi-definite operators. Hence,  
\begin{align}
    P_{C} \prod_{i \in I} (\mathds{1} - \Pi_{S_{i}})  P_{C} \geq 0.
\end{align}
Expand the product and notice that $P_{\mathcal{C}} \prod_{j \in J} \Pi_{S_{j}} P_{\mathcal{C}} = 0$ whenever $|J| > 1$. This along with Eq.~\ref{eq: S cond corr} imply the desired inequality. \end{proof}

Another observation to be made is the following. If $S$ contains one element, then recovery or lack thereof is deterministic. An intuitive explanation for this fact is that there is not enough room in a single qudit for the encoded quantum information \textit{and} the classical information about whether recovery is possible. This is a consequence of the following lemma.

\begin{lem}
\label{lem: cp reovery for authomorphisms}
A completely positive map $\mathcal{M}: \mathcal{B} (\mathcal{H}) \rightarrow \mathcal{B} (\mathcal{H})$ has a completely positive inverse if and only if $\mathcal{M} \sim \{M\}$ for some invertible operator $M \in \mathcal{B}(\mathcal{H})$.
\end{lem}
\begin{proof}
    It is clear that the map $M (\cdot) M^{*}$ has the completely positive inverse $M^{-1} (\cdot) (M^{-1})^{*}$ whenever $M$ is invertible. For the other direction, suppose that $\mathcal{R} \sim \{ R_{\ell} \}_{\ell = 1}^{r}$ is a completely positive inverse of $\mathcal{M} \sim \{ M_{k} \}_{k=1}^{m}$. Without loss of generality, $M_{k} \neq 0$ for all $k \in [m]$ and $R_{\ell} \neq 0$ for all $\ell \in [r]$. Notice that for each unit vector $\ket{\alpha} \in \mathcal{H}$, 
    \begin{align}
        \ket{\alpha}\bra{\alpha} = \sum_{\ell = 1}^{r} \sum_{k = 1}^{m} R_{\ell} M_{k} \ket{\alpha}\bra{\alpha} M_{k}^* R_\ell^*.
    \end{align}
    This is a sum of positive semi-definite operators that equals a rank-1 projector. Hence, for all $\ell \in [r]$ and $k \in [m]$, there exists a complex coefficient $c_{k, \ell, \alpha}$ such that $R_{\ell} M_{k} \ket{\alpha} = c_{k, \ell, \alpha} \ket{\alpha}$. This implies that the operator $R_{\ell} M_{k}$ is normal and diagonal in every basis. Hence, it must be proportional to the identity. That is, $ c_{k, \ell, \alpha}$ does not depend on $\alpha$. Therefore, for each $k \in [m]$, $M_{k}$ is proportional to an inverse of $R_{1}$. Since inverses are unique, $M_{k} \propto M_{k'}$ for all $k, k' \in [m]$.  \end{proof}

\begin{prop}
\label{prop: single qudit recovery is deterministic}
If $S \subseteq [n]$ is a singleton, then $\gamma_{S} \in \{0,1\}$.
\end{prop}
\begin{proof}
 Without loss of generality, let $S = \{1\}$. If $\gamma_{1} > 0$, then the channel $\mathcal{M}_{1} := \tr_{ [n] \setminus \{1\}} \circ \; \mathcal{U}_{\mathcal{C}} : \mathcal{B} (\mathcal{H}) \rightarrow \mathcal{B} (\mathcal{H})$ satisfies the conditions in Eq.~\ref{eq: cond error correction cond} for some nonzero recovery operator $\Pi \in \mathcal{B} (\mathcal{H} )$, where the code in this instance is $\mathcal{H}$. $\mathcal{M}_{1}$ has a completely positive inverse. Since it is also trace-preserving, then it is unitary by Lem~\ref{lem: cp reovery for authomorphisms}. Hence, $\gamma_{1} = 1$ and $\mathcal{C} = \mathcal{H} \otimes \mathbb{C} \ket{\alpha}$ for some $\ket{\alpha} \in \mathcal{H}^{\otimes n-1}$.       
\end{proof}

We can use this proposition along with Prop.~\ref{prop: partitioning no-cloning.} to show that erasure channels with erasure probability at least a half cannot be improved with four uses. In particular, for an encoding into $4$ qudits where the information is encoded into one qudit, then the optimal recovery probability is exactly $(1 - q)$. Otherwise, for $i \in [4]$, $\gamma_{i} = 0$ by Prop.~\ref{prop: single qudit recovery is deterministic}. In such a case, the probability of recovery can be bounded from above as follows 

\begin{align}
 \label{eq: probability of recovery for 4 qudits}
r_{4} (q) &= q^4 (0) + q^{3} (1 - q) (0) + q^{2} (1 - q)^{2} (3) + q (1-q)^3 (4) + (1-q)^{4}, \\
 &= (1-q)^2 (3 q^2 + 4 q (1-q) + (1-q)^2)
\end{align}

where we used the fact that $\gamma_{\emptyset} = 0$, the fact that $\gamma_{i} = 0$, and Prop.~\ref{prop: partitioning no-cloning.} to bound from the above the coefficient of $q^{2} (1 - q)^{2}$. For the last point, there are $3$ distinct partitions of a set of $4$ elements into two sets of two elements each. The zeros of the polynomial $r_{4} (q) - (1-q)$ are $0$, $\frac{1}{2}$, and $1$. It is negative when $q$ is close to $1$. Hence, it is non-positive on $[\frac{1}{2}, 1]$. This implies the following corollary.
\begin{cor}
    For $n \leq 4$ and $q \in [\frac{1}{2}, 1]$, if $\mathcal{R} \circ \mathcal{E}_{q}^{\otimes n} \circ \mathcal{U}_{\mathcal{C}} = \mathcal{E}_{p}$, then $p \geq q$. 
\end{cor}

In order to progress further, we consider the problem for a special class of quantum error correcting codes. We say that a code is \textit{deterministic} if the associated tuple of recovery probabilities $( \gamma_{S} )_{S \subseteq [n]}$ satisfies $\gamma_{S} \in \{0, 1\}$ for all erasure patterns $S^c \subseteq [n]$. In \cite{hastings2014notes}, Hastings considers a slightly different restriction. He restricted the decoding procedure to output an error flag whenever an erasure pattern $S^c \subseteq [n]$ occurs where the probability of recovery is less than 1. Restricting the encoding procedure is more natural as the optimal decoding procedure is known for any given encoding (see Prog.~\ref{prog: semidefinite}). Moreover,  the class of deterministic codes includes ubiquitously studied codes such as stabilizer codes.

\begin{prop}
\label{prop: stab is det}
Let $\mathcal{C} \leq \mathcal{H}^{\otimes n}$ be a stabilizer code. If $\gamma_{S} < 1$, then $\gamma_{S} = 0$.  
\end{prop}
\begin{proof}
Let $G_{\mathcal{C}}$ denote the stabilizer group of the code. Since $\gamma_{S} < 1$, the code $\mathcal{C}$ is not $(\tr_{S^c}, 1)$-correcting. This implies that there is an element of $\text{Centraliser}(G_{\mathcal{C}}) \setminus G_{\mathcal{C}}$ of the form $\mathds{1}_{S} \otimes E_{S^{c}}$. This is a non-identity logical operator for $\mathcal{C}$. Hence, there exist two distinct $\ket{\phi}, \ket{\psi} \in \mathcal{C}$ such that $\ket{\phi} = \mathds{1}_{S} \otimes E_{S^{c}} \ket{\psi}$. This implies
\begin{align}
\tr_{S^{c}} (\ket{\phi} \bra{\phi}) \otimes (\frac{\mathds{1}}{d})^{\otimes S^c} &\propto 
\int \mu_{U} \mathds{1}_{S} \otimes U_{S} \ket{\phi} \bra{\phi} \mathds{1}_{S} \otimes U_{S}^* \\
&=  \int \mu_{U} \mathds{1}_{S} \otimes U_{S} E_{S} \ket{\psi} \bra{\psi}   \mathds{1}_{S} \otimes (U_{S} E_{S})^*\\
    &= \: \int \mu_{U}  \mathds{1}_{S} \otimes U_{S} \ket{\psi} \bra{\psi} \mathds{1}_{S} \otimes U_{S}^*,
\end{align}
where the $\mu_{U}$ is the Haar measure of the unitary group on the systems in $S$. Therefore, $\tr_{S^{c}} \otimes (\frac{\mathds{1}}{d})^{\otimes S^c}$ is not one-to-one on $\mathcal{C}$ and so $\gamma_{S} = 0$. \end{proof}

For a code $\mathcal{C}$, denote the contributing family $\mathcal{F}_{\mathcal{C}} := \{ S \subseteq [n] \; \big| \; \gamma_{S} (\mathcal{C}) > 0 \}$. From Prop.~\ref{prop: partitioning no-cloning.}, the family $\mathcal{F}_{\mathcal{C}, 1} := \{ S \subseteq [n] \; \big| \; \gamma_{S} (\mathcal{C}) = 1 \}$ is intersecting. Moreover, any $T \in \mathcal{F}_{\mathcal{C}}$ must intersect with every member of $\mathcal{F}_{\mathcal{C}, 1}$. For a deterministic code $\mathcal{C}$, $\mathcal{F}_{\mathcal{C}} = \mathcal{F}_{\mathcal{C}, 1}$. The following two propositions are based on the observation that intersecting families of small sets cannot have too many members. This is made precise by the Erdös-Ko-Rado theorem \cite{EKR_1961} for intersecting families. It says that an intersecting family of $\ell$-sets, where $\ell \leq \lfloor \frac{n}{2} \rfloor$, has at most $\binom{n - 1}{\ell - 1}$ members.

 \begin{prop}
 \label{prop: k erasure of n qudits}
 Let $\mathcal{C} \leq \mathcal{H}^{\otimes n}$ be a deterministic code. For $\ell \in [n]$, consider the channel 
 \begin{align}
 \label{eq: equiprobablity erasure}
 \mathcal{N}_{\ell} (\cdot) = \frac{1}{\binom{n}{\ell}} \sum_{S \in \binom{[n]}{\ell}} \tr_{S^{c}} (\cdot) \otimes \ket{e}\bra{e}^{\otimes S^{c}}.
 \end{align}
 If $\ell \leq \lfloor \frac{n}{2} \rfloor$ and $\mathcal{C}$ is $(\mathcal{N}_{\ell}, t)$-correcting, then $t \leq \frac{\ell}{n}$.
 \end{prop}
 \begin{proof}
     Since $\mathcal{F}_{\mathcal{C}}$ is intersecting and $\ell \leq \lfloor \frac{n}{2} \rfloor$, 
     $\frac{1}{\binom{n}{\ell}} \sum_{S \in \binom{[n]}{\ell}} \gamma_{S} \leq \frac{1}{\binom{n}{\ell}} \binom{n-1}{\ell-1} = \frac{\ell}{n}.$ \end{proof}

By appealing to a typicality argument, we can lift this result to show a similar upper bound for $\mathcal{E}_{q}^{\otimes n}$. 
\begin{thm}
\label{thm: deterministic codes do not help}
Let $(\mathcal{C}_{n} ) _{n \in \mathbb{N}}$ be a sequence of deterministic codes. If $q \in (\frac{1}{2}, 1)$, then the asymptotic recovery probability is bounded from above as follows
\begin{align}
    \lim_{n \rightarrow \infty} \sum_{S \subseteq [n]}  q^{n - |S|} (1 - q)^{|S|} \gamma_{S} (\mathcal{C}_{n}) \leq (1 - q).
\end{align}

\end{thm}

\begin{proof}
Set $\delta = \frac{1}{2} ( \frac{1}{2} - (1 - q))$ and consider the intersecting families 
\begin{align}
\mathcal{F}_{\mathcal{C}_{n}} ^{(\ell)}=\mathcal{F}_{\mathcal{C}_{n}} \cap \binom{[n]}{\ell}, 
\end{align}
where $\ell \in T^{(n)}_{1-q, \delta}$ (see Def~\ref{def: typical weights}). Denote the atypical summand of the recovery probability by $\varepsilon(n)$ and consider
\begin{align}
\sum_{S \subseteq [n]}  q^{n - |S|} (1 - q)^{|S|} \gamma_{S}  &= \sum_{\ell \in T^{(n)}_{1-q, \delta}} \sum_{|S| = \ell} q^{n - |S|} (1 - q)^{|S|} \gamma_{S} + \varepsilon(n) \\
&=
\sum_{\ell \in T^{(n)}_{1-q, \delta}} \sum_{S \in \mathcal{F}_{\mathcal{C}_{n}} ^{(\ell)}} q^{n - |S|} (1 - q)^{|S|} + \varepsilon(n) \\
& \leq \sum_{\ell \in T^{(n)}_{1-q, \delta}}  q^{n - \ell} (1 - q)^{\ell} \binom{n - 1}{\ell - 1} + \varepsilon(n) \\
& \leq (1 - q) + \varepsilon(n).
\end{align}
The claim follows from the fact that $\lim_{n \rightarrow \infty} \varepsilon(n) = 0$. \end{proof}

\section{Concluding remarks}
\label{ch5: sec 7}
We have provided a systematic study of exact simulation of quantum erasure channels. We showed that it is closely related to the notion of probabilistic recovery of quantum information. We showed that the best erasure channel simulable by a given resource channel and a given code can be ascertained by solving a simple semi-definite program Prog.~\ref{prog: semidefinite}. We showed that Conj.~\ref{conj: Hastings} holds in the case where $n \leq 4$ and the case where the encoding channel is an isometric encoding into some deterministic code. 

We speculate about a possible path towards a resolution of Conj.~\ref{conj: Hastings}. Suppose that for a code $\mathcal{C} \leq \mathcal{H}^{\otimes n}$ it holds that $\gamma_{S} + \gamma_{S^{c}} = 1$ for some $S \subseteq [n]$. If both $\gamma_{S}$ and $\gamma_{S^{c}}$ are nonzero, then for $\ket{c} \in \mathcal{C}$, it holds that $\ket{c} = \ket{c_{S}} + \ket{c_{S^{c}}}$, where $\ket{c_{S}}$ is contained in the support of $\Pi_{S}$ and $\ket{c_{S^{c}}}$ is contained in the support of $\Pi_{S^{c}}$. The following is a simple example of such an encoding.
\begin{align}
 \label{eq: example of non det code}
 \mathcal{H} \ni \ket{\alpha} \mapsto  \frac{1}{\sqrt{2}} ( \ket{\alpha} \otimes \ket{0} \otimes \ket{0} \otimes \ket{0} + \ket{0} \otimes \ket{1} \otimes \ket{\alpha} \otimes \ket{1}),
\end{align}
where the second and fourth qudit carry the classical information about where the quantum information was encoded. The probability of recovery is zero if both of these qubits are erased. In a code that is not deterministic, classical information is encoded along with every erasure pattern where recovery is not deterministic. In the example above, there's a whole bit encoded along with the quantum information. Perhaps there is not enough room for both the quantum and the classical information if there are too many such erasure patterns. Even more simply, can we bound from above the cardinality of the set 
$\{ S \in \binom{[n]}{k} \; \big| \;  \gamma_{S} > 0 \}$ for $k \leq \lfloor \frac{n}{2} \rfloor$ by appealing to a packing argument? It is clear from the example above that it can exceed $\binom{n - 1}{k - 1}$. Still, an upper bound might come with some insights that would aid in proving Conj.~\ref{conj: Hastings}.

%%%%%%%%%   then the Bibliography, if any   %%%%%%%%%
\bibliographystyle{plain}	% or "siam", or "alpha", etc.
\nocite{*}		% list all refs in database, cited or not
\bibliography{refs}		% Bib database in "refs.bib"

%%%%%%%%%   then the Appendices, if any   %%%%%%%%%
%\appendix

\end{document}